\documentclass[a4paper,fleqn,usenatbib]{mnras}

\usepackage[T1]{fontenc}

\usepackage{graphicx}	
\usepackage{amsmath}	
\usepackage{amssymb}	
\usepackage{pdflscape}
\usepackage{threeparttable}

\newcommand       \mum          {\,{\rm \mu m}}

\newcommand       \simali       {{\sim}\,}

\newcommand       \simlt        {\lesssim}

\newcommand       \gtsim        {\gtrsim}

\newcommand       \g            {\,{\rm g}}
\newcommand       \Msun         {\,{M_\odot}}
\newcommand       \cm           {\,{\rm cm}}

\newcommand       \km           {\,{\rm km}}
\newcommand       \s            {\,{\rm s}}
\newcommand       \yr           {\,{\rm yr}}

\newcommand       \K            {\,{\rm K}}
\newcommand       \Mloss        {\dot{M}}
\newcommand       \Teff         {T_\star}
\newcommand       \Lstar        {L_{\star}}
\newcommand       \rmin         {r_{\rm min}}
\newcommand       \rmax         {r_{\rm max}}

\newcommand       \Mdustloss  {\dot{M}_{\rm dust}}

\newcommand       \fmc  {\eta_{\rm csi,m}}
\newcommand       \ffc  {\eta_{\rm csi,f}}
\newcommand       \ffcp {\eta^{\prime}_{\rm csi,f}}
\newcommand       \Mcsi {M_{\rm csi}}
\newcommand       \Masi {M_{\rm asi}}
\newcommand       \Fcsi {P_{\rm csi}}
\newcommand       \Fasi {P_{\rm asi}}
\newcommand       \Fcont {P_{\rm con}}

\newcommand       \gastodust {M_{\rm gas}/M_{\rm dust}}
\newcommand       \lambdamean {\langle \lambda \rangle}
\newcommand       \lambdamin {\lambda_{\rm min}}
\newcommand       \lambdamax {\lambda_{\rm max}}
\newcommand       \gammamean {\langle \gamma\lambda \rangle}
\newcommand       \gammamin {(\gamma\lambda)_{\rm min}}
\newcommand       \gammamax {(\gamma\lambda)_{\rm max}}
\title[]{On the Silicate Crystallinities of Oxygen-Rich Evolved Stars
and Their Mass Loss Rates}

\author[Liu, Jiang, Li \& Gao]{
Jiaming Liu$^{1,2}$,
B.W.~Jiang$^{1}$\thanks{bjiang@bnu.edu.cn},
Aigen Li$^{2}$\thanks{lia@missouri.edu}
and Jian Gao$^{1}$
\\
$^{1}$Department of Astronomy, Beijing Normal University, Beijing 100875, China\\
$^{2}$Department of Physics and Astronomy, University of Missouri, Columbia, MO 65211, USA
}

\date{Accepted XXX. Received YYY; in original form ZZZ}

\pubyear{2016}

\begin{document}
\label{firstpage}
\pagerange{\pageref{firstpage}--\pageref{lastpage}}
\maketitle

\begin{abstract}
For decades ever since the early detection in the 1990s
of the emission spectral features of crystalline silicates
in oxygen-rich evolved stars, there is a long-standing debate
on whether the crystallinity of the silicate dust
correlates with the stellar mass loss rate.
To investigate the relation between the silicate
crystallinities and the mass loss rates of evolved stars,
we carry out a detailed analysis of
28 nearby oxygen-rich stars.
%
We derive the mass loss rates of these sources
by modeling their spectral energy distributions
from the optical to the far infrared.
Unlike previous studies in which the silicate crystallinity
was often measured in terms of the crystalline-to-amorphous
silicate mass ratio,
we characterize the silicate crystallinities
of these sources with the flux ratios of
the emission features of crystalline silicates
to that of amorphous silicates.
This does not require the knowledge of
the silicate dust temperatures
which are the major source of uncertainties
in estimating the crystalline-to-amorphous
silicate mass ratio.
With a Pearson correlation coefficient of $\simali$$-0.24$,
we find that the silicate crystallinities
and the mass loss rates of these sources
are not correlated.
This supports the earlier findings
that the dust shells of low mass-loss rate stars
can contain a significant fraction of
crystalline silicates
without showing the characteristic features
in their emission spectra.
\end{abstract}




\section{Introduction}

Astronomical silicates consist predominantly of
silicate minerals made up of cations and silicic
acid radical ions (SiO$_{4}^{4-}$) or (SiO$_{3}^{2-}$),
with the main metals being
magnesium (Mg) and iron (Fe).
According to their chemical structure,
silicates can be classified as olivine
(Fe$_{2x}$Mg$_{2(1-x)}$SiO$_{4}$) and
pyroxene (Mg$_{(1-x)}$Fe$_{x}$SiO$_{3}$)
where $0 \leq x \leq1$.
Meanwhile, silicates can also be divided into
crystalline (ordered structure) and
amorphous (unordered structure)
based on their lattice structure.
Laboratory studies show that both amorphous
and crystalline silicates resonate
in the infrared (IR)
due to the Si--O stretching
and O--Si--O bending modes originating
from the silica tetrahedra.
These vibrational modes of silicates
dominate the emission or absorption spectra
of oxygen-rich stars.
More specifically, amorphous silicates reveal
their presence in O-rich stars through
the broad, smooth and featureless bands
at 9.7 and 18$\mum$
arising from the Si--O stretch
and the O--Si--O bend, respectively,
while crystalline silicates exhibit
various distinct narrow sharp bands
at $\simali$10--60$\mum$
(Henning 2010, Molster \& Kemper 2005,
Liu \& Jiang 2014).

The 9.7 and 18$\mum$ amorphous silicate features
of evolved stars were first detected
in emission in M stars
(Woolf \& Ney 1969, Treffers \& Cohen 1974)
and then in absorption in heavily obscured stars
with an extended circumstellar dust shell
(Jones \& Merrill 1976, Bedijn 1987).
The detection of crystalline silicates in evolved stars
was first made by Waters et al.\ (1996) who reported
the $\simali$12--45$\mum$ emission spectra of
six oxygen-rich evolved stars obtained with
the {\it Short Wavelength Spectrometer} (SWS)
on board the {\it Infrared Space Observatory} (ISO).
Thanks to {\it ISO}/SWS
and the {\it Infrared Spectrograph} (IRS)
aboard the {\it Spitzer Space Telescope},
crystalline silicates have now been seen in
all evolutionary stages of evolved stars:
red giants and supergiants,
asymptotic giant branch (AGB) stars,
post-AGB stars, planetary nebulae (PNe),
and luminous blue variable (LBV) stars
(see Jiang et al.\ 2013 and references therein).

Molster et al.\ (2002a,b,c) systematically investigated
a sample of 17 evolved stars based on the {\it ISO}/SWS
and LWS ({\it Long Wavelength Spectrometer}) spectra
in the $\simali$2.4--195$\mum$ wavelength range.
They identified about 50 narrow bands of crystalline silicates,
with distinct emission complexes
at approximately 10, 18, 23, 28, 33.6, 40 and 69$\mum$.
The strengths and peak wavelengths of these features
of crystalline silicates are experimentally shown
to be affected by their iron (Fe) contents
(e.g., Koike et al.\ [2003] found that the 33.6$\mum$
feature of crystalline Mg$_2$SiO$_4$ becomes weaker
and its peak wavelength shifts to longer wavelength
up to $\simali$38.9$\mum$ as the content of
Fe elements increases to 100\% [i.e., Fe$_2$SiO$_4$];
Olofsson et al.\ [2012] found that,
with a tiny increase of 5\% of Fe content
from Mg$_2$SiO$_4$ to Mg$_{1.9}$Fe$_{0.1}$SiO$_4$,
the 69$\mum$ crystalline olivine feature shifts to
a peak wavelength longer than 70$\mum$).
The sensitivity of the peak wavelengths and strengths of
the crystalline silicate spectral features to the Fe content
convincingly shows that the crystalline silicate minerals
in evolved stars are nearly Fe free
as they very often show the features at 33.6 and 69$\mum$,
i.e.,  the dominant crystalline silicate species
in evolved stars are Mg$_{2}$SiO$_{4}$ and MgSiO$_{3}$
(Molster \& Kemper 2005, Henning 2010,
Sturm et al.\ 2013, Jiang et al.\ 2013, Blommaert et al. 2014).
%
%

The mechanism of silicate crystallization
in evolved stars is still unclear.
Gail \& Sedlmayr (1999) modeled
the dust condensation in the circumstellar envelopes
of O-rich mass-losing stars.
They  found that, due to its high condensation temperature,
Fe-free crystalline olivine could directly condense
out of the stellar winds.
Alternatively, a conversion from amorphous form
through thermal annealing has also been frequently
invoked to explain the presence of crystalline silicates
in evolved stars
(Molster \& Kemper 2005, Henning 2010, Liu \& Jiang 2014).

Apparently, the possible correlation of the silicate
crystallinity with some stellar parameters would shed
light on the origin of crystalline silicates in evolved stars
and their crystallization mechanism.
By silicate crystallinity we mean the mass fraction
of silicate dust in crystalline form
(e.g., see Jiang et al.\ 2013).
Egan \& Sloan (2001) speculated that
the compositions of the dust condensed
in the shells around evolved stars
could be related to their mass loss history
(e.g., episodic vs. continuous) and the structure
of the shells (e.g., geometrically thin vs.
geometrically thick).
Molster et al.\ (2002c) argued that the silicate
crystallinity of evolved stars is usually around
10--15\% and can be much higher in systems
with the presence of a disk
(e.g., IRAS~09425-6040, a post-AGB star
with a disk around it,
has a crystallinity of $\simali$60--80\%,
see Molster et al.\ 2001).
Jiang et al.\ (2013) reported a silicate crystallinity
of $\simali$97\% in IRAS~16456-3542, a planetary nebula,
the highest to date ever reported for
crystalline silicate sources.
%
%

The stellar mass loss rates $\Mloss$
of the mass-losing evolved stars
are often considered to be one of
the most important factors which
determine whether or not crystalline silicates
would be present in their circumstellar envelopes.
Theoretical calculations have shown that amorphous
silicates cannot be crystallized in stars
with a low $\Mloss$ because the dust cannot be heated
to temperatures high enough for crystallization,
and that crystalline silicates can only form in
stars undergoing substantial mass losses
with a critical value of $\Mloss\gtsim10^{-5}\Msun\yr^{-1}$
and having high dust column densities
(e.g., see Tielens et al.\ 1998,
Gail \& Sedlmayr 1999, Sogawa \& Kozasa 1998).
The early detection of crystalline silicates
in evolved stars by Molster et al.\ (2002,a,b,c)
appeared to be consistent with this argument:
those with crystalline silicates detected all
have rather high mass loss rates,
i.e., $\Mloss > 10^{-5}\Msun\yr^{-1}$.
More recently, Jones et al.\ (2012) analyzed
the {\it Spitzer}/IRS spectra of 315 evolved stars
and found that the mass loss rates of the stars
exhibiting the crystalline silicate features
at 23, 28 and 33$\mum$ span over 3 dex,
down to $\simali$$10^{-9}\Msun\yr^{-1}$,
although for most of the stars
$\Mloss > 10^{-6}\Msun\yr^{-1}$.
Jones et al.\ (2012) investigated the
 possible correlation between $\Mloss$
and the silicate crystallinity
by examining the relation of $\Mloss$ with
the strengths of the 23, 28 and 33$\mum$ features
measured as their equivalent widths.
They found no correlation, except a general tendency
that stars with a high mass loss rate would have a higher
probability of displaying crystalline silicate features.
Kemper et al.\ (2001) performed an extensive
radiative transfer calculations of the model
IR emission spectra for O-rich AGB stars of
mass loss rates ranging from
$\Mloss = 10^{-7}\Msun\yr^{-1}$
to $10^{-4}\Msun\yr^{-1}$
and of a wide range of crystallinities
up to 40\%. They found that crystallinity
is not necessarily a function of mass-loss rate.
They argued that,
due to the temperature difference between
amorphous and crystalline silicates which is
caused by the lower visual/near-IR absorptivity
of the latter because of its very low Fe content,
it is possible to allow for a crystallinity of
up to 40\% in the circumstellar dust shell,
without its IR emission spectra showing
the characteristic spectral features of
crystalline silicate minerals.

In this work we revisit the relation
between $\Mloss$ and the silicate
crystallinities of evolved stars
by taking an alternative approach:
unlike previous studies in which
the silicate crystallinity was often
measured in terms of the crystalline-to-amorphous
silicate mass ratio,
we characterize the silicate crystallinities
with the flux ratios of
the emission features of crystalline silicates
to that of amorphous silicates.
This does not require the knowledge of
the silicate dust temperatures
which are the major source of uncertainties
in estimating the crystalline-to-amorphous
silicate mass ratio.
This paper is organized as follows:
\S\ref{sec:sample} describes the sample of
28 nearby O-rich stars selected for
this $\Mloss$--crystallinity correlation study.
In \S\ref{sec:SED} we model
the spectral energy distributions (SEDs) of
the selected sources from the optical to the far-IR
and derive their mass loss rates.
We derive the silicate crystallinity in \S\ref{sec:CSi}.
The results are presented and discussed
in \S\ref{sec:CSi_vs_dotM} and
summarized in \S\ref{sec:summary}.

\section{Sample Stars}
\label{sec:sample}
We select a sample of 28 O-rich evolved stars
(see Table~\ref{tab:fluxdata})
based on the following criteria:
(1) the {\it ISO}/SWS spectra of
     most of these sources exhibit
    prominent crystalline silicate emission features;
(2) they exhibit distinguished 10 and 18$\mum$
    amorphous silicate emission features;
(3) they are relatively ``local'' 
    with a distance of less than 5\,kpc
    to remove the influence of extinction
    and metallicity.\footnote{%
       In an examination of the {\it Spitzer}/IRS
       and {\it ISO}/SWS spectra of
       217 O-rich AGB stars and 98 red supergiants,
       Jones et al.\ (2012) noticed a possible change
       of crystalline silicate mineralogy with metallicity,
       with enstatite seen increasingly at low metallicity
       while forsterite becomes depleted.
       }

The {\it ISO}/SWS spectra are taken from the archive
and the $\simali$2.4--45$\mum$ wavelength range covers
6 out of 7 of the crystalline silicate feature complexes
summarized by Molster et al.\ (2002a),
i.e., the 10, 18, 23, 28, 33.6 and 40$\mum$ complexes
except the 69$\mum$ complex.
 These archival data have already been processed
by Sloan et al.\ (2003) in a uniform manner.

\subsection{Mass Loss Rates}
\label{sec:SED} 

To derive the mass loss rate of each source,
we employ the ``2-DUST'' radiative transfer code
of Ueta \& Meixner (2003) to model its SED
from the optical to the far-IR.
Although the ``2-DUST'' code was developed
for dusty axisymmetric systems
and is capable of dealing with
layered dust shells formed
during the AGB mass-loss phase
and the subsequent post-AGB superwind phase,
it can be simplified so that it is also applicable
to spherical shells around AGB stars.\footnote{%

   Ueta \& Meixner (2003) designed a 2-dimensional
   density distribution which has
   (i) a spherical outer shell
       -- the remnant of the AGB wind,
   (ii) a spheroidal mid-region, and
   (iii) an inner toroidal core
   created during the superwind phase
   -- a rather brief period of
   equatorially-enhanced mass-loss
   near the end of the AGB mass-loss phase.
   Such a density distribution
   is a function of the radius
   of the dust shell $r$,
   the latitudinal angle $\Theta$,
   and five geometric parameters
   ($A$, $B$, $C$, $D$, and $E$):
{\tiny
\begin{eqnarray}
\nonumber
\rho(r,\Theta) & = &
\rho_{\rm min}\left(r/r_{\rm min}\right)^{-B\left\{1+C\sin^{F}{\Theta}
   \times\exp\left[-\left(r/r_{\rm sw}\right)^{D}\right]/
   \exp\left[-\left(r_{\rm min}/r_{\rm sw}\right)^{D} \right]\right\}}
   \nonumber \\
&& \times
\left\{1+A\left(1-\cos\Theta\right)^{F}
   \times \exp\left[-\left(r/r_{\rm sw}\right)^{E}\right]/
   \exp\left[-\left(r_{\rm min}/r_{\rm sw}\right)^{E} \right]\right\} ~~,
\end{eqnarray}
}
  where $\rho(r,\Theta)$ is the dust mass density
  at radius $r$ and latitude $\Theta$,
  $\rho_{\rm min}$ is the dust mass density on
  the polar axis at the inner edge of the shell,
  $r_{\rm min}$ is the inner radius of the shell,
  $r_{\rm max}$ is the outer radius of the shell,
  $r_{\rm sw}$ is the radius of the superwind
  between $r_{\rm min}$ and the AGB wind
  which defines the ``thickness'' of the inner,
  axisymmetric region of the shell.
  By simply setting $A=C=D=E=F=0$ and $B=2$
  we obtain a density function of
  $\rho(r) = \rho_{\rm min}\left(r/r_{\rm min}\right)^{-2}$
  for AGB stars
  which have a spherical shell
  and undergo constant outflow.
  }
In this work we are mainly concerned with stars
which have not yet evolved to the post-AGB phase.
Their circumstellar shells are commonly assumed
to be spherical. If we assume a constant outflow,
the dust density function has a simple form of
$\rho(r) = \rho_{\rm min}\left(r/r_{\rm min}\right)^{-2}$,
where $\rho(r)$ is the dust mass density at radius $r$,
$\rho_{\rm min}$ is the dust mass density
at the inner edge of the shell, and
$r_{\rm min}$ is the inner radius of the shell.
Therefore, the dust shell is described by
three parameters: $\rho_{\rm min}$,
$\rmin$, and the outer radius of the shell $\rmax$.
The total dust mass in the shell is given by
$M_{\rm dust} = 4\pi\rho_{\rm min}\rmin^2\left(\rmax-\rmin\right)$.

The input stellar parameters required by
the 2DUST code are the stellar effective temperature $\Teff$
and the stellar luminosity $\Lstar$
(see Table~\ref{tab:sedmod}).
For each source, we treat $\Teff$ and $\Lstar$
as free parameters but requiring
$2,000 \simlt \Teff \simlt 4,000\K$
and $10^{3} \simlt \Lstar/L_\odot \simlt 10^{6}$
which are reasonable values for AGB stars
(see Sargent et al.\ 2011).

For the dust composition, we only consider
amorphous silicate for accounting for
the 9.7 and 18$\mum$ amorphous silicate
emission features and the continuum emission.
We take the optical constants of amorphous olivine
MgFeSiO$_4$ measured by Dorschner et al.\ (1995).
We adopt a mass density of
$\rho_{\rm sil} = 3.20\g\cm^{-3}$
for amorphous MgFeSiO$_4$.
For the dust size distribution,
we take a MRN-type power-law
distribution function of $dn/da \propto a^{-3.5}$
(Mathis et al.\ 1977)
for $a_{\rm min} < a < a_{\rm max}$,
where $a$ is the spherical radius of the dust
(we assume the dust to be spherical)
with a lower and upper cutoff of
$a_{\rm min}=0.01\mum$ and
$a_{\rm max}=1\mum$
(see Sargent et al.\ 2010).
%

\begin{figure}
\centering
\includegraphics[width=0.65\textwidth,bb=50 150 400 600]{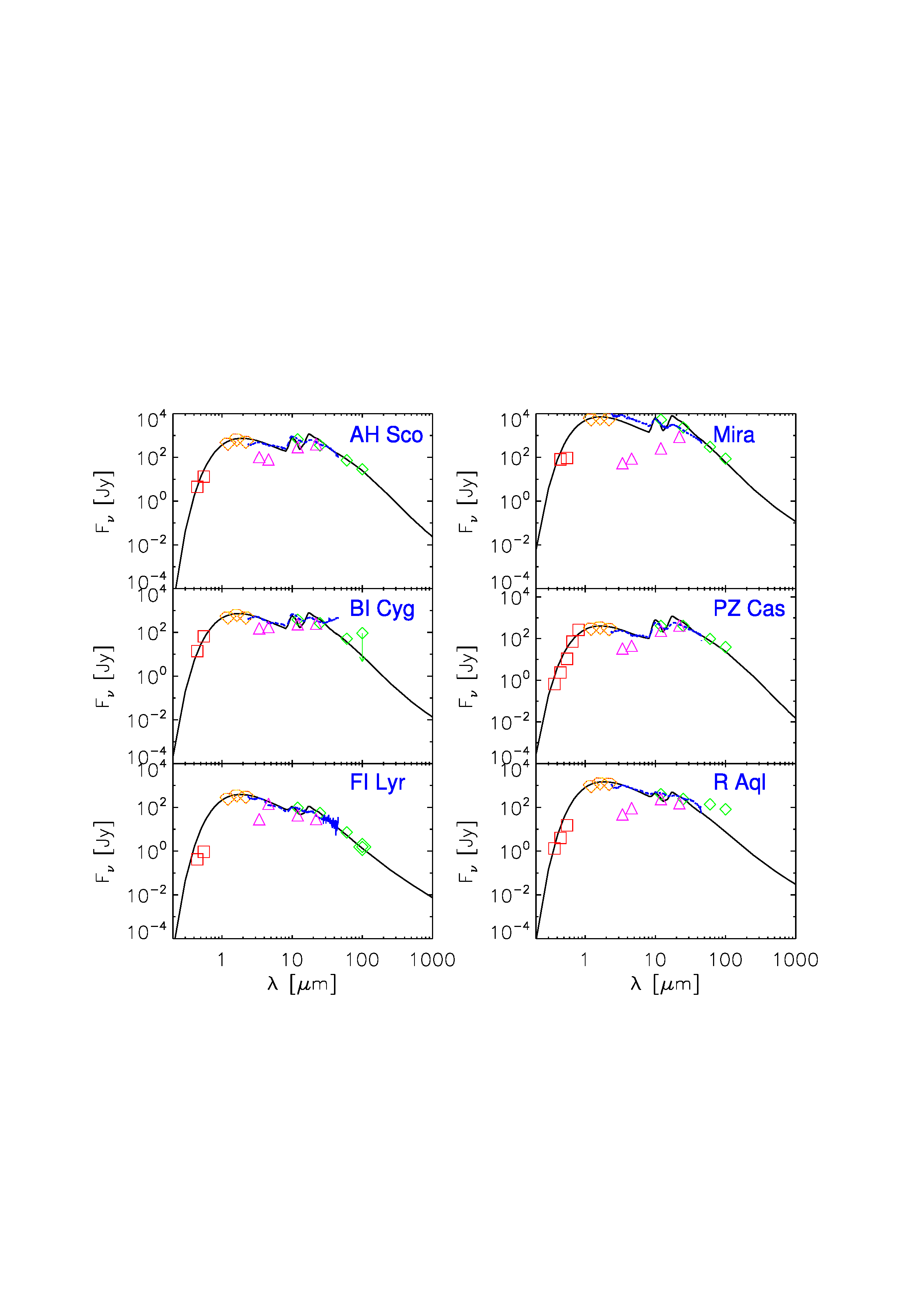}
\caption{Comparison of the model SEDs (black lines)
         calculated from 2DUST
         with the {\it ISO}/SWS spectra (blue lines),
         and the Johnson {\it UBVRI} photometry (red rectangles),
         the JHK {\it 2MASS} photometry (orange diamonds),
         the {\it WISE} photometry (pink triangles)
         as well as the {\it IRAS} photometry (green rhombuses)
         for AH Sco, BI Cyg, FI Lyr, FP Aqr, PZ Cas and R Aql.
         Green arrow means that the photometric flux
         is an upper limit,
         while doubled rhombus means that the photometry
         has a large uncertainty.
         }
\label{fig:sedmod1}
\end{figure}

The input photometric data (see Table~\ref{tab:fluxdata})
are compiled from the literature,
including the Johnson {\it UBVRI} photometry,
the $J$ (1.22\,$\mu$m), $H$ (1.63\,$\mu$m),
$K$ (2.19\,$\mu$m) {\it 2MASS} photometry
({\it 2MASS}),
the 4-band {\it WISE} photometry
(Cutri et al.\ 2012) at W1 (3.4\,$\mu$m),
W2 (4.6\,$\mu$m), W3 (12.0\,$\mu$m),
and W4 (22.0\,$\mu$m),
and the {\it Infrared Astronomical Satellite}
(IRAS) photometry at 12, 25, 60 and 100$\mum$
(Beichman et al.\ 1988). These broadband
photometric data from the near-ultraviolet
to the far-IR are supplemented
to the {\it ISO}/SWS spectra
(see Figure~\ref{fig:sedmod1}).

To correct for the interstellar extinction along
the line of sight toward the stars considered here,
we assume the wavelength-dependence of the extinction
to be that of the Galactic average extinction law of
$R_V=3.1$, where $R_V$ is the total-to-selective
extinction ratio. Let $A_V$ be the visual extinction
and $A_\lambda$ be the extinction at wavelength $\lambda$
(which corresponds to frequency $\nu =c/\lambda$
where $c$ is the speed of light).
We restore the unobscured, ``true'' flux density $F_\nu$
from the observed, reddened flux density $F_\nu^{\rm obs}$
as follows:
\begin{equation}
F_\nu = F_\nu^{\rm obs}
\exp\left\{\frac{A_V}{1.086}
\left(\frac{A_\lambda}{A_V}\right)\right\} ~~,
\end{equation}
where $A_\lambda/A_V$ is taken to
be the $R_V=3.1$ extinction law
(see Cardelli, Clayton \& Mathis 1989).
We search for the visual extinction $A_V$
for each source
from the {\it NASA/IPAC Extragalactic Database} ({\it NED}).
If not available or uncertain in {\it NED}(when the Galactic latitude is larger than 10 degree),
we derived $A_V$ from the observed color index
of each source:
\begin{equation}
A_V = E(J-K) \left[\frac{A_V}{E(J-K)}\right]~~,
\end{equation}
where $E(J-K)\equiv A_J-A_K$ is the color excess
between the $J$ and $K$ bands
and $A_V/E(J-K)\approx 5.88$
is the extinction-to-color excess ratio
(see Wang, Li \& Jiang 2015).
For a given source, the color excess is
obtained from
\begin{equation}
E(J-K) = \left(J-K\right)_{\rm obs}
       - \left(J-K\right)_{\rm int} ~~,
\end{equation}
where $\left(J-K\right)_{\rm obs}\equiv J_{\rm obs}-K_{\rm obs}$
and $\left(J-K\right)_{\rm int}$ are
respectively the observed and intrinsic
color indices of the stars,
and $J_{\rm obs}$ and $K_{\rm obs}$ are
respectively the {\it 2MASS} $J$ and $K$
photometric fluxes ({\it 2MASS}).
In Table~\ref{tab:Av}
we list the visual extinction for each source,
obtained either from {\it NED}
or from the observed color index.

\begin{figure}
\centering
\includegraphics[width=0.65\textwidth,bb=50 150 400 650]{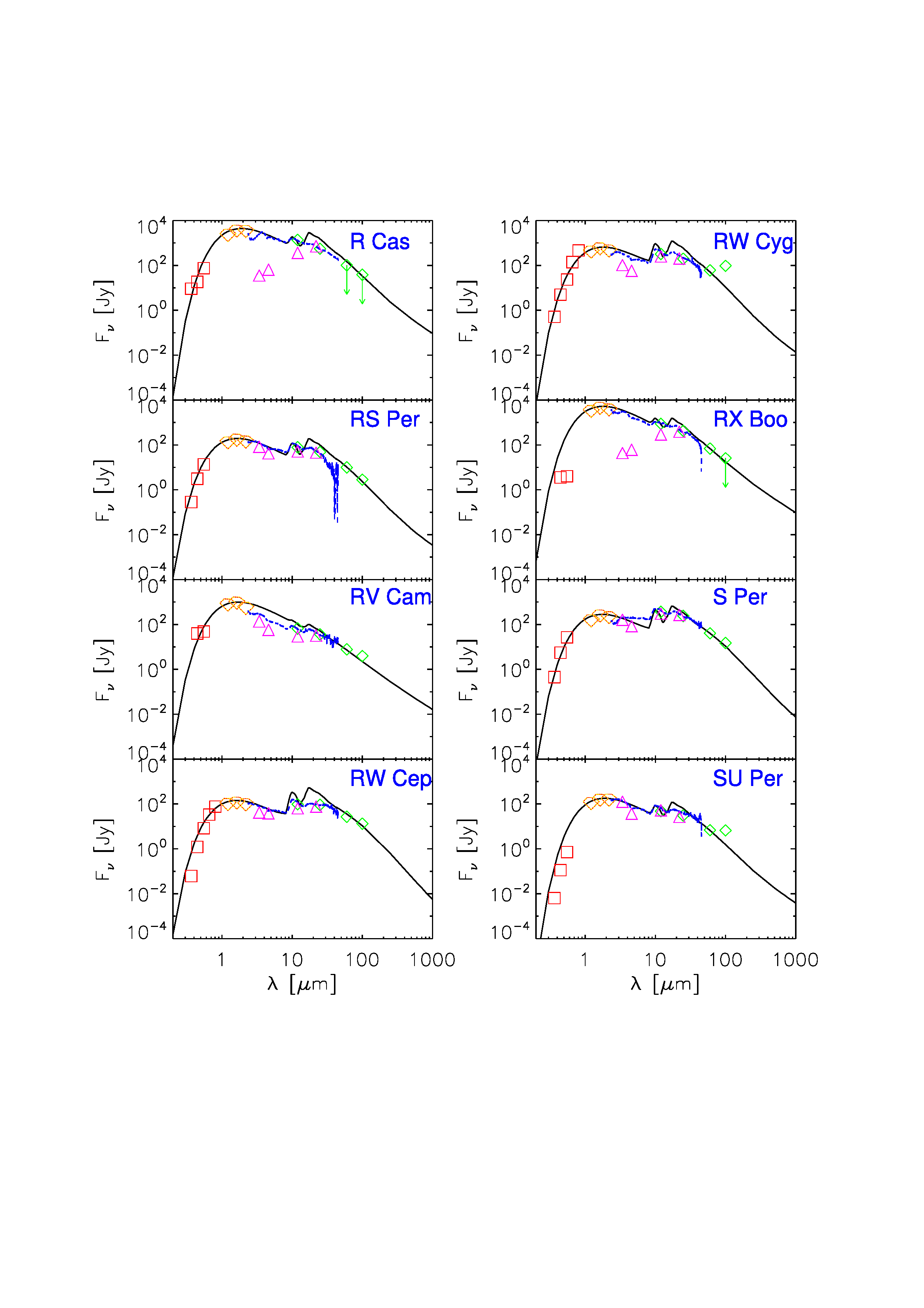}
\caption{         Same as Figure~\ref{fig:sedmod1}
         but for R Cas, RS Per, RV Cam,
         RW Cep, RW Cyg, RX Boo, S Per and SU Per.
         }
\label{fig:sedmod2}
\end{figure}

\begin{figure}
\centering
\includegraphics[width=0.65\textwidth,bb=50 150 400 650]{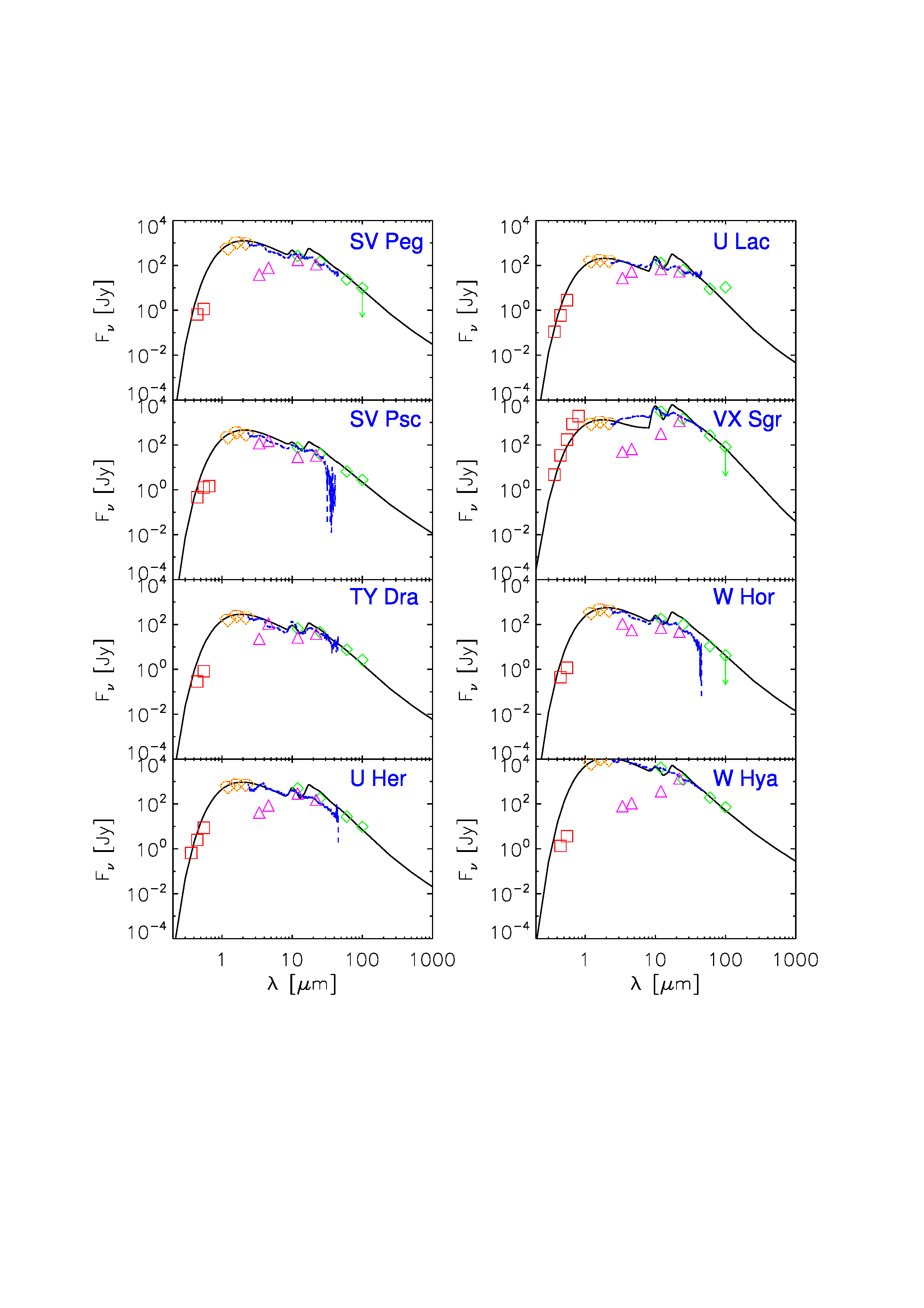}
\caption{
         Same as Figure~\ref{fig:sedmod1}
         but for SV Peg, SV Psc, TY Dra,
         U Her, U Lac, VX Sgr, W Hor and W Hya.
         }
\label{fig:sedmod3}
\end{figure}

\begin{figure}
\centering
\includegraphics[width=0.65\textwidth,bb=50 150 400 600]{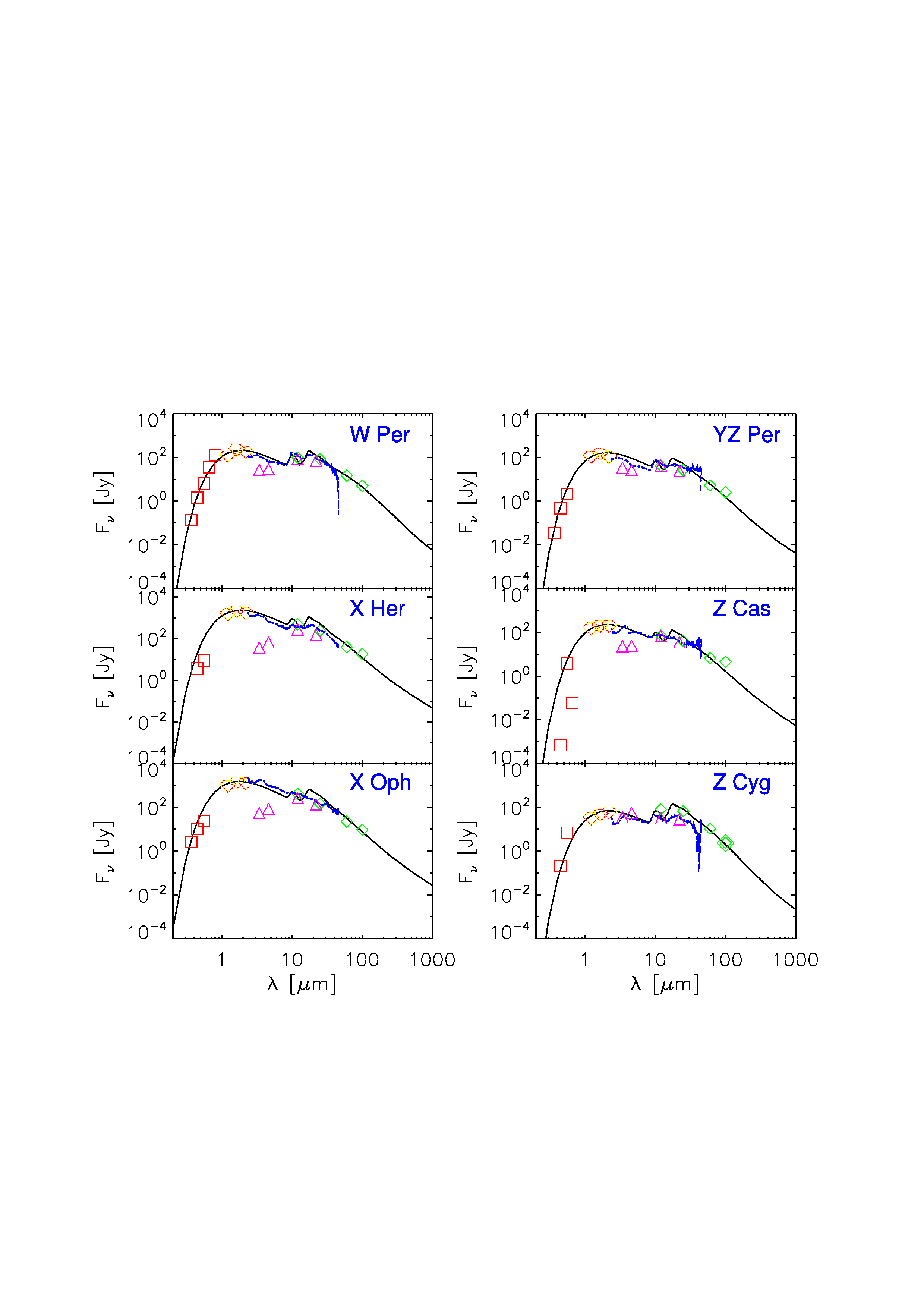}
\caption{ Same as Figure~\ref{fig:sedmod1}
         but for W Per, X Her,
         X Oph, YZ Per, Z Cas and Z Cyg.
         }
\label{fig:sedmod4}
\end{figure}

Because the selected sources are bright,
saturation can be a problem
in the {\it 2MASS} and {\it WISE} photometry.
The {\it WISE}/W1 and W2 bands are the most seriously affected
by saturation due to their high sensitivity and the high fluxes
of some sources in these two bands. Even the improved data quality
 in ALLWISE (Cutri et al. 2013) does not alleviate the problem,
while we adopted the version of the WISE All-Sky data release
(Cutri et al. 2012). Fortunately, the ISO/SWS spectrum covers
the waveband of WISE, which has no problem of saturation and
consequently a more reliable flux measurement and is used to
judge the model SED fitting.
For the {\it IRAS} photometry, some data have a high uncertainty.
We label these data with ``:'' in Table~\ref{tab:fluxdata}
and in Figures~\ref{fig:sedmod1}--\ref{fig:sedmod4}
they are plotted as doubled-rhombus.
Some measurements are just an upper limit of the flux.
We label these fluxes with ``U'' in Table~\ref{tab:fluxdata}
and in Figures~\ref{fig:sedmod1}--\ref{fig:sedmod4}
they are plotted with a downward arrow.
In the SED fitting process,
the saturated {\it WISE} bands
and the uncertain {\it IRAS} 100$\mum$ band
are not considered.
For illustrative purposes,
we also plot them in the figures.

We select the best-fit model by eye.
In Table~\ref{tab:MlossComp} we compare
the mass loss rates derived here with that
reported in the literature.
They are generally consistent.

Finally, assuming an outflow velocity
of $v_{\rm exp}=10\km\s^{-1}$
(Habing \& Olofsson 2003),
we calculate the dust mass loss rate
$\dot{M}_{\rm dust}$
for each source from the SED modeling
(see Table~\ref{tab:sedmod}).

\section{Crystallinity}
\label{sec:CSi}
In the literature, the silicate crystallinity is
 usually defined as the {\it mass} fraction of crystalline
silicate, $\fmc\equiv\Mcsi/\left(\Mcsi+\Masi\right)$,
where $\Mcsi$ and $\Masi$ are respectively the mass
of the crystalline and amorphous silicate components.
The silicate dust masses are often derived by fitting
the observed IR emission flux density ($F_\nu$)
under the assumption of an optically-thin dust shell
in the IR:
$F_\nu = \Sigma\left[B_\nu(T_{i})\times\kappa_\nu^{i}\times m_{i}\right]$
which sums the contribution from dust species $i$
with a mass absorption coefficient of $\kappa_\nu^{i}$,
mass $m_{i}$ and temperature $T_{i}$
(e.g., see Molster et al.\ 2002c, Jones et al.\ 2012,
Gielen et al.\ 2008, Jiang et al.\ 2013),
where $B_\nu(T)$ is the Planck function
at temperature $T$ and frequency $\nu$.
This approach is also widely adopted in
modeling the silicate emission spectra of
protoplanetary and debris disks
(e.g., see Sargent et al.\ 2006, 2009a,b;
Lisse et al.\ 2007, 2009, 2012).
One tends assume three dust species:
amorphous silicate, crystalline forsterite,
and crystalline enstatite.
For each species, a cold component
and a warm component are often assumed.
The crystalline silicate mass $\Mcsi$
is obtained by summing over
four components
(i.e., warm crystalline forsterite,
cold crystalline forsterite,
warm crystalline enstatite, and
cold crystalline enstatite),
while the amorphous silicate mass $\Masi$
is the sum of warm and cold amorphous silicates
(Suh \ 2004, Gielen et al.\ 2008, Jiang et al.\ 2013).
The problem associated with this method
is that the derived silicate crystallinity $\fmc$
is highly sensitive to the fitted dust temperatures
which are often uncertain. If the temperature is
uncertain by 20\%, the derived dust mass and $\fmc$
will be uncertain by a factor of $\simali$2.
Also, it is not physical to treat the temperatures
of crystalline silicates as free parameters
since the observed peak wavelengths and relative
strengths of the crystalline silicate features
already contain clues about their temperatures
(Koike et al.\ 1993, 1999; J\"ager et al.\ 1998).

To avoid the temperature uncertainty,
we take an alternative approach to
characterize the silicate crystallinity:
we propose to measure the degree of crystallinity,
$\ffc$, as the ratio of the {\it fluxes} emitted
in the crystalline silicate emission features
to that in the amorphous silicate features.
To this end, we decompose the $\simali$2.4--45$\mum$
{\it ISO}/SWS spectrum of each source into
four components: (i) a stellar continuum,
(ii) a dust thermal emission continuum,
(iii) two broad emission bands of amorphous silicates
      at 10 and 18$\mum$, and
(iv) a number of sharp emission features
     of crystalline silicates.

The decomposition is carried out
with the PAHFIT software of Smith et al.\ (2007)
which was originally developed for decomposing
the polycyclic aromatic hydrocarbon (PAH)
emission features.

We modify the PAHFIT code by considering
(i) a Planck black-body $B_\nu(T_\star)$ of temperature
$T_\star\simali$2,000--5,000$\K$ for the stellar continuum,
(ii) a warm, modified black-body $\nu^2 B_\nu(T_1)$
of $T_1\simali$150--400$\K$
and a cold, modified black-body $\nu^2 B_\nu(T_2)$
of $T_2\simali$80--150$\K$
for the dust continuum,
(iii) two opacity ($\kappa_{\rm abs}$)-based profiles
for the 9.7 and 18$\mum$ amorphous silicate
emission features, and
(iv) $N$ sharp Drude profiles
for the crystalline silicate emission features
with the $j$-th Drude profile\footnote{%

  Drude profiles are expected for
  classical damped harmonic oscillators
  (see Li 2009). They closely resemble
  Lorentzian profiles and are more extended
  than Gaussian profiles
  in the blue- and red-wing regions.
  }
peaking at wavelength $\lambda_j$ and having
a FWHM of $\gamma_j\lambda_j$:

\begin{eqnarray}
\label{eq:Fnu}
\nonumber
F_\nu & = & \{A_\star B_\nu(T_\star) +
             A_1 \nu^2 B_\nu(T_1) + A_2 \nu^2 B_\nu(T_2)\\
& &  + A_W \kappa_{\rm abs}(\nu) B_\nu(T_W)
+A_C \kappa_{\rm abs}(\nu) B_\nu(T_C)\\
\nonumber
& &  + \sum_{j=1}^{N} \frac{I_j\gamma_j^2}
{\left(\lambda/\lambda_j-\lambda_j/
\lambda\right)^2+\gamma_j^2} \}/d^2 ~~,
\end{eqnarray}
where $F_\nu$ is the model flux density,
$A_\star$, $A_1$, $A_2$, $A_W$,
and $A_C$ are constants,
and $I_j$ is the central flux density
of the $j$-th Drude profile.

\begin{figure}
\centering
\includegraphics[width=0.5\textwidth,bb=100 400 600 650]{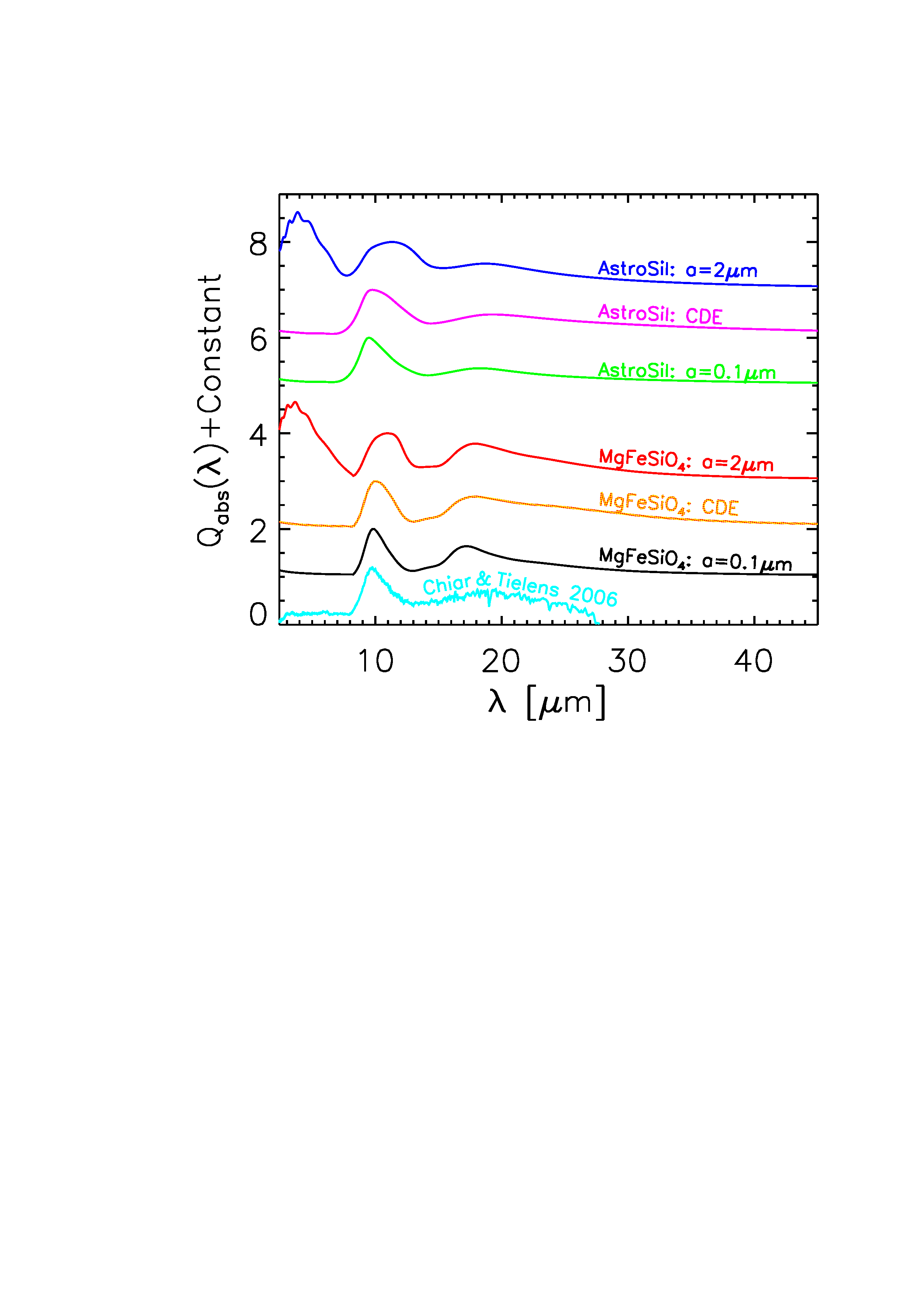}
\caption{ Absorption efficiency $Q_{\rm abs}(a,\,\lambda)$
         of amorphous silicate dust calculated from
         spherical amorphous olivine MgFeSiO$_4$ of
         Dorschner et al.\ (1995) of radii $a=0.1\mum$ and $a=2\mum$,
         spherical ``astronomical silicate'' of Draine \& Lee (1984)
         of radii $a=0.1\mum$ and $a=2\mum$,
         silicate dust of CDE shapes of amorphous olivine MgFeSiO$_4$
         and  ``astronomical silicate''.
         Also shown is the observed absorption profile
         of the Galactic diffuse ISM along the line of sight toward WR\,98a
         (Chiar \& Tielens 2006).
        }
\label{fig:asi_opacity}
\end{figure}

To approximate the 9.7 and 18$\mum$ amorphous
silicate emission features, we consider the following
opacity profiles:
(1) the absorption efficiency $Q_{\rm abs}(\lambda,\,a)$
     of spherical amorphous olivine MgFeSiO$_4$
     (Dorschner et al.\ 1995) of radii $a=0.1\mum$;
(2) the absorption efficiency $Q_{\rm abs}(\lambda,\,a)$
     of spherical amorphous olivine MgFeSiO$_4$
     (Dorschner et al.\ 1995) of radii $a=2\mum$;
(3) the absorption efficiency $Q_{\rm abs}(\lambda,\,a)$
     of amorphous olivine MgFeSiO$_4$
     (Dorschner et al.\ 1995) of shapes of continuous
     distributions of ellipsoids (CDE; Bohren \& Huffman 1983);
(4) the absorption efficiency $Q_{\rm abs}(\lambda,\,a)$
     of spherical  ``astronomical silicate''
     (Draine \& Lee 1984) of radii $a=0.1\mum$;
(5) the absorption efficiency $Q_{\rm abs}(\lambda,\,a)$
     of spherical ``astronomical silicate''
     (Draine \& Lee 1984) of radii $a=2\mum$;
(6) the absorption efficiency $Q_{\rm abs}(\lambda,\,a)$
     of ``astronomical silicate''
     (Draine \& Lee 1984) of CDE shapes; and
 (7) the silicate absorption profile of the diffuse ISM
      sightline toward the WC-type Wolf-Rayet star WR\,98a
      (Chiar \& Tielens 2006).
Here the absorption efficiency $Q_{\rm abs}(\lambda,\,a)$
is related to the opacity $\kappa_{\rm abs}(\lambda)$ through
$\kappa_{\rm abs}(\lambda) = 3 Q_{\rm abs}(\lambda,\,a)/4a\rho$
for spherical grains of radii $a$ and mass density of $\rho$,
where $Q_{\rm abs}(\lambda,\,a)$ is calculated from Mie theory
(Bohren \& Huffman 1983) using the refractive indices
of amorphous olivine measured by Dorschner et al.\ (1995)
or ``astronomical silicate'' synthesized by Draine \& Lee (1984).
In Figure~\ref{fig:asi_opacity} we compare the absorption profiles
of these seven types of amorphous silicates.
Most appreciably, both the 9.7 and 18$\mum$ silicate features
become substantially wider as the dust size increases from
$a=0.1\mum$ to $a=2\mum$.
For the same size, the 9.7 and 18$\mum$ features
of amorphous olivine MgFeSiO$_4$ are narrower than
that of ``astronomical silicate''.
The 9.7 and 18$\mum$ features of silicate grains of
CDE shapes are intermediate between that of
$a=0.1\mum$ and that of $a=2\mum$.
We find that the one calculated from amorphous olivine
MgFeSiO$_4$ of CDE shapes best fit the {\it ISO}/SWS spectra.
In the following, we will adopt the opacity profile of
amorphous olivine MgFeSiO$_4$ of CDE shapes.
In Figures~\ref{fig:Sloan1}--\ref{fig:Sloan4}
we show the modeled silicate emission spectra
for our 28 sample stars. The results for the peak wavelengths,
FWHMs, and emitted fluxes of the decomposed silicate emission
features are listed in Tables~\ref{tab:FeatList1}--\ref{tab:FeatList9}.

To make sure that we have picked out
all the features of crystalline silicates
and these features were not contaminated
by molecular lines and spectral features of
other dust species, such as aluminum oxide,
magnesium-iron oxide, metallic iron, and melilite
which could be present in the circumstellar
envelopes of oxygen-rich evolved stars
(e.g., see Lorentz \& Pompeia 2000,
Fabian et al.\ 2001,
Posch et al.\ 1999, 2002,
Sloan et al.\ 2003,
Heras \& Hony 2005,
Verholest et al.\ 2009,
de Vries et al.\ 2010,
Zeidler et al.\ 2013,
Nowotny et al. \ 2015),
we consult the crystalline silicate
features compiled by Molster et al.\ (2002b).
In Table~\ref{tab:lambdaminmax}
we compare the wavelength and FWHM ranges
derived in this work with that of Molster et al.\ (2002b)
and it is found that they are generally consistent.
The possible effects on the relation between
the silicate crystallinity and $\Mdustloss$
caused by the ``contamination'' of amorphous
Al$_{2}$O$_{3}$, Mg$_{x}$Fe$_{(1-x)}$O
($0 \leq x \leq1$) and spinel (MgAl$_2$O$_4$)
will be discussed in \S\ref{sec:CSi_vs_dotM}.

Let $\Fasi$ be the wavelength-integrated fluxes of
the 10 and 18$\mum$ amorphous silicate emission features,
$\Fcsi$ be the sum of the wavelength-integrated fluxes of
all of the crystalline silicate emission features, and
$\Fcont$ be the wavelength-integrated flux of
the dust continuum.\footnote{%
   In determining $\Fcont$, we exclude
   the $\simali$2.4--8$\mum$ wavelength range
   as the continuum emission in this wavelength range
   is likely due to iron grains, not silicates
   (see Kemper et al.\ 2002).
   }
If we assume that the dust continuum
is predominantly emitted by amorphous silicates,
we define the flux-based silicate crystallinity
to be $\ffc\equiv \Fcsi/\left(\Fcsi + \Fasi+ \Fcont\right)$.
In Table~\ref{tab:sedmod} we tabulate the derived $\ffc$
for each source. In Figure~\ref{fig:hist_csi} we
show the silicate crystallinity ($\ffc$) histrogram.
For the majority of our sources, $\ffc<20\%$.

We note that the flux-based silicate crystallinity $\ffc$
may differ from the mass-based crystallinity $\fmc$.
If the amorphous silicate dust component is richer in
its iron content than the crystalline silicate component
and if both components have more or less the same
spatial distribution, we expect $\ffc < \fmc$
since the iron-richer amorphous silicate component
is more absorptive in the UV/visual/near-IR wavelength range
and by implication, would emit more energy in the IR
on a per unit mass basis.

\section{Crystallinities versus $\Mdustloss$:
         Results and Discussion}
\label{sec:CSi_vs_dotM}
With the dust mass loss rate $\Mdustloss$ (see \S\ref{sec:SED})
and the silicate crystallinity $\ffc$ (see \S\ref{sec:CSi})
calculated for each source, we now explore
whether $\Mdustloss$ and $\ffc$ are correlated.
As illustrated in Figure~\ref{fig:csi_mloss1},
with a Pearson correlation coefficient of $r\approx-0.24$,
$\Mdustloss$ and $\ffc$ are apparently not correlated.
This supports the proposition of Kemper et al.\ (2001)
but contradicts the earlier findings that the detections
of crystalline silicate emission features in evolved stars
appear to be restricted to objects with
$\Mloss \gtsim10^{-5}\Msun\yr^{-1}$
which corresponds to
$\Mdustloss \gtsim5\times10^{-8}\Msun\yr^{-1}$
for $\gastodust=200$
(e.g., see Molster et al.\ 2002a,b,c).
\footnote{%
   Jones et al.\ (2012) reported the detection of
   crystalline silicate emission features in stars
   with mass loss rates as low as
   $\Mloss\approx10^{-9}\Msun\yr^{-1}$.
   We note that they derived $\Mloss$
   from fitting the observed SEDs with a large number
   of template SEDs created by the ``Grid of Red supergiant
   and Asymptotic giant branch Models''
   (GRAMS, Sargent et al.\ 2011).
   The GRAMS model was designed to fit the SED
   from the optical to the mid-IR,
   with the {\it Spitzer}/MIPS 24$\mum$ band
   being the longest wavelength.
   One would expect the GRAMS model to underestimate
   the actual dust mass loss rates since a substantial
   quantity of dust would emit in the far-IR
   at $\lambda$\,$>$\,60--100$\mum$.
   However, Jones et al.\ (2015) found that AGB stars
   do not produce a significant quantity of dust in the far-IR,
   so using the models optimized for the mid-IR will not
   significantly affect the measured dust production rate.
   }
In Figure~\ref{fig:histogram} we plot the histogram
of the dust mass loss rates of our sample. It is seen that
the dust mass loss rates of our sample are distributed
over 2 dex ranging
 from $\simali$$1.47\times 10^{-9}\Msun\yr^{-1}$
to $\simali$$1.22\times 10^{-7}\Msun\yr^{-1}$,
with $\simali$50\% of the stars
having $\Mdustloss <10^{-8}\Msun\yr^{-1}$.
There appears to exist a tendency that the number
of stars decrease as the dust mass loss rate increases.
Whether this is a real tendency of evolved stars
that possess crystalline silicates, or a result of
sample capacity, remains to be discussed.

So far we have assumed that the dust thermal continuum
emission is predominantly attributed to amorphous silicates.
To relax this assumption, we also define the silicate
crystallinity to be $\ffcp\equiv \Fcsi/\left(\Fcsi + \Fasi\right)$
which only compares the fluxes emitted in the spectral
features. The derived $\ffcp$ ratio is also tabulated
in Table~\ref{tab:sedmod}. Figure~\ref{fig:csi_mloss2}
shows that $\ffcp$ does not correlate with $\Mdustloss$.

Finally, we also show that the silicate crystallinity $\ffc$
does not correlate with the stellar temperature
(see Figure~\ref{fig:csi_Tstar})
or the stellar luminosity (see Figure~\ref{fig:csi_Lstar}),
in agreement with Gielen et al.\ (2008).
The difference between this work and
that of de Vries et al.\ (2010) and Jones et al.\ (2012)
is that they were more concerned with the relation
between the mass loss rates and individual features
(e.g., the 11.3 and 33.6$\mum$ emission features
[de Vries et al.\ 2010]) or the 23, 28 and 33$\mum$
complex features, while we mainly focus on
the relation between the entire flux ratios
and the mass loss rates.

De Vries et al.\ (2010) showed that the strength of
the 11.3$\mum$ feature does show some correlation
with the mass loss rate, while Jones et al.\ (2012)
found that at least the 23, 28 and 33$\mum$ complexes
do not seem to correlate with the mass loss rates of the stars.

\begin{figure}
\centering
\includegraphics[width=0.5\textwidth,bb=50 200 300 650]{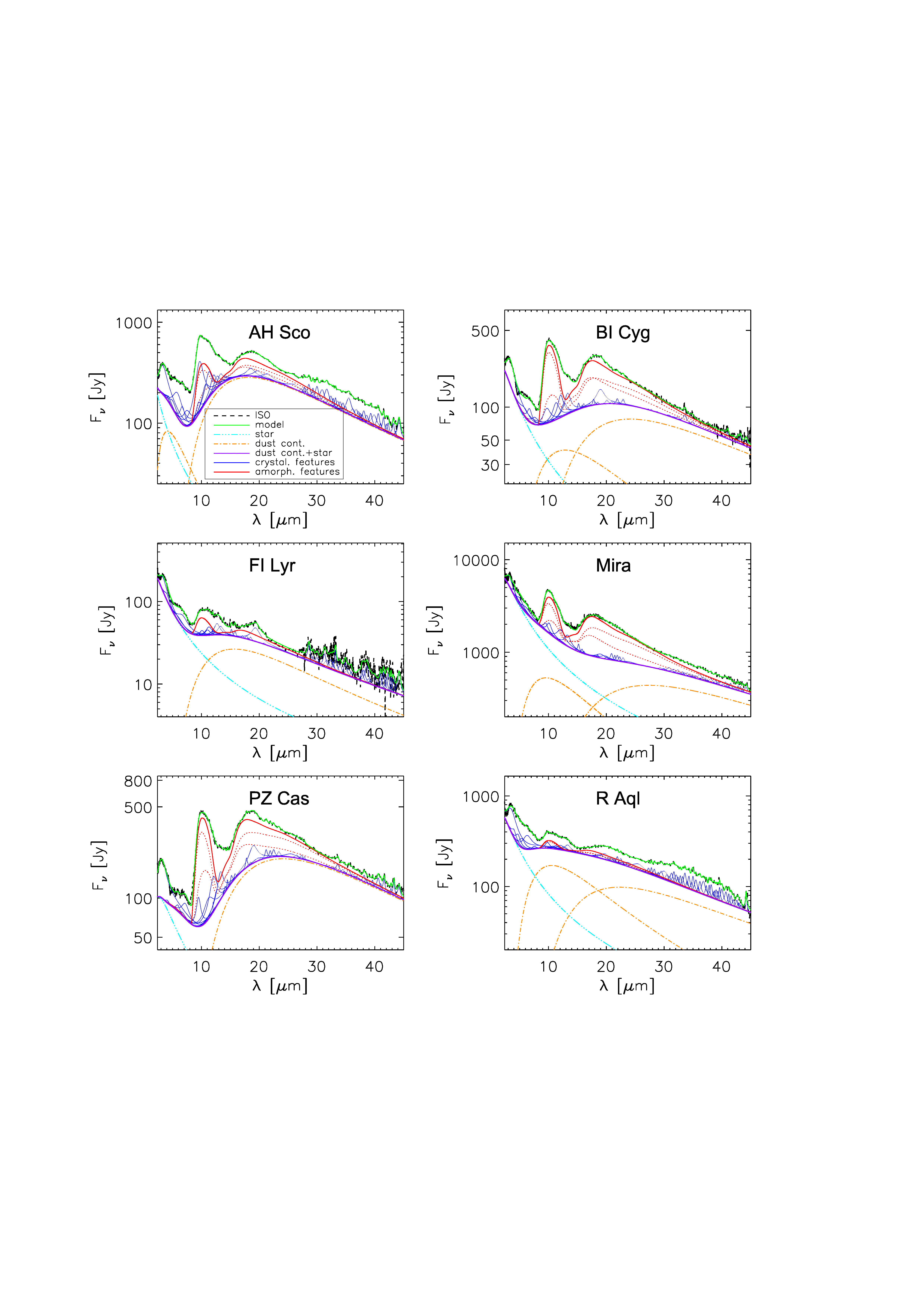}
\caption{Decomposing the {\it ISO}/SWS spectra (dashed black lines)
         of AH Sco, FI Lyr, Mira, BI Cyg, PZ Cas and R Aql
         into a stellar continuum (cyan lines),
         two dust thermal continua (orange lines),
         and individual dust spectral features (blue lines).
         Purple lines plot the summed continuum
         (i.e., the stellar continuum plus the two dust continua),
         red lines indicate the contribution of amorphous silicates,
         while green lines show the fitted spectra.
         }
\label{fig:Sloan1}
\end{figure}

\begin{figure}
\centering
\includegraphics[width=0.5\textwidth,bb=50 120 300 700]{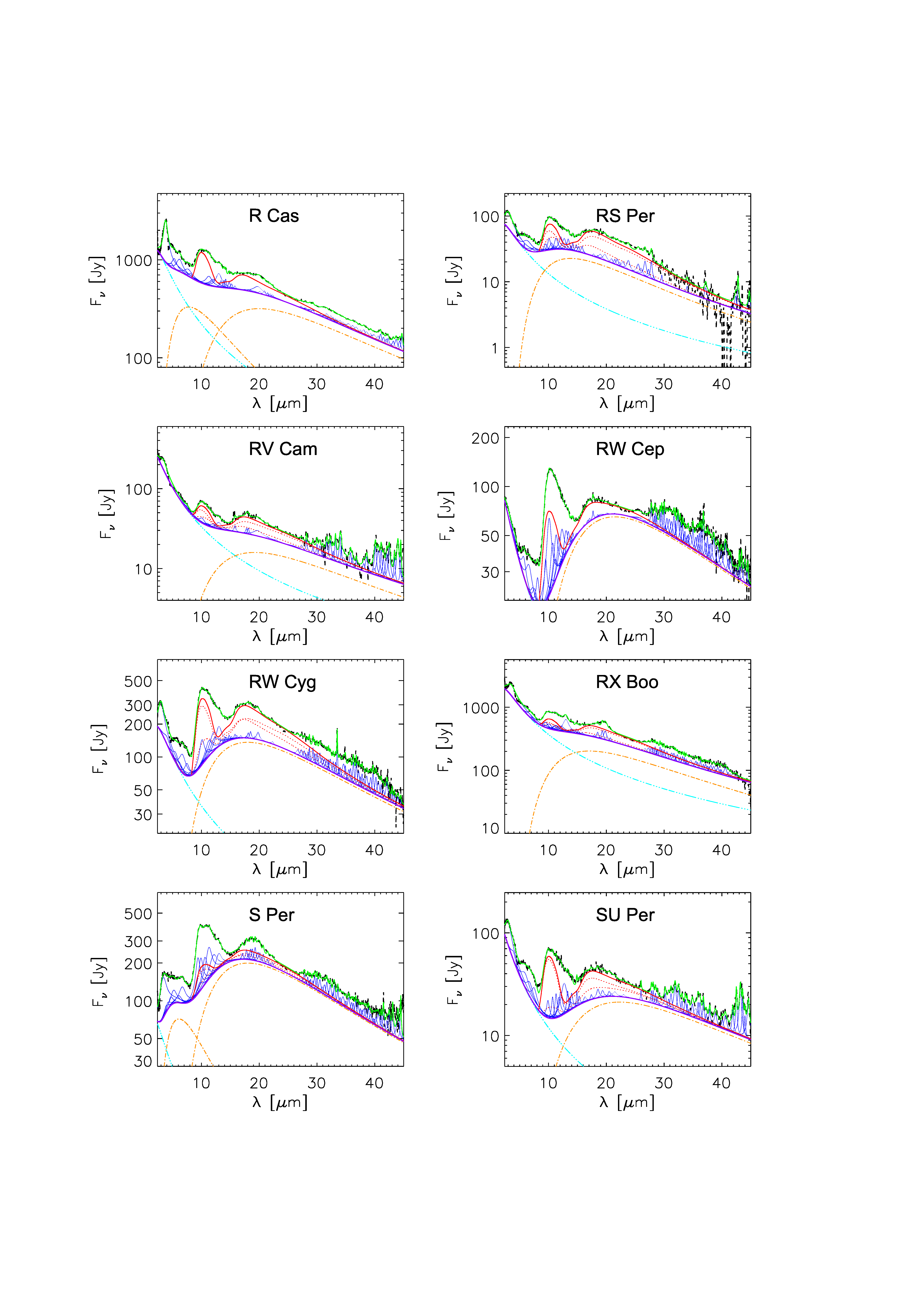}
\caption{ Same as Figure~\ref{fig:Sloan1}
         but for R Cas, RS Per, RV Cam,
         RW Cep, RW Cyg, RX Boo, S Per and SU Per.
         }
\label{fig:Sloan2}
\end{figure}

\begin{figure}
\centering
\includegraphics[width=0.5\textwidth,bb=50 120 300 700]{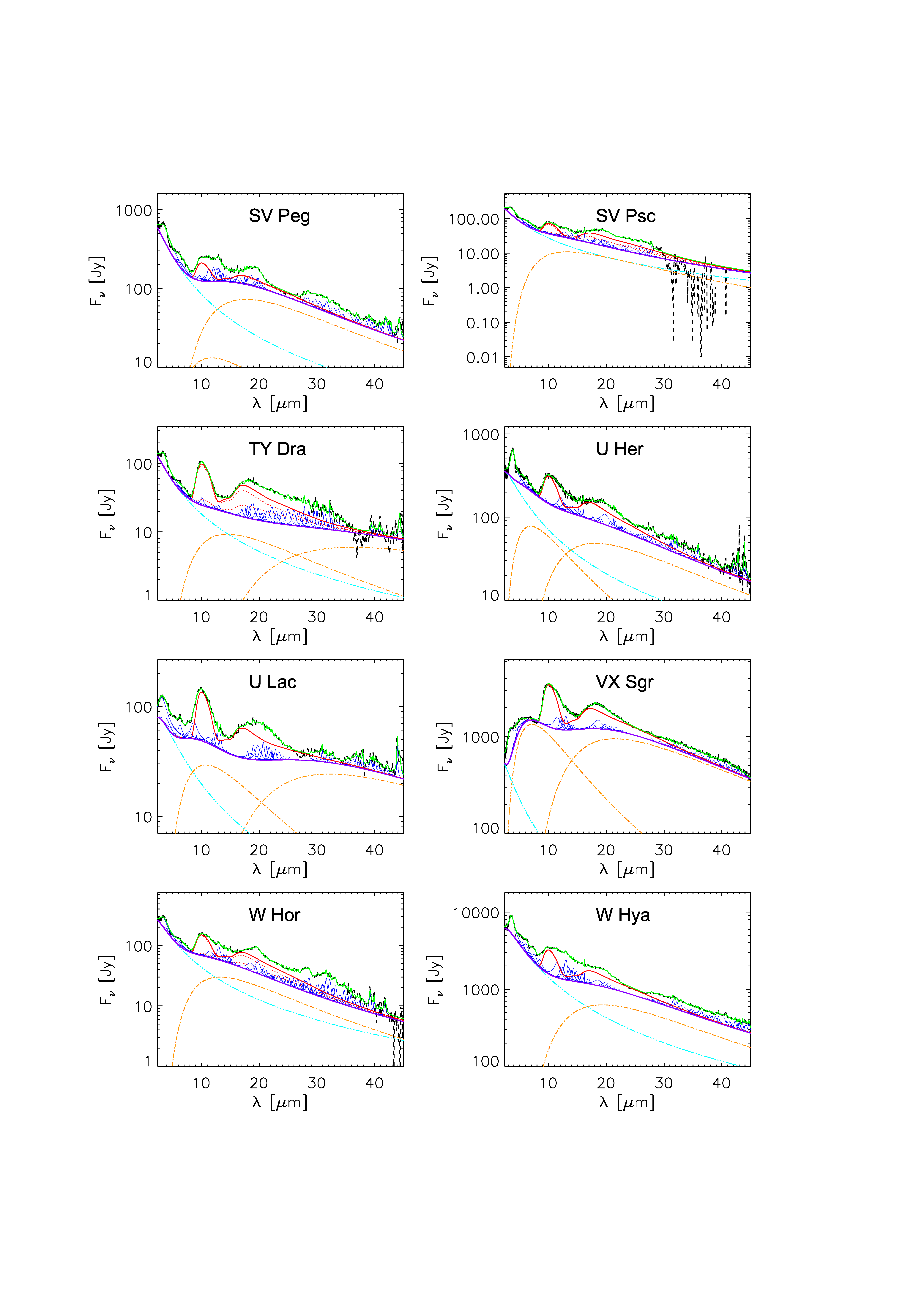}
\caption{    Same as Figure~\ref{fig:Sloan1}
         but for SV Peg, SV Psc, TY Dra,
         U Her, U Lac, VX Sgr, W Hor and W Hya.
         }
\label{fig:Sloan3}
\end{figure}

\begin{figure}
\centering
\includegraphics[width=0.5\textwidth,bb=50 200 300 650]{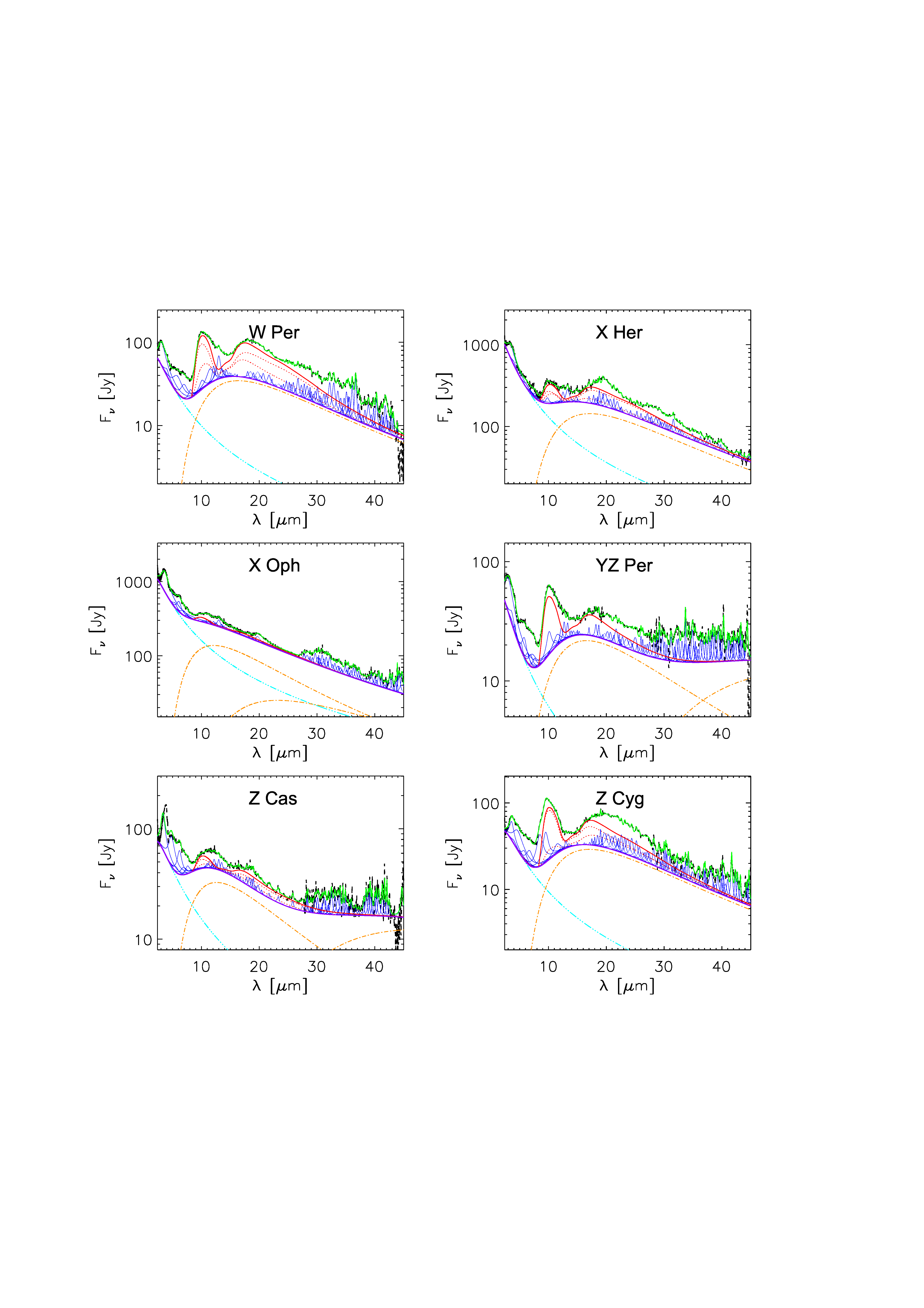}
\caption{  Same as Figure~\ref{fig:Sloan1}
         but for W Per, X Her,
         X Oph, YZ Per, Z Cas and Z Cyg.
         }
\label{fig:Sloan4}
\end{figure}

\begin{figure}[htb]
\centering
\includegraphics[width=0.40\textwidth,bb=100 200 400 600]{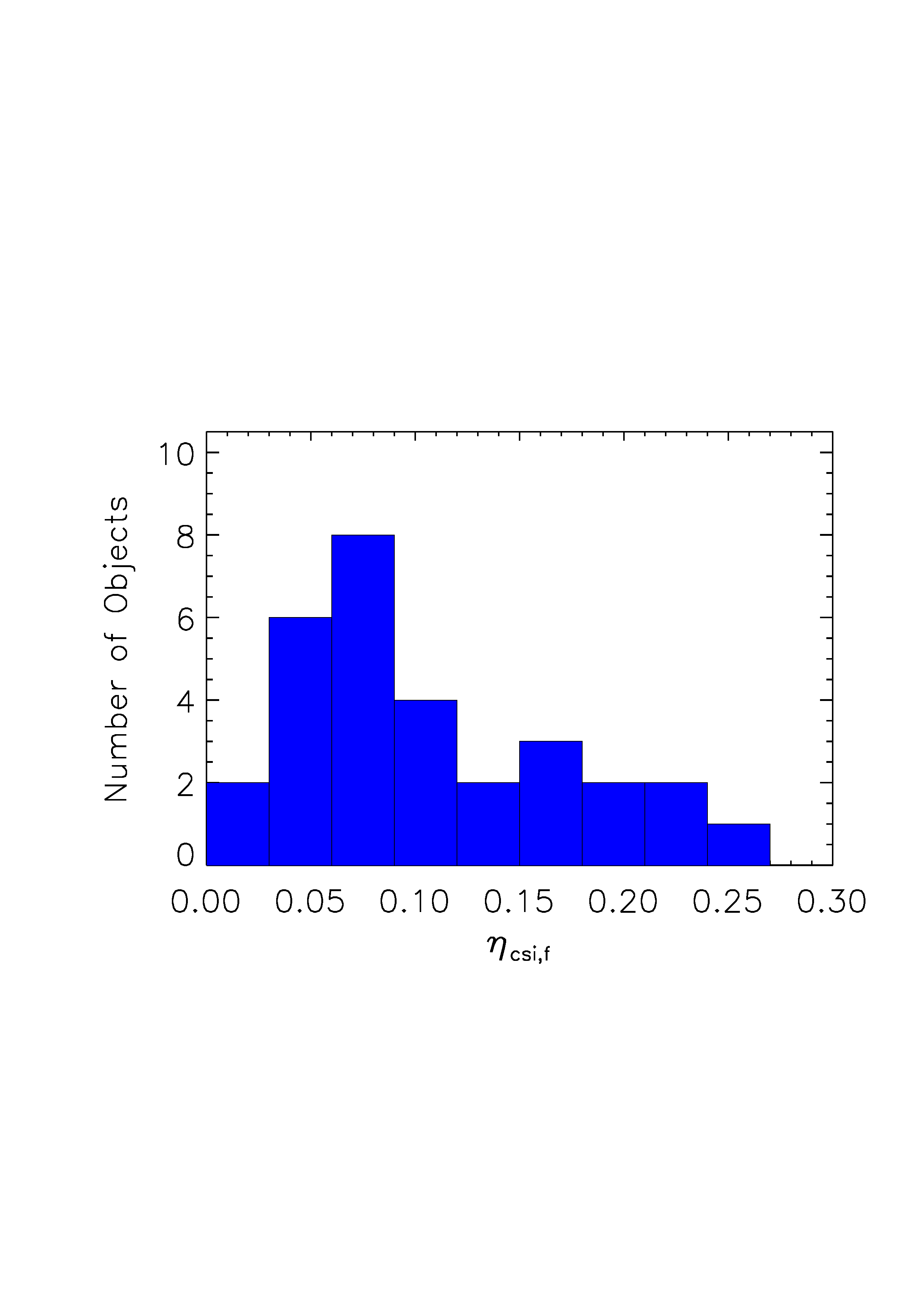}
 \caption{Histogram of the flux-based
  silicate crystallinity ($\ffc$).
         }
\label{fig:hist_csi}
\end{figure}

\begin{figure}[htb]
\centering
\includegraphics[width=0.4\textwidth,bb=100 100 400 350]{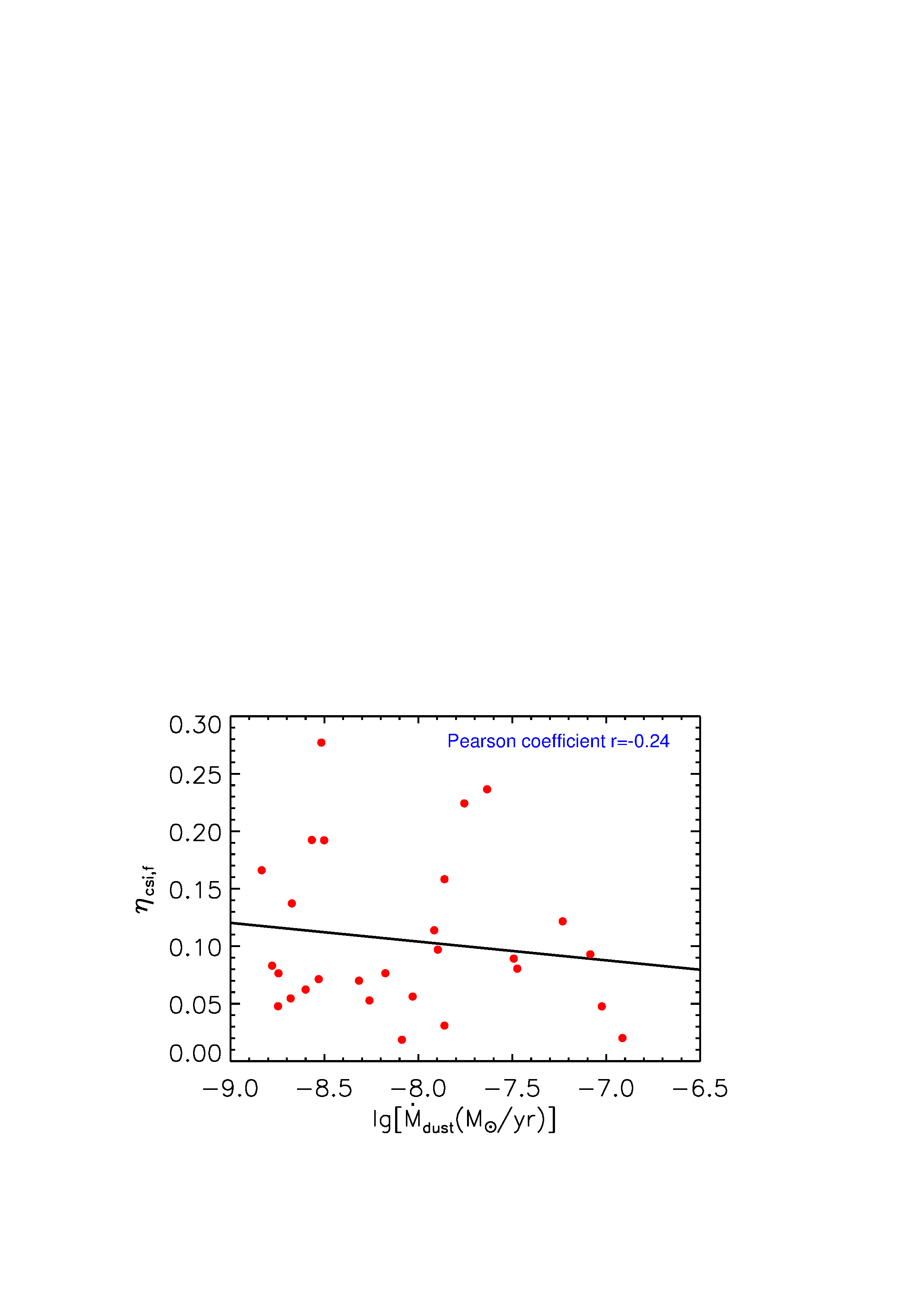}
 \caption{   Correlation of the silicate crystallinity $\ffc$
         with the dust mass loss rate $\Mdustloss$.
         }
\label{fig:csi_mloss1}
\end{figure}

\begin{figure}
\centering
\includegraphics[width=0.4\textwidth,bb=100 300 400 750]{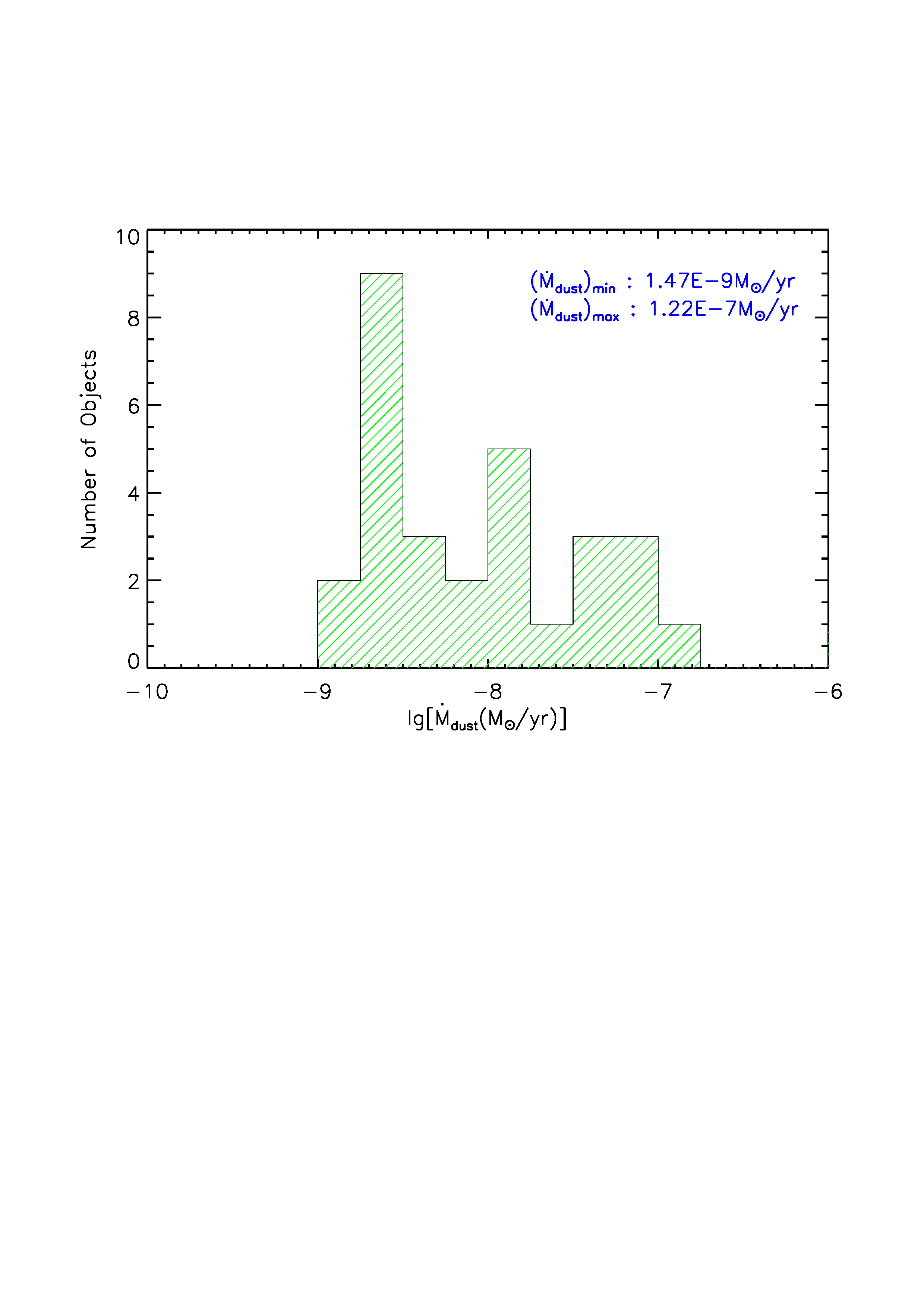}
\caption{Histogram of the dust mass loss rates $\Mdustloss$.
        }
 \label{fig:histogram}
\end{figure}

\begin{figure}
\centering
\includegraphics[width=0.4\textwidth,bb=100 100 400 550]{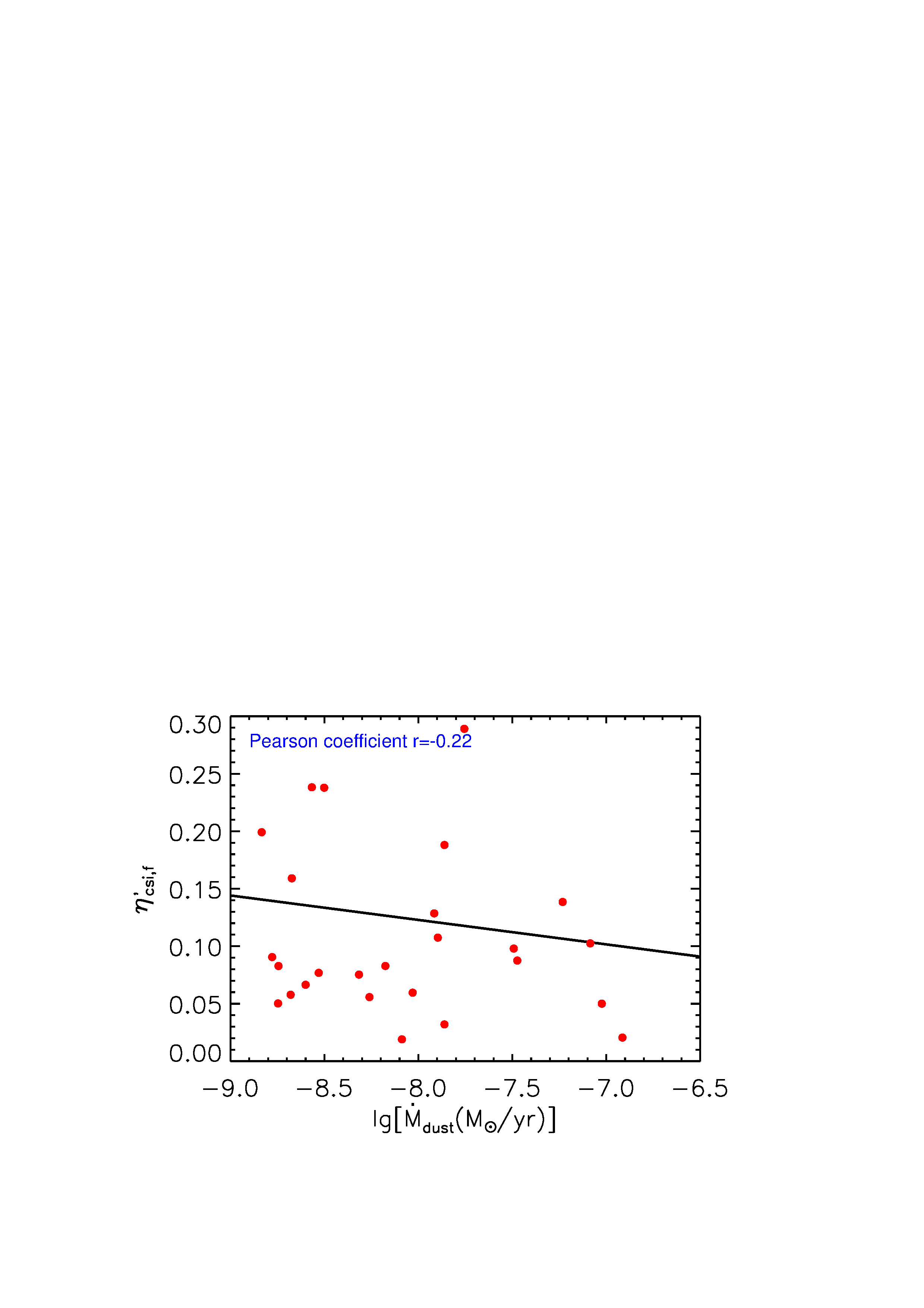}
 \caption{
                Correlation of the silicate crystallinity $\ffcp$
        with the dust mass loss rate $\Mdustloss$.
        }
\label{fig:csi_mloss2}
\end{figure}

\begin{figure}
\centering
\includegraphics[width=0.4\textwidth,bb=100 100 400 550]{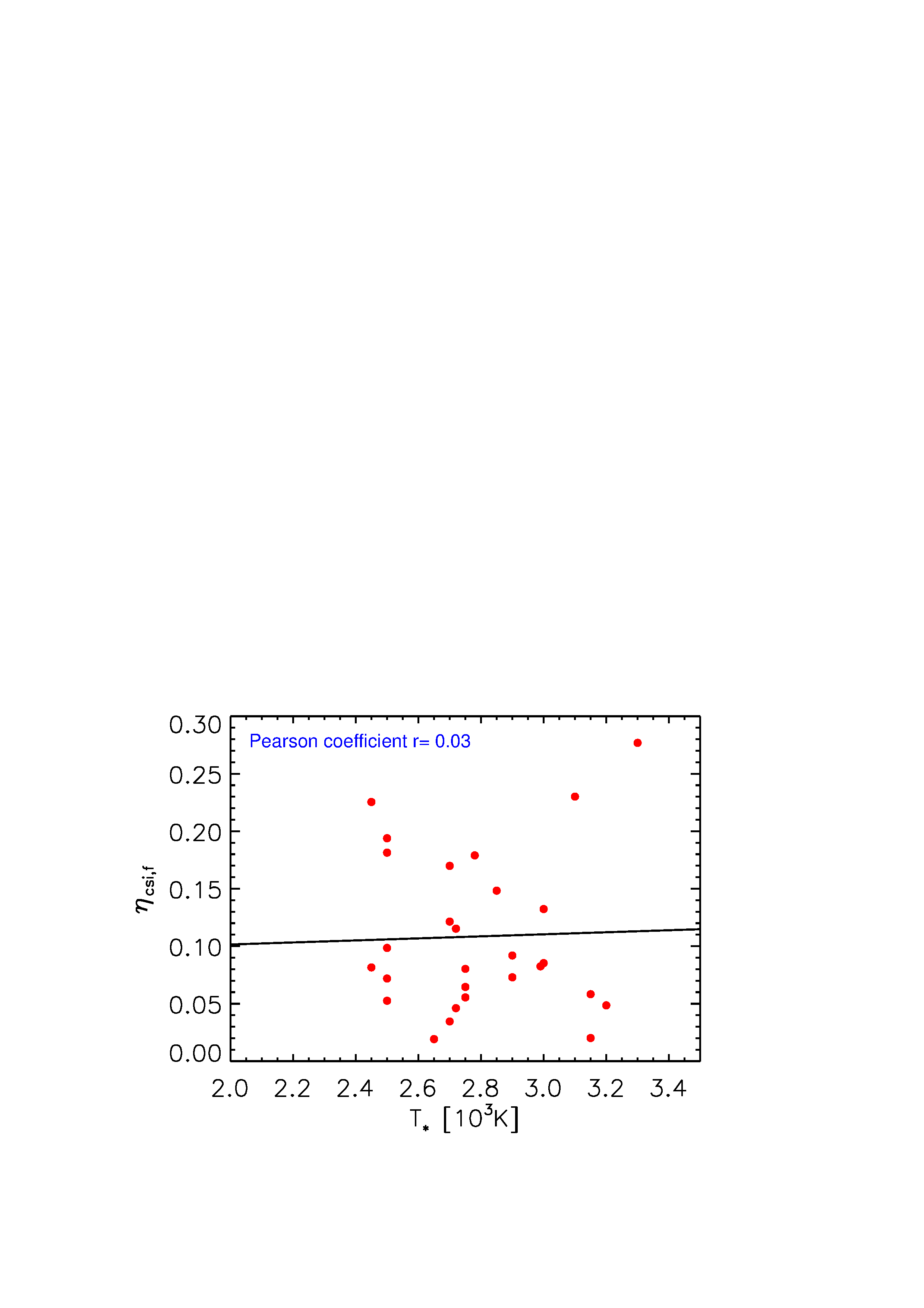}
 \caption{
                Correlation of the silicate crystallinity $\ffc$
        with the stellar effective temperature $T_\star$.
        }
\label{fig:csi_Tstar}
\end{figure}

\begin{figure}
\centering
\includegraphics[width=0.4\textwidth,bb=100 100 400 550]{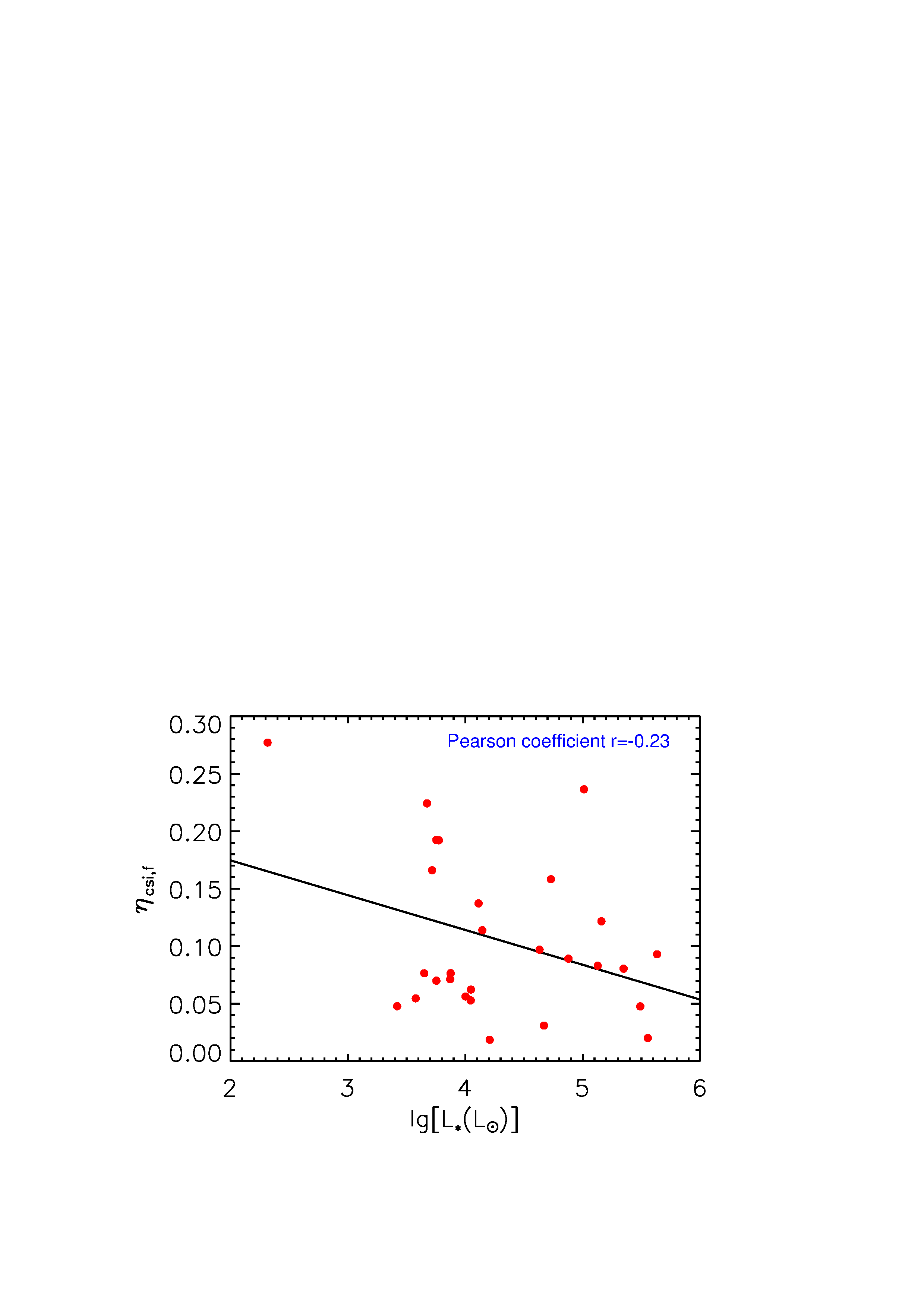}
 \caption{
                Correlation of the silicate crystallinity $\ffc$
        with the stellar luminosity $L_\star$.
        }
\label{fig:csi_Lstar}
\end{figure}

\begin{figure}
\centering
\includegraphics[width=0.4\textwidth,bb=100 100 400 550]{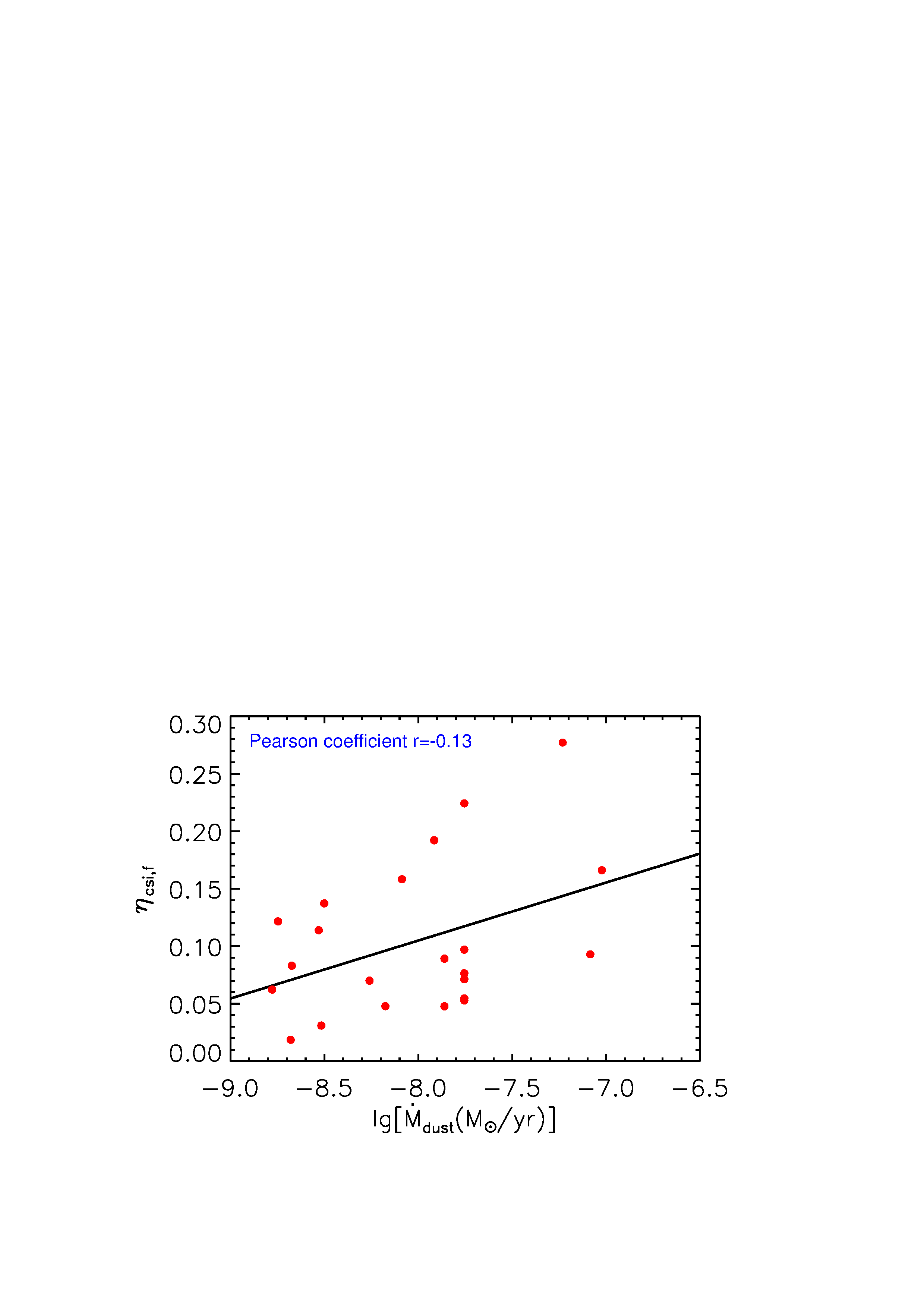}
 \caption{ Same as Figure~\ref{fig:csi_mloss1}
        but with BI Cyg, Mira, RS Per, SV Psc, VX Sgr and W Hya excluded.
        }
\label{fig:csi_mloss1a}
\end{figure}

\begin{figure}
\centering
\includegraphics[width=0.4\textwidth,bb=100 100 400 550]{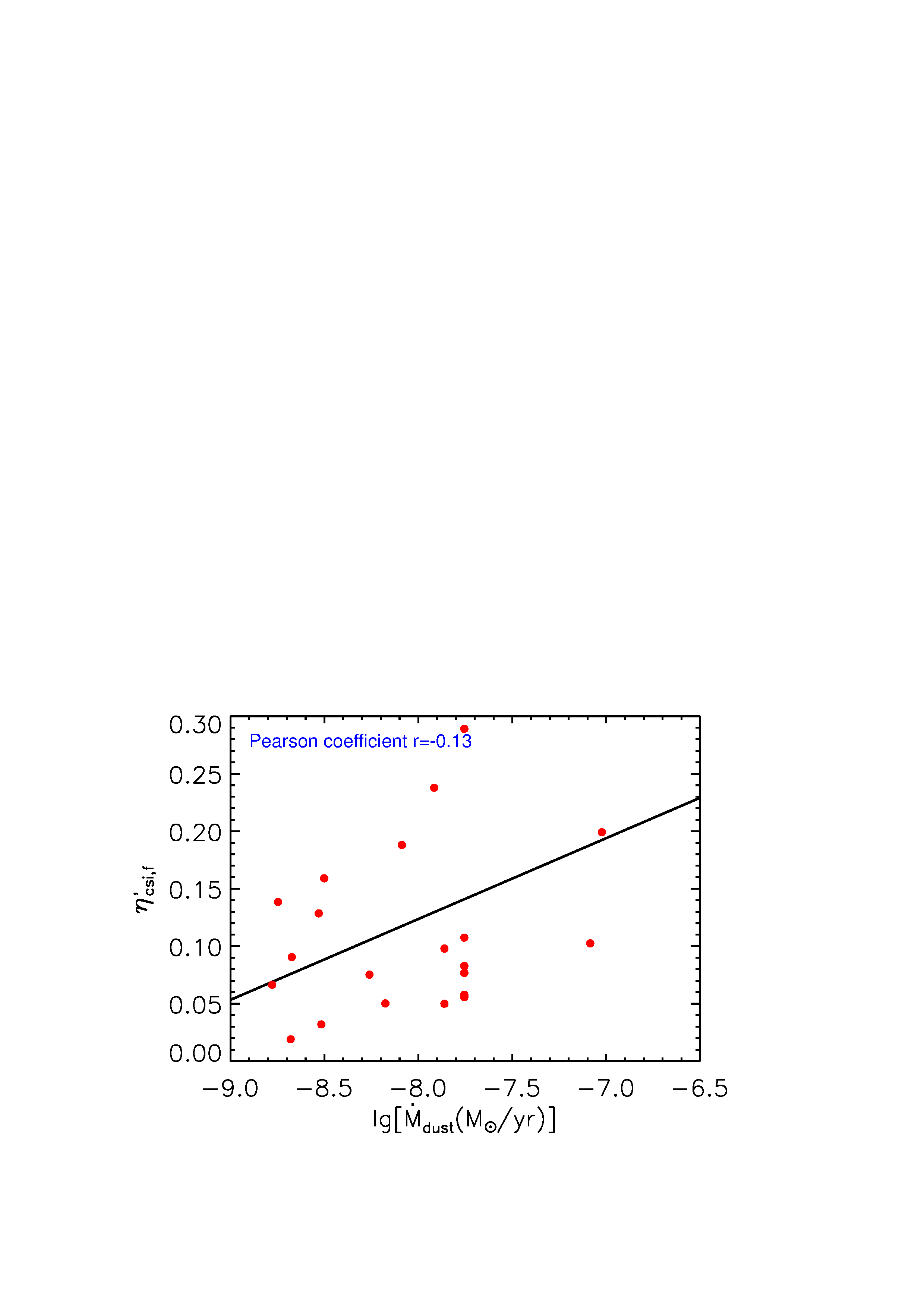}
 \caption{ Same as Figure~\ref{fig:csi_mloss2}
             but with BI Cyg, Mira, RS Per, SV Psc, VX Sgr and W Hya excluded.
             }
\label{fig:csi_mloss2a}
\end{figure}

\begin{figure}
\centering
\includegraphics[width=0.4\textwidth,bb=100 100 400 550]{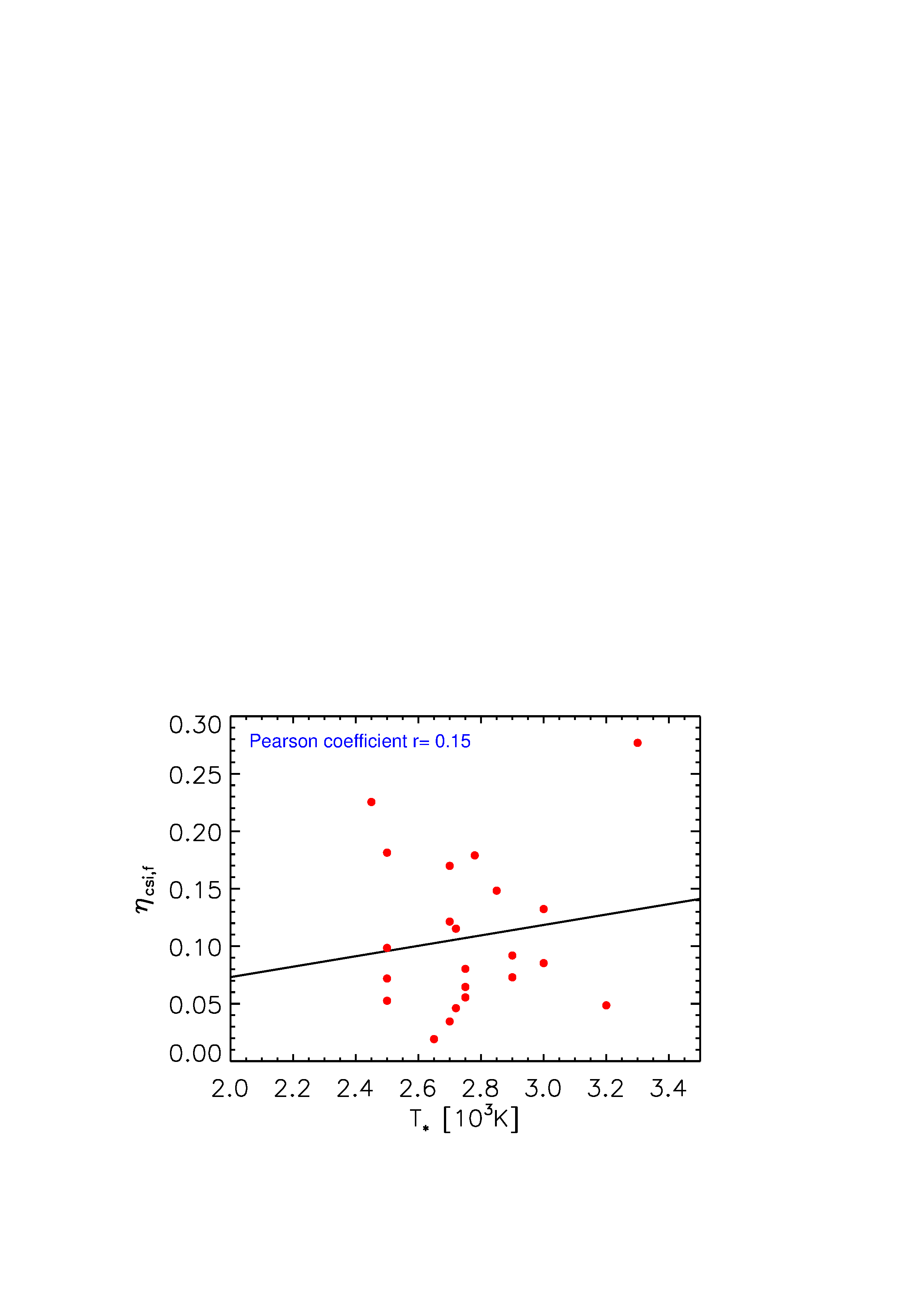}
 \caption{Same as Figure~\ref{fig:csi_Tstar}
             but with BI Cyg, Mira, RS Per, SV Psc, VX Sgr and W Hya excluded.
             }
\label{fig:csi_Tstara}
\end{figure}

\begin{figure}
\centering
\includegraphics[width=0.4\textwidth,bb=100 100 400 550]{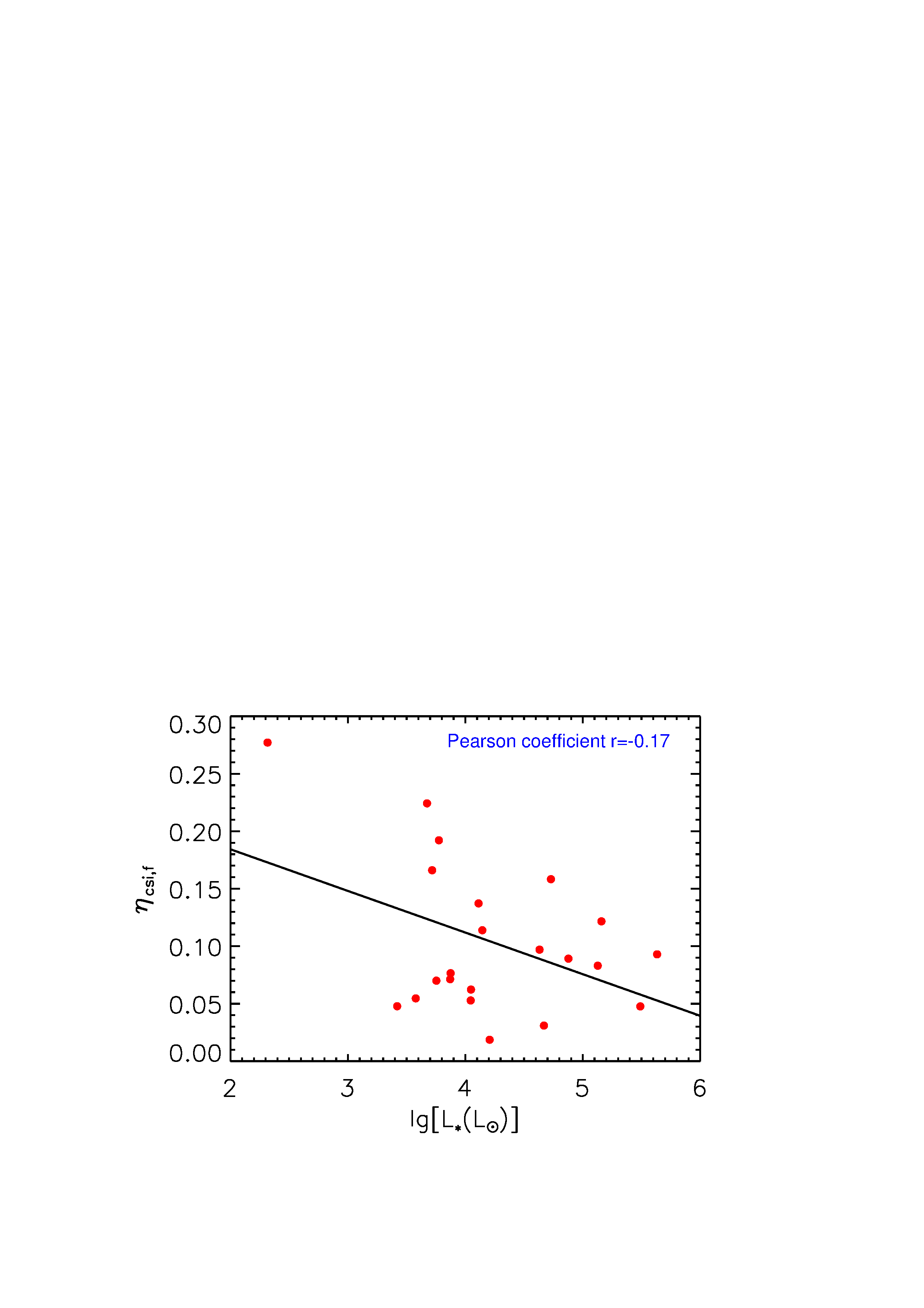}
 \caption{ Same as Figure~\ref{fig:csi_Lstar}
              but with BI Cyg, Mira, RS Per, SV Psc, VX Sgr and W Hya excluded.
              }
 \label{fig:csi_Lstara}
\end{figure}

We note that among our sample,
the {\it ISO}/SWS spectra of the following six stars
are rather smooth and show little crystalline silicate emission:
BI Cyg (see Figure~\ref{fig:Sloan1}),
Mira (see Figure~\ref{fig:Sloan1}),
RS Per (see Figure~\ref{fig:Sloan2}),
SV Psc (see Figure~\ref{fig:Sloan3}),
VX Sgr (see Figure~\ref{fig:Sloan3}), and
W Hya (see Figure~\ref{fig:Sloan3}).
With these six sources excluded, we have also performed correlation studies
for $\ffc$ with $\Mdustloss$ (see Figure~\ref{fig:csi_mloss1a})
and for $\ffcp$ with $\Mdustloss$ (see Figure~\ref{fig:csi_mloss2a})
as well as for $\ffc$ with $T_\star$ (see Figure~\ref{fig:csi_Tstara})
and $L_\star$ (see Figure~\ref{fig:csi_Lstara}).
Similar to the results for the full sample of 28 objects,
no correlation is found.

A visual inspection of the {\it ISO}/SWS spectra
of our 28 sample stars shows that the solid-state
spectral features of oxides are not apparent
in the observed spectra.
While the 11$\mum$ feature of amorphous Al$_{2}$O$_{3}$
and the 19.5$\mum$ feature of Mg$_{x}$Fe$_{1-x}$O
(Posch et al.\ 2002) may be confused
with the identification of some crystalline
silicate features, the 13$\mum$ feature
of spinel (Posch et al.\ 1999, Fabian et al.\ 2001)
does not seem to be present in most of
our sources unless it blends with
the broad 10$\mum$ amorphous silicate feature
and the sharp crystalline silicate features
at 10.5--11.5$\mum$
(see Figures~\ref{fig:Sloan1}--\ref{fig:Sloan4}).
Nevertheless, we have explored the effects of
the possible presence of oxides in the dust shells
of our sample stars on the correlation between
$\ffc$ and $\Mdustloss$
by deriving the upper limits of the fluxes of
the 11, 13, and 19.5$\mum$ features of oxides.
To this end, we take the emission
profiles of amorphous Al$_{2}$O$_{3}$ at 11$\mum$,
spinel at 13$\mum$ (together with
a minor feature at 16.5$\mum$),
and Mg$_{x}$Fe$_{1-x}$O at 19.5$\mum$
observed in the prototypical low-mass-loss-rate
AGB star g Her (see Figure~3 of Waters 2004).
We add these emission profiles in the decompositional
fitting scheme (see eq.\,\ref{eq:Fnu})
described in \S\ref{sec:CSi}
and once again decompose the {\it ISO}/SWS
spectra of all 28 sources. This allows us to
derive the upper limits of the total fluxes
emitted by these oxide dust species
and consequently, also to obtain
a new set of $\Fcsi$ and $\Fasi$
which are expected to be somewhat
reduced compared to that derived
earlier in \S\ref{sec:CSi}.
We then estimate the silicate crystallinity
$\ffc$ from the newly-determined
$\Fcsi$ and $\Fasi$ and correlate
it with the dust mass loss rate.
Similar to that derived earlier
in this section, no correlation
is found between $\ffc$ and $\Mdustloss$.
This demonstrates that our conclusion
of no correlation between the silicate crystallinity
and the dust mass loss rate is not affected
by ignoring the possible presence of oxides
in our sample sources.

\begin{figure}
\centering
\includegraphics[width=0.5\textwidth,bb=50 300 550 700]{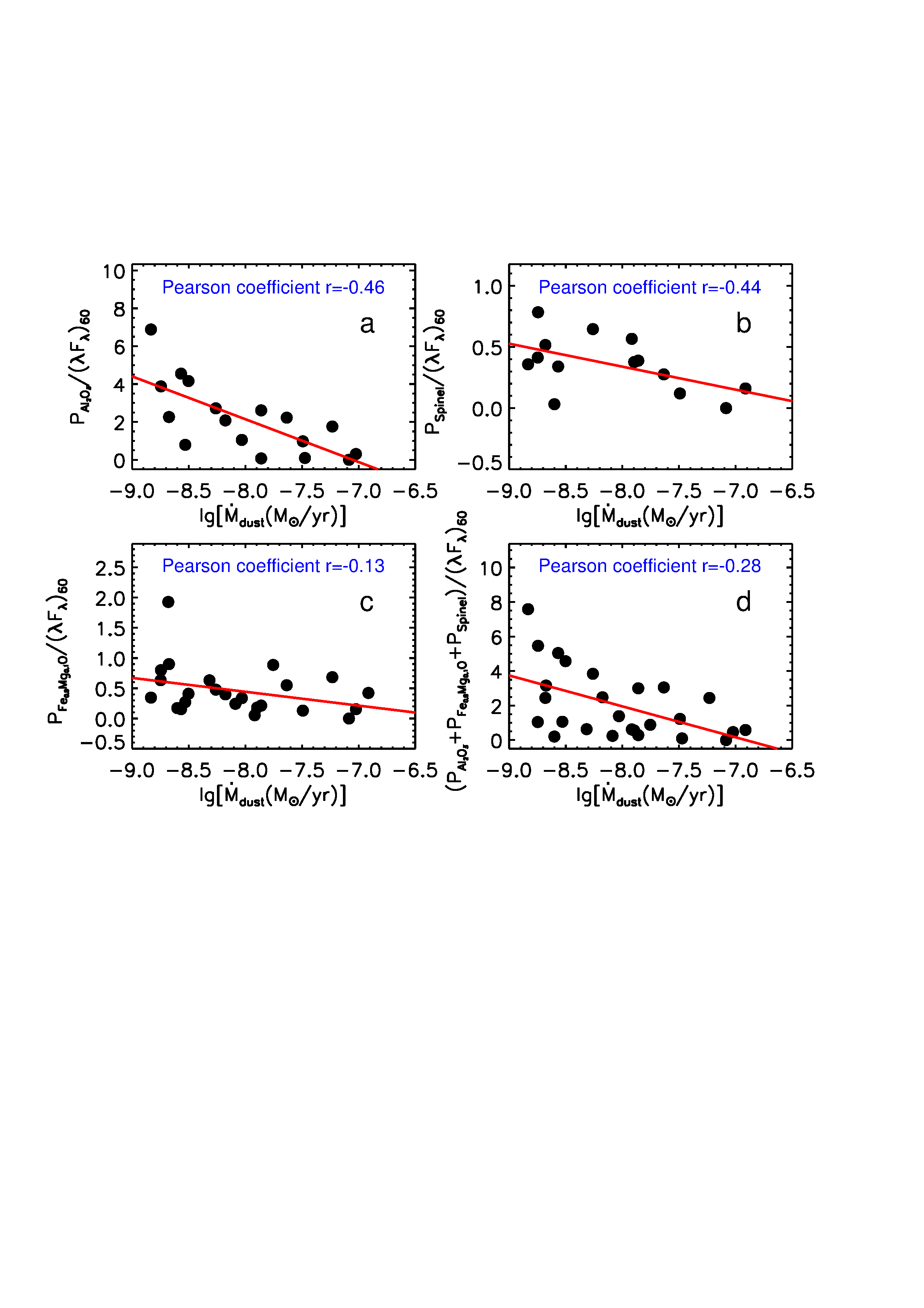}
 \caption{Correlation of $\Mdustloss$
               with the wavelength-integrated fluxes
               (normalized to the {\it IRAS} 60$\mum$ emission)
               of the 11, 13, and 19.5$\mum$ emission features
               of amorphous Al$_{2}$O$_{3}$ (a), spinel (b), Mg$_{x}$Fe$_{(1-x)}$O (c),
                and all three oxide species as a whole (d).
               }
\label{fig:oxides}
\end{figure}

\begin{figure}
\centering
\includegraphics[scale=0.25,bb=0 0 1200 700,angle=90]{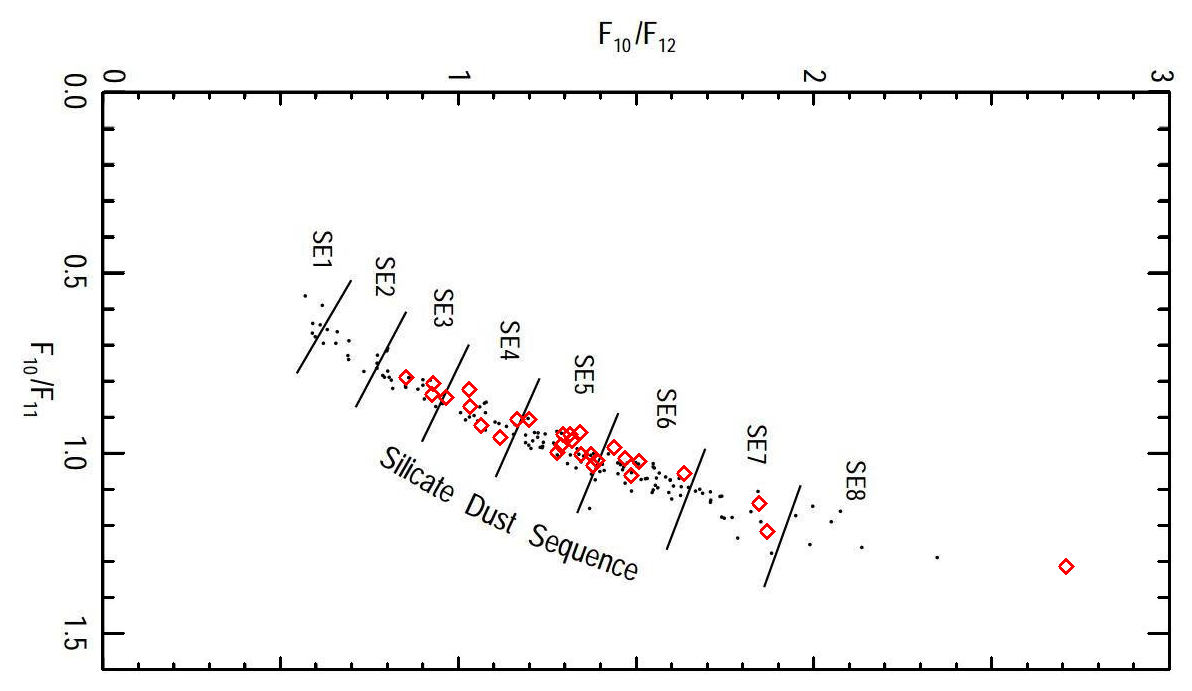}
\caption{The ``silicate dust sequence'', as defined by
               the flux ratios $F_{10}/F_{11}$ and $F_{10}/F_{12}$ of
               hundreds of oxygen-rich evolved stars
               (black dots; Egan \& Sloan 2001).
               The sequence is divided into eight segments
               (SE1, SE2, ..., SE8). Most (25/28) of our sources
               (red diamonds) fall in SE3--SE6.
               }
\label{fig:sds}
\end{figure}

Egan \& Sloan (2001) and Nowotny et al.\ (2015)
argued for a correlation between the mass loss rates
of oxygen-rich AGB stars and the mineralogical
composition of their dust shells; particularly,
they suggested that Al- and MgAl-oxides are
predominantly seen in the shells of AGB stars
with low mass-loss rates.
We have also investigated the possible
correlation between $\Mdustloss$ and
the wavelength-integrated fluxes
(normalized to the {\it IRAS} 60$\mum$ emission)
of the 11, 13 and 19.5$\mum$ features
of amorphous Al$_{2}$O$_{3}$, spinel, and Mg$_{x}$Fe$_{1-x}$O.
As shown in Figure~\ref{fig:oxides},
they do not appear to be correlated.
At most, amorphous Al$_{2}$O$_{3}$ and Mg$_{x}$Fe$_{1-x}$O
show a weak trend of decreasing with $\Mdustloss$.
We have also considered all three oxide species
as a whole by summing up the wavelength-integrated
fluxes of all three features.
Again, no correlation is found
(see Figure~\ref{fig:oxides}d).
However, we note that our sample may be biased to stars
of relatively high mass loss rates, too high to allow
an appreciable amount of oxide dust to be present
in their dust shells.
To examine this possibility,
we derive the stellar-continuum-subtracted fluxes
($F_{10}$, $F_{11}$, and $F_{12}$) of the dust shells of
our sample stars at 10, 11, and 12$\mum$ and
then plot their flux ratios $F_{10}/F_{11}$ and
$F_{10}/F_{12}$ in the ``silicate dust sequence''
diagram (see Figure~\ref{fig:sds}).
Sloan \& Price (1995, 1998) found that
the flux ratios $F_{10}/F_{11}$ and $F_{10}/F_{12}$
of hundreds of the oxygen-rich sources
detected by {\it IRAS}  through
the {\it Low Resolution Spectrometer}
(LRS) fall along one smooth progression,
which they called the ``silicate dust sequence''.

As shown in Figure~\ref{fig:sds},
Sloan \& Price (1995, 1998) divided
the silicate dust sequence into
eight segments (SE1, SE2, ..., SE8)
and classified these segments into three groups:
SE1--SE2 at the bottom end of
the silicate dust sequence
are dominated by stars rich in
amorphous alumina,
SE3--SE6 stand for stars
showing ``structured'' silicate emission
(i.e., crystalline silicates),
and SE7--SE8 on the upper part of
the silicate dust sequence
are for stars mostly containing
amorphous silicate grains
(see Egan \& Sloan 2001).
As illustrated in Figure~\ref{fig:sds},
while 25/28 of our sources fall in
the crystalline-silicate-rich SE3--SE6 class,
none of our sources fall in
the alumina-rich SE1--SE2 class.
The former confirms the source-selection
criterion of ``most of our sources exhibiting
prominent crystalline silicate emission features''
(see \S\ref{sec:sample}).
The latter confirms that oxides are not important in
the dust shells around our sources, justifying
the neglect of oxides in deriving $\Fcsi$
(see \S\ref{sec:CSi}).
Finally, we note that our results do not necessarily
falsify Nowotny et al.\ (2015) who argued for
a potential dependence of the dust composition
on the mass loss rates as our sample stars do not
cover the alumina-rich, low-$\Mdustloss$
SE1 and SE2 segments.
In future work, it will be useful to
carefully select a sample of evolved stars
of which the oxide spectral features are present
and exhibit a wide range of intensities.

Finally, we hypothesize that the silicate crystallinity
may be related to the stellar mass loss history:
one may expect more crystalline silicates
in an episodic mass loss event due to
the local density enhancement in regions
close the star where amorphous silicates
can be annealed. For stars experiencing
episodic mass losses, the dust mass loss rate
is not  a good description of the actual
mass loss history since $\Mdustloss$ is
obtained from dividing the total dust mass
of the shell by the dust outflow timescale
under the assumption of
a continuous mass loss process.
Therefore, one would not expect
a tight correlation between the silicate
crystallinity and $\Mdustloss$.
One way to test this hypothesis
is to perform high spatial resolution
mid-IR imaging observations of AGB stars
of high crystallinity. If silicate crystallinity
is indeed associated with episodic mass
loss events, one expects to see layered
structures or local clumps
with an enhanced dust density
in the mid-IR images of the dust shells
around these AGB stars.

\section{Summary}
\label{sec:summary}
We have selected 28 O-rich evolved stars
in the solar neighbourhood to explore
the relation between the silicate crystallinity
and the stellar mass loss rate.
The SED of each source
from the near-UV to the far-IR
has been fitted with the 2DUST model
to derive its mass loss rate.
Assuming the silicate crystallinity
to be the ratio of the fluxes
emitted in the crystalline silicate features
to that in the amorphous silicate features,
we have determined the silicate crystallinity
for each source.
With a Pearson correlation coefficient
of {\bf $r\approx-0.24$}, it is found that
the silicate crystallinity does not appear
to correlate with the mass loss rate.
Moreover, the silicate crystallinity
does not correlate with the stellar temperature
or the stellar luminosity.


\section*{Acknowledgements}
We dedicate this paper to the 80th birthday
of Prof. Jingyao Hu who first detected
the 11.3$\mum$ crystalline silicate feature
outside of the solar system.
We thank A.~Mishra, S.~Wang, Y.X.~Xie,  Z.Z.~Shao
and the referee
for very helpful suggestions and comments.
This work is supported by NSFC through
Projects 11173007, 11373015, 11533002,
and 973 Program 2014CB845702.
%









%
%

\begin{landscape}
\begin{table}
\caption{Photometric Data of Our Sample Stars}
\label{tab:fluxdata}
\begin{tabular}{lllllllllllllllllllll}
\hline\hline
Star & \multicolumn{2}{c}{J2000} & Type & $d^{0}$ & \multicolumn{5}{c}{F$_{\nu}$ (Johnson): mag} & \multicolumn{3}{c}{F$_{\nu}$ (2MASS): mag} & \multicolumn{4}{c}{F$_{\nu}$ (WISE): mag} & \multicolumn{4}{c}{F$_{\nu}$ (IRAS): Jy}\\
\cline{2-3}\cline{6-10}\cline{11-13}\cline{14-17}\cline{18-21}
         &   ra  &   dec   &       & (pc) & U & B & V & R & I & J & H & K & W1 & W2 & W3 & W4 & 12$\mum$ & 25$\mum$ & 60$\mum$ & 100$\mum$\\
\hline
AH Sco   & 17 11 17.02 & -32 19 30.71 & RSG & 2600$^{1}$ & -- &10.0 & 8.10 &-- & -- & 1.88& 0.70  & 0.30 & 1.26 & 0.87 & -2.45 & -4.16 & 629.7 & 349.6 & 73.3 & 28.9 \\
BI Cyg   & 20 21 21.88  & +36 55 55.77  & RSG & 1580$^{8}$ & -- &11.5 & 8.40 &-- & -- & 2.35 & 1.15 & 0.62 & 0.97 & 0.09 & -2.11 & -3.62 & 334.6 & 244.9 & 51.3 &92.9U\\
FI Lyr   & 18 42 04.83 & +28 57 29.81 & RSG & 410$^{2}$  & -- &10.8 & 9.58 &-- & -- & 2.04 & 1.12 & 0.71 & 2.60 & 0.19 & -0.42 & -1.35 & 93.4 & 54.9 & 7.23 & 1.57:\\
Mira     & 02 19 20.79 & -02 58 39.50 & AGB & 120$^{3}$  & -- &7.63 & 6.54&-- & -- & -0.73 & -1.57 &-2.21 & 1.87 & 0.71 &-2.38 & -5.06 & 4881.0 & 2261.0 &300.8&88.4\\
PZ Cas   & 23 44 03.28 & +61 47 22.18 & RSG & 2710$^{1}$ & 12.8 & 11.5 & 8.90&6.08 &3.90 &2.42& 1.53 & 1.00 & 2.44 & 1.45 & -2.27 & -4.26 & 373.0 & 398.2 & 96.5 & 39.3\\
R Aql    & 19 06 22.25 & +08 13 48.01 & AGB & 310$^{1}$  & 8.06 &7.69 &6.09&-- & -- &0.46 &-0.36 & -0.83 & 2.11 & 0.72 & -2.23 & -3.15 & 401.7 & 244.6 & 139.7L & 83.1U\\
R Cas    & 23 58 24.87 & +51 23 19.70 & AGB & 359$^{1}$  & 6.71 & 6.63 & 4.80 &-- & -- & -0.40 &-1.40 &-1.92 &2.36 &1.07& -2.73 & -4.83 & 1341.0 & 554.6 & 102.8 & 38.9\\
RS Per   & 02 22 24.30   & +57 06 34.36   & RSG & 2350$^{4}$ & 12.4 & 10.1 & 7.82 &-- & -- & 3.08 & 2.12 & 1.68 & 1.50 & 1.56 & -0.56 & -1.81 & 74.4 & 47.8 & 9.93 & 2.86\\
RV Cam 	 & 04 30 41.68 & +57 24 42.26 & AGB & 1219$^{5}$ & -- & 9.65 & 8.20 &-- & -- & 1.67 & 0.60 & 0.41 & 1.06 & 1.27 & 0.12 & -1.39 & 58.7 & 34.6 & 7.71 & 3.95\\
RW Cep   & 22 23 07.02 & +55 57 47.62 & RSG & 115$^{2}$  & 11.3 & 8.87 & 6.65 &4.94 &3.78 & 2.83 & 2.22 & 1.88 & 2.17 & 1.63 &-0.86 &-2.41 &97.4&91.6 & 27.4 & 13.5 \\
RW Cyg	 & 20 28 50.59 & +39 58 54.43 & RSG & 1118$^{6}$ & 13.0 & 10.6& 8.00 &5.32&3.25&2.06& 0.94& 0.48 & 1.30 & 1.26 & -2.24 & -3.42 & 298.4 & 189.8 & 60.7 & 97.0U\\
RX Boo	 & 14 24 11.63 & +25 42 13.41 & AGB & 169$^{3}$  & -- & 9.23 & 8.60 &-- & -- & -0.59 & -1.55 & -1.96 & 2.18 & 1.15 & -2.45 & -4.16 & 846.5 & 419.3 & 69.2 & 25.8\\
S Per	 & 02 22 51.71 & +58 35 11.45 & RSG & 2300$^{7}$ & 13.2 & 10.6 & 7.90&-- & --  & 3.14 & 1.85 & 1.33 & 0.84 & 0.84 & -2.44 & -3.70 & 339.4 & 233.2 & 40.6 & 15.0\\
SU Per   & 02 22 06.90  & +56 36 14.87  & RSG & 1900$^{8}$ & 13.9 & 11.6& 9.40 &-- & -- & 2.76 & 1.93 & 1.39 & 0.98 & 1.66 & -0.62 & -1.26 & 48.7 & 30.7 & 6.87 & 6.73U\\
SV Peg   & 22 05 42.08 & +35 20 54.53 & AGB & 250$^{9}$  & -- & 10.1 & 9.20 &-- & -- & 1.11 & -0.09 & -0.55 & 2.26 & 0.87 & -1.93 & -2.83 & 264.7 & 146.2 & 23.6 & 9.94\\
SV Psc	 & 01 46 35.34   & +19 05 04.52   & AGB & 362$^{3}$  & -- & 10.0 & 8.77 &8.41& -- & 2.02 & 1.01 & 0.72 & 1.04 & 0.13 & -0.01 & -1.53 & 76.7 & 39.9 & 6.60 & 2.80\\
TY Dra   & 17 37 00.12  & +57 44 25.30  & AGB & 430$^{8}$  & -- & 10.6 & 9.30 &-- & -- & 2.41 & 1.47 & 1.08 & 2.82 & 0.49 & 0.11 & -1.66 & 66.3 & 45.8 & 7.56 & 2.67\\
U Her    & 16 25 47.47 & +18 53 32.86 & AGB & 461$^{10}$  & 8.85 & 8.23 & 6.70 &-- & -- & 1.01 & 0.23 & -0.27 & 2.19 & 0.76 & -2.51 & -3.18 & 499.8 & 179.5 & 27.2 & 9.70\\
U Lac    & 22 47 43.43 & +55 09 30.30 & RSG & 885$^{5}$  & 13.2 & 11.7 & 9.40 &-- & -- & 2.90 & 2.09 & 1.58 & 2.70 & 1.32 & -0.88 & -2.00 & 124.0 & 61.5 & 9.04 & 10.6U\\
VX Sgr   & 18 08 04.05 & -22 13 26.63 & RSG & 1570$^{7}$ & 11.7 & 9.41 & 6.52 &3.90&2.11& 1.48 & 0.42 & -0.17 & 2.15 & 1.14 & -2.48 & -5.31 & 2738.0 & 1385.0 & 262.7 & 82.3 \\
W Hor    & 02 44 14.75  & -54 18 04.11  & RSG & 364$^{3}$  & -- & 10.0 & 8.84 &-- & -- & 1.63 & 0.71 & 0.32 & 1.14 & 1.20 & -0.97 & -1.90 & 181.0 & 99.9 & 11.0 & 4.22 \\
W Hya    & 13 49 02.00 & -28 22 03.49 & AGB & 139$^{3}$  & -- & 8.97 & 7.70 &-- & -- & -1.59 & -2.56 & -3.05 & 1.49 & 0.51 & -2.76 & -5.54 & 4200.0 & 1189.0 & 195.0 & 72.3\\
W Per    & 02 50 37.89 & +56 59 00.27 & RSG & 650$^{3}$  & 14.8 & 12.2 & 9.62 &6.99& 4.75 & 3.45 & 2.00 & 2.00 & 2.78 & 1.98 & -1.06 & -2.28 & 90.6 & 78.9 & 14.9 & 5.01\\
X Her    & 16 02 39.17 & +47 14 25.28 & AGB & 141$^{6}$  & -- & 7.64 & 6.58 &-- & -- & -0.12 & -0.92 & -1.20 & 2.30 & 1.03 & -2.45 & -3.17 & 485.0 & 241.0 & 39.4 & 18.3 \\
X Oph    & 18 38 21.12 & +08 50 02.75 & AGB & 235$^{3}$  & 8.61 & 7.72 & 6.40 &-- & -- & 0.62 & -0.31 & -0.79 & 1.93 & 0.77 & -2.38 & -3.04 & 409.0 & 146.0 & 22.6 & 9.47\\
YZ Per   & 02 38 25.42 & +57 02 46.18 & RSG & 1850$^{5}$ & 15.0 & 12.4 & 10.0 &-- & -- & 3.32 & 2.30 & 2.03 & 2.47 & 2.09 & -0.36 & -1.11 & 38.9 & 26.1 & 5.28 & 2.52:\\
Z Cas    & 23 44 31.59    & +56 34 52.70    & AGB & 797$^{9}$  & -- & 18.3 & 8.50 &12.6& -- & 2.68 & 1.80 & 1.23 & 2.86 & 2.13 & -0.88 & -1.55 & 69.2 & 40.3 & 6.70 & 4.55U\\
Z Cyg    & 20 01 27.50   & +50 02 32.69   & AGB & 930$^{9}$  & -- & 11.1 & 7.10 &-- & -- & 4.18 & 3.28 & 2.44 & 2.35 & 1.21 & -0.09 & -1.33 & 81.2 & 67.1 & 10.7 & 2.39 \\

\hline
\multicolumn{3}{l}{$^0$ Distance to Earth.}\\
\multicolumn{3}{l}{$^1$ Engels 1979}\\
\multicolumn{3}{l}{$^2$Ammons et al.\ 2006}\\
\multicolumn{3}{l}{$^3$Pickles \& Depagne 2010}\\
\multicolumn{3}{l}{$^4$Frinchaboy \& Majewski 2008}\\
\multicolumn{3}{l}{$^5$McDonald et al.\ 2012}\\
\multicolumn{3}{l}{$^6$Famaey et al.\ 2005}\\
\multicolumn{3}{l}{$^7$Richards et al.\ 2012}\\
\multicolumn{3}{l}{$^8$Jones et al.\ 2012}\\
\multicolumn{3}{l}{$^9$Kim et al.\ 2014}\\
\multicolumn{3}{l}{$^{10}$Palagi et al.\ 1993}\\
\end{tabular}
\end{table}
\end{landscape}
\begin{landscape}
\begin{table}
\begin{center}
\caption{Stellar and Circumstellar Parameters
         and the Dust Mass Loss Rates Derived
         from 2DUST as well as the Silicate
         Crystallinities $\ffc$ and $\ffcp$
         Derived from PAHFIT
         \label{tab:sedmod}}
\begin{tabular}{l c c c c c c c c c }
\hline\hline
Star & $T_\star$  & $L_\star$	& $R_\star$
     & $\rmin$ & $\rmax$ & $\rho_{\rm min}$
     & $\dot{M}_{\rm dust}$
     & $\ffc$$^{1}$
     & $\ffcp$$^{2}$ \\
     & [K] & [$L_\odot$]
     & [cm] & [$R_\star$] & [$R_\star$]
     & [$\g\cm^{-3}$] & [$M_\odot\yr^{-1}$]
     &  & \\
\hline
  AH Sco   &   2700   &  4.29E+5 &  2.1E+14  &	33.7	&	14697.0`	&	8.41E-21&	8.22E-8  &  0.093 &  0.102  \\
  BI Cyg   &   2990   &  2.22E+5 &  1.2E+14  &	40.2	&	2614.0		&	7.02E-21&	3.36E-8  &  0.080 &  0.088  \\
  FI Lyr   &   2780   &  5.21E+3 &  2.2E+13  &	36.0	&	1009.1		&	1.22E-20&	1.47E-9  &  0.166 &  0.199  \\
    Mira   &   3150   &  1.00E+4 &  2.3E+13  &	40.0	&	1614.6		&	3.58E-20&	9.32E-9  &  0.056 &  0.060  \\
  PZ Cas   &   3200   &  3.09E+5 &  1.3E+14  &	48.4	&	25631.5`	&	1.29E-20&	9.47E-8  &  0.048 &  0.050  \\
   R Aql   &   2750   &  1.12E+4 &  3.2E+13  &	34.7	&	2779.4		&	1.00E-20&	2.51E-9  &  0.062 &  0.066  \\
   R Cas   &   2700   &  4.67E+4 &  6.9E+13  &	43.1	&	1293.7		&	7.91E-21&	1.38E-8  &  0.031 &  0.032  \\
  RS Per   &   3100   &  1.02E+5 &  7.7E+13  &	54.9	&	3074.2		&	6.54E-21&	2.32E-8  &  0.236 &  0.310  \\
  RV Cam   &   3000   &  1.34E+5 &  9.4E+13  &	48.5	&	3396.4		&	4.02E-22&	1.67E-9  &  0.083 &  0.091  \\
  RW Cep   &   3300   &  2.07E+2 &  3.1E+12  &	56.4	&	22573.6	    &	5.16E-19&	3.05E-9  &  0.277 &  0.383  \\
  RW Cyg   &   2900   &  7.54E+4 &  7.5E+13  &	33.3	&	1995.1		&	2.56E-20&	3.22E-8  &  0.089 &  0.098  \\
  RX Boo   &   2850   &  1.29E+4 &  3.2E+13  &	35.2	&	1054.6		&	8.21E-21&	2.12E-9  &  0.137 &  0.159  \\
   S Per   &   3000   &  1.44E+5 &  9.7E+13  &	42.4	&	8474.8		&	1.72E-20&	5.86E-8  &  0.122 &  0.138  \\
  SU Per   &   2700   &  5.36E+4 &  7.3E+13  &	38.7	&	1937.1		&	8.55E-21&	1.38E-8  &  0.158 &  0.188  \\
  SV Peg   &   2500   &  5.95E+3 &  2.8E+13  &	30.1	&	1778.5		&	2.14E-20&	3.15E-9  &  0.192 &  0.238  \\
  SV Psc   &   2450   &  4.47E+3 &  2.6E+13  &	31.5	&	1261.8		&	1.37E-20&	1.80E-9  &  0.076 &  0.083  \\
  TY Dra   &   2720   &  2.63E+3 &  1.6E+13  &	40.6	&	2028.8		&	2.13E-20&	1.79E-9  &  0.048 &  0.050  \\
   U Her   &   2650   &  1.61E+4 &  4.2E+13  &	32.9	&	1120.6		&	2.16E-20&	8.18E-9  &  0.019 &  0.019  \\
   U Lac   &   2720   &  1.39E+4 &  3.7E+13  &	27.3	&	927.4		&	6.01E-20&	1.21E-8  &  0.114 &  0.129  \\
  VX Sgr   &   3150   &  3.58E+5 &  1.4E+14  &	31.2	&	6867.2		&	3.24E-20&	1.22E-7  &  0.020 &  0.021  \\
   W Hor   &   2500   &  5.67E+3 &  2.8E+13  &	33.3	&	997.5		&	2.83E-20&	4.83E-9  &  0.070 &  0.075  \\
   W Hya   &   2500   &  5.67E+3 &  2.8E+13  &	25.8	&	567.4		&	2.64E-20&	2.71E-9  &  0.192 &  0.238  \\
   W Per   &   2750   &  7.49E+3 &  2.6E+13  &	36.8	&	16544.7		&	3.54E-20&	6.67E-9  &  0.076 &  0.083  \\
   X Her   &   2750   &  3.78E+3 &  1.9E+13  &	33.4	&	1337.4		&	2.66E-20&	2.09E-9  &  0.055 &  0.058  \\
   X Oph   &   2900   &  7.42E+3 &  2.4E+13  &	44.6	&	1337.1		&	1.33E-20&	2.95E-9  &  0.071 &  0.077  \\
  YZ Per   &   2500   &  4.28E+4 &  7.7E+13  &	36.2	&	2171.1		&	8.29E-21&	1.27E-8  &  0.097 &  0.107  \\
   Z Cas   &   2500   &  1.11E+4 &  3.9E+13  &	30.6	&	1376.6		&	1.94E-20&	5.49E-9  &  0.053 &  0.056  \\
   Z Cyg   &   2450   &  4.72E+3 &  2.6E+13  &	37.9	&	1628.6		&	8.77E-20&	1.76E-8  &  0.224 &  0.289  \\

\hline
\multicolumn{3}{l}{$^{1}$ $\ffc\equiv \Fcsi/\left(\Fcsi+\Fasi+\Fcont\right)$ }\\
\multicolumn{3}{l}{$^{2}$ $\ffcp\equiv \Fcsi/\left(\Fcsi+\Fasi\right)$ }\\
\end{tabular}
\end{center}
\end{table}
\end{landscape}

\begin{table}
\begin{center}
\caption{The Visual Extinction $A_V$
         and the $J-K$ Color Index
         of Each Object
         \label{tab:Av}
         }
\begin{tabular}{lcccccl}
\hline\hline
Star   & $J_{\rm obs}$ & $K_{\rm obs}$
       & $\left(J-K\right)_{\rm obs}$
       & $\left(J-K\right)_{\rm int}$
       & $A_V$& Sources  \\
\hline
AH Sco   &  1.88	&0.30	& 1.58	&  1.24    &    2.02   &  {\it 2MASS}      \\
BI Cyg   &  2.35	&0.62	& 1.73 	&  1.04    &    4.06    &  {\it 2MASS}      \\
FI Lyr   &  2.04	&0.72	& -{-}	&  -{-}    &    0.62     &  {\it NED}      	  	       \\
Mira     &  -0.73	&-2.21	& -{-}	&  -{-}    &    0.075     &  {\it NED}      		       \\
PZ Cas   &  2.42	&1.00	& 1.42 	&  0.98    &    2.59    &  {\it 2MASS}      \\
R Aql    &  0.46	&-0.83	& 1.29 	&  1.27   &    0.15     &  {\it 2MASS}      \\
R Cas    &  -0.40 &-1.92	& -{-} 	&  -{-}    &    0.60     &  {\it NED}      		       \\
RS Per   &  3.08	&1.68	& 1.4  	&  1.04    &    2.12    &  {\it 2MASS}      \\
RV Cam   &  1.67	&0.41	& 1.26	&  1.18    &    0.44     &  {\it 2MASS}      \\
RW Cep   &  2.83	&1.88	& 0.95 	&  0.65    &    1.76     &  {\it 2MASS}      \\
RW Cyg   &  2.06	&0.48	& 1.58 	&  0.98    &    3.53     &  {\it 2MASS}      \\
RX Boo   &  -0.59	&-1.96	& -{-} 	&  -{-}    &    0.059     &  {\it NED}      		       \\
S Per    &  3.14	&1.33	& 1.81 	&  1.07    &    4.35    &  {\it 2MASS}      \\
SU Per   &  2.76	&1.39	& 1.37 	&  0.98    &    2.29    &  {\it 2MASS}      \\
SV Peg   &  1.11	&-0.55	& -{-}	&  -{-}    &    0.45     &  {\it NED}      		       \\
SV Psc   &  2.02	&0.72	& -{-}	&  -{-}    &    0.14     &  {\it NED}      		       \\
TY Dra   &  2.41	&1.08	& -{-}	&  -{-}    &    0.20     &  {\it NED}      		       \\
U Her    &  1.01	&-0.27	& 1.28 	&  1.27   &    0.16      &  {\it NED}      		       \\
U Lac    &  2.90		&1.58	& 1.32 	&  1.04    &    1.65    &  {\it 2MASS}      \\
VX Sgr   &  1.48	&-0.17	& 1.65 	&  1.1     &    3.23     &  {\it 2MASS}      \\
W Hor    &  1.63	&0.32	& -{-}	&  -{-}    &    0.12    &  {\it NED}      		       \\
W Hya    &  -1.59 &-3.05	& -{-} 	&  -{-}    &    0.22     &  {\it NED}      		       \\
W Per    &  3.45	&2.00	& 1.45 	&  0.98    &    2.76    &  {\it 2MASS}      \\
X Her    &  -0.12	&-1.32	& -{-}	&  -{-}    &    0.046     &  {\it NED}      		       \\
X Oph    &  0.62	&-0.79	& 1.41 	&  1.02    &    2.29    &  {\it 2MASS}      \\
YZ Per   &  3.32	&2.03	& 1.29 	&  0.81    &    2.82    &  {\it 2MASS}      \\
Z Cas    &  2.68	&1.23	& 1.45 	&  1.27    &    1.06    &  {\it 2MASS}      \\
Z Cyg    &  4.18	&2.44	& -{-}	&  -{-}    &    0.32     &  {\it NED}      		       \\

\hline
\end{tabular}
\end{center}
\end{table}
\clearpage
\begin{table}
\footnotesize
\begin{center}
\caption{ Comparison of the Dust Moss Loss Rates ($\Mdustloss$)
             Derived in This Work with That Reported in the Literature.
             }
\label{tab:MlossComp}
\begin{tabular}{c c c l c c }
\hline\hline
Object   & $\Mdustloss$ &  References  &   Methodology   &   Type    &   $\gastodust$\\
        & ($\Msun$/yr)&              &               \\
\hline
AH Sco   & 8.22$\times10^{-8}$   & This work  & IR SED: $\lambda\leq100\mum$ (2DUST)    &  -{-} & -{-}  \\
         & 3.87$\times10^{-8}$   & Jura \& Kleinmann (1990)     & $F_{\rm \nu}(60\mum)$ (JK90)$^{a}$    & RSG    &  200   \\
BI Cyg   & 3.36$\times10^{-8}$   & This work  & IR SED: $\lambda$\textless $100\mum$ (2DUST)    &  -{-} & -{-}    \\
         & 2.30$\times10^{-8}$   & Mauron \& Josselin (2011)    & $F_{\rm \nu}(60\mum)$ (JK90)$^{a}$    & RSG    &  200   \\
         & 3.00$\times10^{-8}$   & Jura \& Kleinmann (1990)     & $F_{\rm \nu}(60\mum)$ (JK90)$^{a}$    & RSG    &  200   \\
FI Lyr   & 1.47$\times10^{-9}$   & This work  & IR SED: $\lambda\leq100\mum$ (2DUST)    &  -{-} &  -{-}   \\
         & 7.55$\times10^{-9}$   & Heras \& Hony (2005)     & IR SED: $2.38-45.2\mum$ (DUSTY)    & AGB    & 47.7    \\
Mira     & 9.32$\times10^{-9}$   & This work  & IR SED: $\lambda\leq100\mum$ (2DUST)    &  -{-} &  -{-}   \\
         & 6.96$\times10^{-9}$   & Heras \& Hony (2005)     & IR SED: $2.38-45.2\mum$ (DUSTY)    & AGB    & 306.1     \\
         & 1.21$\times10^{-9}$   & Suh (2004)       & IR SED: $2.38-197\mum$ (CSDUST3)   & AGB    &  -{-}      \\
         & 2.85$\times10^{-9}$   & Knapp \& Morris (1985)     & CO: $J=1\rightarrow 0$        & AGB    &  200      \\
PZ Cas   & 9.47$\times10^{-8}$   & This work  & IR SED: $\lambda\leq100\mum$ (2DUST)    &  -{-} &  -{-}    \\
         & 1.30$\times10^{-7}$   & Mauron \& Josselin (2011)    & $F_{\rm \nu}(60\mum)$ (JK90)$^{a}$    & RSG    &  200   \\
         & 5.00$\times10^{-8}$   & Jura \& Kleinmann (1990)     & $F_{\rm \nu}(60\mum)$ (JK90)$^{a}$    & RSG    &  200   \\
R Aql    & 2.51$\times10^{-9}$   & This work  & IR SED: $\lambda$\textless$  60\mum$ (2DUST)    &  -{-} &  -{-}   \\
         & 4.00$\times10^{-9}$   & Lane et al. (1987)       & CO: $J=1\rightarrow 0$        & AGB    &  200      \\
R Cas    & 1.38$\times10^{-8}$   & This work  & IR SED: $\lambda\leq100\mum$ (2DUST)    &  -{-} &  -{-}   \\
         & 2.25$\times10^{-9}$   & Knapp \& Morris (1985)     & CO: $J=1\rightarrow 0$        & AGB    &  200      \\
RS Per   & 2.32$\times10^{-8}$   & This work  & IR SED: $\lambda\leq100\mum$ (2DUST)    &  -{-} &  -{-}    \\
         & 1.00$\times10^{-8}$   & Mauron \& Josselin (2011)    & $F_{\rm \nu}(60\mum)$ (JK90)$^{a}$    & RSG    &  200   \\
RV Cam   & 1.67$\times10^{-9}$   & This work  & IR SED: $\lambda\leq100\mum$ (2DUST)    &  -{-} &  -{-}   \\
         & 1.42$\times10^{-9}$   & Heras \& Hony (2005)     & IR SED: $2.38-45.2\mum$ (DUSTY)    & AGB    &  91.8   \\
         & 1.25$\times10^{-9}$   & Olofsson  et al. (2002) & CO: $v=1$ (Monte Carlo)        & AGB    &  200   \\
RW Cep   & 3.05$\times10^{-9}$   & This work  & IR SED: $\lambda\leq100\mum$ (2DUST)    &  -{-} &  -{-}   \\
RW Cyg   & 3.22$\times10^{-8}$   & This work  & IR SED: $\lambda$\textless$ 100\mum$ (2DUST)    &  -{-} &   -{-}   \\
         & 1.60$\times10^{-8}$   & Mauron \& Josselin (2011)    & $F_{\rm \nu}(60\mum)$ (JK90)$^{a}$    & RSG    &  200   \\
         & 5.00$\times10^{-8}$   & Jura \& Kleinmann (1990)     & $F_{\rm \nu}(60\mum)$ (JK90)$^{a}$    & RSG    &  200   \\
RX Boo   & 2.12$\times10^{-9}$   & This work  & IR SED: $\lambda\leq100\mum$ (2DUST)    &  -{-} &  -{-}   \\
         & 4.73$\times10^{-9}$   & Heras \& Hony (2005)     & IR SED: $2.38-45.2\mum$ (DUSTY)    & AGB    &  69.7   \\
         & 2.50$\times10^{-9}$   & Olofsson  et al. (2002) & CO: $v=1$ (Monte Carlo)        & AGB    &  200   \\
         & 1.65$\times10^{-9}$   & Knapp \& Morris (1985)     & CO: $J=1\rightarrow 0$        & AGB    &  200      \\
S Per    & 5.86$\times10^{-8}$   & This work  & IR SED: $\lambda\leq100\mum$ (2DUST)    &  -{-} & -{-}    \\
         & 3.40$\times10^{-8}$   & Mauron \& Josselin (2011)    & $F_{\rm \nu}(60\mum)$ (JK90)$^{a}$    & RSG    &  200   \\
         & 3.50$\times10^{-8}$   & Jura \& Kleinmann (1990)     & $F_{\rm \nu}(60\mum)$ (JK90)$^{a}$    & RSG    &  200   \\
SU Per   & 1.38$\times10^{-8}$   & This work  & IR SED: $\lambda$\textless$ 100\mum$ (2DUST)    &  -{-} &  -{-}   \\
         & 3.85$\times10^{-9}$   & Mauron \& Josselin (2011)    & $F_{\rm \nu}(60\mum)$ (JK90)$^{a}$    & RSG    &  200   \\
         & 3.02$\times10^{-8}$   & Jura \& Kleinmann (1990)     & $F_{\rm \nu}(60\mum)$ (JK90)$^{a}$    & RSG    &  200   \\
\hline
\end{tabular}
\end{center}
\end{table}

\clearpage
\begin{table}
\footnotesize
\begin{center}
\contcaption{A table continued from the previous one.}
\begin{tabular}{c c c l c c}
\hline\hline
Object    & $\Mloss$ &  References  &   Methodology   &   Type & $\gastodust$ \\
        & ($\Msun$/yr)&              &               \\
\hline
SV Peg   & 3.15$\times10^{-9}$   & This work  & IR SED: $\lambda\leq100\mum$ (2DUST)    &  -{-} &  -{-}   \\
         & 2.51$\times10^{-9}$   & Heras \& Hony (2005)     & IR SED: $2.38-45.2\mum$ (DUSTY)    & AGB    &  75.7   \\
         & 1.50$\times10^{-9}$   & Olofsson  et al. (2002) & CO: $v=1$ (Monte Carlo)        & AGB    &  200   \\
SV Psc   & 1.80$\times10^{-9}$   & This work  & IR SED: $\lambda\leq100\mum$ (2DUST)    &  -{-} & -{-}    \\
         & 3.31$\times10^{-9}$   & Heras \& Hony (2005)     & IR SED: $2.38-45.2\mum$ (DUSTY)    & AGB    &  48.4   \\
TY Dra   & 1.79$\times10^{-9}$   & This work  & IR SED: $\lambda\leq100\mum$ (2DUST)    &  -{-} & -{-}    \\
         & 1.20$\times10^{-9}$   & Olofsson  et al. (2002) & CO: $v=1$ (Monte Carlo)        & AGB    &  500   \\
U Her    & 8.18$\times10^{-9}$   & This work  & IR SED: $\lambda\leq100\mum$ (2DUST)    &  -{-} & -{-}    \\
         & 1.30$\times10^{-8}$   & Lane et al. (1987)      & CO: $J=1\rightarrow 0$        & AGB    &  200      \\
U Lac    & 1.21$\times10^{-8}$   & This work  & IR SED: $\lambda$\textless$ 100\mum$ (2DUST)    &  -{-} & -{-}    \\
         & 1.15$\times10^{-8}$   & Mauron \& Josselin (2011)    & $F_{\rm \nu}(60\mum)$ (JK90)$^{a}$    & RSG    &  200   \\
         & 1.00$\times10^{-8}$   & Jura \& Kleinmann (1990)     & $F_{\rm \nu}(60\mum)$ (JK90)$^{a}$    & RSG    &  200   \\
VX Sgr   & 1.22$\times10^{-7}$   & This work  & IR SED: $\lambda\leq100\mum$ (2DUST)    &  -{-} & -{-}     \\
         & 1.00$\times10^{-7}$   & Mauron \& Josselin (2011)    & $F_{\rm \nu}(60\mum)$ (JK90)$^{a}$    & RSG    &  200   \\
         & 2.00$\times10^{-7}$   & Jura \& Kleinmann (1990)     & $F_{\rm \nu}(60\mum)$ (JK90)$^{a}$    & RSG    &  200   \\
W Hor    & 4.83$\times10^{-9}$   & This work  & IR SED: $\lambda\leq100\mum$ (2DUST)    &  -{-} & -{-}    \\
W Hya    & 2.71$\times10^{-9}$   & This work  & IR SED: $\lambda\leq100\mum$ (2DUST)    &  -{-} & -{-}    \\
         & 6.18$\times10^{-9}$   & Heras \& Hony (2005)     & IR SED: $2.38-45.2\mum$ (DUSTY)    & AGB    &  142.5   \\
         & 3.50$\times10^{-10}$  & Olofsson  et al. (2002) & CO: $v=1$ (Monte Carlo)        & AGB    &  200   \\
W Per    & 6.67$\times10^{-9}$   & This work  & IR SED: $\lambda\leq100\mum$ (2DUST)    &  -{-} & -{-}     \\
         & 1.05$\times10^{-8}$   & Mauron \& Josselin (2011)    & $F_{\rm \nu}(60\mum)$ (JK90)$^{a}$    & RSG    &  200   \\
X Her    & 2.09$\times10^{-9}$   & This work  & IR SED: $\lambda\leq100\mum$ (2DUST)    &  -{-} & -{-}    \\
         & 3.00$\times10^{-10}$  & Olofsson  et al. (2002) & CO: $v=1$ (Monte Carlo)        & AGB    &  500   \\
X Oph    & 2.95$\times10^{-9}$   & This work  & IR SED: $\lambda\leq100\mum$ (2DUST)    &  -{-} & -{-}    \\
         & 4.70$\times10^{-7}$   & Heras \& Hony (2005)     & IR SED: $2.38-45.2\mum$ (DUSTY)    & AGB    &  130.6   \\
YZ Per   & 1.27$\times10^{-8}$   & This work  & IR SED: $\lambda\leq100\mum$ (2DUST)    &  -{-} & -{-}    \\
         & 3.25$\times10^{-9}$   & Mauron \& Josselin (2011)    & $F_{\rm \nu}(60\mum)$ (JK90)$^{a}$    & RSG    &  200   \\
Z Cas    & 5.49$\times10^{-9}$   & This work  & IR SED: $\lambda$\textless$ 100\mum$ (2DUST)    &  -{-} & -{-}    \\
Z Cyg    & 1.76$\times10^{-8}$   & This work  & IR SED: $\lambda\leq100\mum$ (2DUST)    &  -{-} &  -{-}   \\
         & 8.45$\times10^{-10}$  & Suh (2004)       & IR SED: $2.38-197\mum$ (CSDUST3)   & AGB    &  -{-}      \\
\hline
\multicolumn{6}{l}{$^{a}$ JK90 refers to the empirical formula of Jura \& Kleinmann (1990)
                        which estimates the mass loss rates from the IRAS 60$\mum$ photometry.}\\
\multicolumn{6}{l}{$^{ }$ This formula is a function of the stellar effective temperature
                        $T_{\rm eff}$, the stellar luminosity $L_{\star}$, and the stellar mass $M_{\star}$}

\end{tabular}
\end{center}
\end{table}


\clearpage
\begin{table}
\begin{center}
\caption{The Peak Wavelength ($\lambda$ in $\mu$m),
         FWHM ($\gamma\lambda$ in $\mu$m),
         and Total Emitted Fluxes
         ($P\equiv \int \Delta F_\nu\,d\nu$
          in ${\rm W\,m^{-2}}$)
         of Each Silicate Feature
         for AH Sco, BI Cyg and FI Lyr.
         The Total Emitted Flux ($P$)
          Is Only Tabulated for Those Features
          with $P$ Exceeding $10^{-21}\,{\rm W\,m^{-2}}$.
         Those Labeled with $\star$ Are for Amorphous Silicates.
         }
\label{tab:FeatList1}
\begin{tabular}{ c c c c c c c c c }
\hline\hline
            & AH Sco&             &          & BI Cyg &             &          & FI Lyr  &            \\
\hline
  $\lambda$ &  $\gamma\lambda$ &  $P$   &  $\lambda$ &   $\gamma\lambda$ &    $P$    &   $\lambda$ &   $\gamma\lambda$  &   $P$    \\
       &     &        &      &     &      &  9.14&  0.69 &        \\
    9.50 & 0.74  &  1.65E-12  &   9.50 &  0.74 &      &  9.50&  0.71 & 3.25E-13   \\
    9.70 & 1.21  &  1.02E-11  &   9.70 &  1.21 & 2.08E-12 &      &     &        \\
   10.10$^{\star}$ & 2.52  &  3.37E-11  &  10.10$^{\star}$ &  2.52 & 2.24E-11 & 10.10$^{\star}$&  2.52 & 1.64E-12   \\
   10.70 & 1.02  &  2.10E-12  &  10.70 &  1.02 & 2.59E-13 & 10.50&  0.94 & 2.39E-13   \\
   11.05 & 0.89  &        &  11.05 &  0.89 &      & 11.05&  0.83 & 2.29E-13   \\
   11.30 & 1.08  &  5.17E-12  &  11.30 &  1.08 & 1.44E-12 & 11.30&  1.02 & 4.42E-13   \\
   15.10 & 0.03  &        &  15.20 &  0.30 & 7.56E-14 & 15.20&  0.30 & 3.56E-14   \\
       &     &        &  15.80 &  0.28 & 3.43E-14 & 15.80&  0.28 & 3.41E-14   \\
       &     &        &  16.10 &  0.29 & 4.67E-14 & 16.10&  0.29 &        \\
   16.80 & 0.04  &  3.31E-14  &  16.50 &  0.25 &      & 16.50&  0.25 &        \\
   17.40 & 0.03  &  7.05E-15  &  17.80 &  0.27 & 6.42E-14 & 17.80&  0.27 & 1.60E-14   \\
   17.80$^{\star}$ &11.76  &  2.40E-12  &  17.80$^{\star}$ & 11.76 & 1.04E-11 & 17.80$^{\star}$& 11.76 &        \\
       &     &        &  18.10 &  0.27 & 3.75E-14 & 18.10&  0.27 & 8.71E-15   \\
   18.98 & 1.42  &        &  18.98 &  1.42 & 7.03E-13 & 18.98&  1.42 & 3.08E-14   \\
   19.36 & 1.45  &  2.47E-13  &  19.36 &  1.45 &      & 19.36&  1.45 & 2.30E-13   \\
       &     &        &  22.30 &  0.30 & 1.50E-14 & 22.30&  0.30 &        \\
   22.71 & 1.74  &  2.45E-13  &  22.90 &  0.38 & 4.15E-14 & 22.90&  0.38 &        \\
   23.81 & 1.82  &  3.19E-15  &      &     &      &      &     &        \\
       &     &        &  23.81 &  1.82 &      & 23.81&  1.82 &        \\
   25.51 & 1.95  &  1.77E-13  &  25.51 &  1.95 &      & 25.51&  1.95 &        \\
   27.80 & 0.25  &  5.02E-14  &  27.80 &  0.25 & 3.52E-15 & 27.80&  0.25 &        \\
   28.10 & 0.46  &  6.90E-14  &  28.10 &  0.46 &      & 28.10&  0.46 & 1.87E-15   \\
   28.60 & 0.43  &  1.05E-13  &  28.80 &  0.30 & 2.34E-15 & 28.80&  0.30 &        \\
   29.30 & 0.26  &  1.30E-14  &  29.30 &  0.26 &      & 29.30&  0.26 & 4.56E-15   \\
   30.90 & 0.70  &  5.96E-14  &  30.90 &  0.70 & 6.29E-15 & 30.90&  0.70 &        \\
   31.30 & 0.47  &  2.47E-14  &  31.30 &  0.47 &      & 31.30&  0.47 & 1.58E-14   \\
   32.40 & 0.53  &  9.85E-14  &  32.40 &  0.53 & 8.85E-15 & 32.40&  0.53 & 1.35E-14  \\
   32.80 & 0.30  &  4.58E-14  &  32.80 &  0.30 & 3.42E-15 & 32.80&  0.30 & 6.34E-15   \\
       &     &        &  32.80 &  0.30 & 3.42E-15 & 33.20&  0.30 & 1.67E-14  \\
   33.60 & 0.55  &  7.11E-14  &  33.60 &  0.55 & 2.85E-14& 33.60&  0.55 &        \\
   34.00 & 0.51  &  6.75E-14  &  34.00 &  0.51 &      & 34.00&  0.51 & 8.89E-15  \\
   34.90 & 0.42  &  4.38E-14  &  34.90 &  0.42 & 9.81E-15 & 34.90&  0.42 & 8.93E-15   \\
   35.90 & 0.27  &  1.76E-15  &  35.90 &  0.27 & 1.71E-15 & 35.90&  0.27 & 1.85E-15   \\
   36.80 & 0.61  &  5.94E-14  &  36.20 &  0.65 &      & 36.80&  0.61 &        \\
   39.90 & 0.84  &  7.71E-14  &  39.90 &  0.84 & 3.83E-15 & 39.90&  0.84 & 7.46E-15   \\
   40.60 & 0.43  &  4.82E-14  &  40.60 &  0.43 & 3.02E-15 & 40.60&  0.43 &        \\
   42.00 & 0.32  &  1.67E-14  &  42.00 &  0.32 & 2.62E-15 & 42.00&  0.32 & 3.98E-15   \\
   43.20 & 0.32  &  2.07E-14  &  42.80 &  0.19 & 2.63E-15 & 43.20&  0.32 & 3.77E-15  \\
   43.60 & 0.20  &  6.45E-15  &  43.60 &  0.20 & 8.89E-16 & 43.60&  0.20 &        \\
       &     &        &      &     &      & 44.20&  0.46 & 7.47E-15   \\

\hline
\end{tabular}
\end{center}
\end{table}

\clearpage
\begin{table}
\begin{center}
\caption{ Same as Table~\ref{tab:FeatList1}
         but for Mira, PZ Cas, U Her and U Lac
         }
\label{tab:FeatList2}
\begin{tabular}{ c c c c c c c c c c c c}
\hline\hline
         & Mira   &             &          &   PZ Cas&                      & U Her&             &                      & U Lac&             &     \\
\hline
      $\lambda$ &  $\gamma\lambda$ &     $P$   &   $\lambda$ &   $\gamma\lambda$ &    $P$    &   $\lambda$ &   $\gamma\lambda$ &    $P$  &   $\lambda$ &   $\gamma\lambda$ &    $P$  \\
     &    &      &     &    &      &     &       &        &   7.80  &  0.94  &    6.72E-13 \\
     &    &      &     &    &      &      &       &        &   8.70  &  1.04  &    1.41E-12 \\
  9.50 & 0.74 & 1.20E-11 &  9.50 & 0.71 & 1.34E-12 &  9.50  &   0.17  &   1.49E-13 &   9.20  &  0.10  &    3.83E-14\\
 10.10 & 0.81 & 1.49E-11 &     &    &      & 10.10  &   0.51  &        &  10.10  &  0.81  &    2.77E-14\\
 10.10$^{\star}$ & 2.52 & 1.80E-10 & 10.10$^{\star}$ & 2.52 & 2.57E-11 & 10.10$^{\star}$  &   2.52  &   1.58E-11 &  10.10$^{\star}$  &  2.52  &    8.37E-12 \\
 10.50 & 1.00 &      & 10.80 & 1.30 &      &  10.50  &   0.53  &        &  10.50  &  1.00  &         \\
 11.05 & 0.89 &      &     &    &      &  11.05  &   0.56  &   2.42E-13 &  11.05  &  0.89  &         \\
 11.30 & 1.08 &      & 11.30 & 1.70 & 1.95E-12 &  11.40  &   0.57  &   4.52E-13&  11.30  &  1.08  &    2.50E-13 \\
 15.20 & 0.30 & 1.23E-12 & 15.20 & 0.30 & 1.27E-13 &  15.20  &   0.11  &   5.61E-14&  15.10  &  0.27  &         \\
 15.80 & 0.28 & 5.27E-13 & 15.80 & 0.28 & 7.48E-14 &  16.00  &   0.12  &   5.79E-14 &  15.80  &  0.21  &         \\
 16.10 & 0.29 & 2.72E-13 & 16.10 & 0.29 & 5.54E-15 &  16.00  &   0.12  &   5.79E-14 &       &      &         \\
 16.50 & 0.25 &      & 16.50 & 0.25 &      &  16.80  &   0.25  &        &  16.40  &  0.04  &         \\
 17.80 & 0.27 & 1.11E-13 & 17.80 & 0.27 & 1.39E-13 &  17.40  &   0.03  &        &  16.90  &  0.41  &         \\
 17.80$^{\star}$ &11.76 & 1.11E-10 & 17.80$^{\star}$ &11.76 & 1.81E-11 &   17.80$^{\star}$  &  11.76  &        &  17.80$^{\star}$  & 11.76  &         \\
 18.10 & 0.27 & 1.82E-13 & 18.10 & 0.27 & 5.15E-14 &  17.70  &   0.24  &   3.84E-14 &  18.00  &  0.27  &    2.33E-14 \\
 18.98 & 1.42 & 1.68E-12 & 18.50 & 0.28 & 5.14E-14 &  18.80  &   0.23  &   3.12E-14 &  19.00  &  0.43  &    6.70E-14 \\
 19.36 & 1.45 &      & 19.36 & 1.45 &      &  19.40  &   0.26  &   6.24E-16 &  19.40  &  0.26  &    8.69E-15 \\
 22.30 & 0.30 & 3.31E-13 & 22.50 & 0.41 & 1.14E-13 &  22.40  &   0.10  &   7.54E-15 &  22.40  &  0.17  &    8.95E-15 \\
 22.90 & 0.38 & 3.28E-13 & 23.40 & 0.46 & 7.25E-14 &  22.80  &   0.10  &   7.26E-15 &  22.80  &  0.31  &    2.39E-14 \\
 23.81 & 1.82 & 3.99E-13 & 23.81 & 1.82 & 2.51E-14 &  23.81  &   0.18  &   5.01E-16 &  23.81  &  0.18  &    3.68E-15 \\
     &    &      &     &    &      &  24.70  &   0.22  &   3.30E-15 &  24.70  &  0.22  &    2.15E-15 \\
 25.51 & 1.95 &      & 25.51 & 1.95 & 1.40E-13 &  25.20  &   0.19  &   5.92E-15 &  25.20  &  0.19  &    2.66E-15 \\
     &    &      & 26.50 & 0.44 & 3.49E-14 &  26.70  &   0.32  &   1.20E-14 &  26.70  &  0.32  &         \\
 27.80 & 0.25 &      & 27.80 & 0.25 & 1.68E-14 &  27.80  &   0.25  &   9.61E-15 &  27.80  &  0.25  &         \\
 28.10 & 0.46 &      & 28.10 & 0.46 &      &  28.10  &   0.46  &   3.60E-14 &  28.10  &  0.46  &         \\
 28.80 & 0.30 &      & 28.80 & 0.30 &      &  28.80  &   0.17  &   4.05E-16 &  28.80  &  0.17  &         \\
 29.30 & 0.26 & 4.24E-14 & 29.30 & 0.26 &      &  29.40  &   0.49  &   1.40E-14 &  29.40  &  0.49  &    2.61E-15 \\
 30.90 & 0.70 & 2.09E-13 & 30.90 & 0.70 &      &  30.50  &   0.37  &   1.41E-14&  30.50  &  0.37  &    6.09E-15 \\
 31.30 & 0.47 &      & 31.30 & 0.47 &      &  31.20  &   0.19  &   1.26E-14 &  31.20  &  0.19  &    3.95E-15 \\
 32.40 & 0.53 & 1.41E-13 & 32.40 & 0.53 & 3.83E-15 &  32.20  &   0.24  &   1.63E-14 &  32.20  &  0.24  &    4.04E-15 \\
 32.80 & 0.30 & 3.65E-14 & 32.80 & 0.30 & 7.78E-15 &  32.80  &   0.30  &        &  32.80  &  0.30  &         \\
     &    &      & 33.20 & 0.30 & 1.07E-14 &  33.20  &   0.30  &   6.65E-15 &  33.20  &  0.30  &    5.88E-16 \\
 33.60 & 0.55 & 8.34E-14 & 33.60 & 0.40 & 3.61E-14 &  33.60  &   0.55  &   1.94E-14 &  33.60  &  0.55  &    2.97E-15 \\
 34.00 & 0.51 & 1.01E-13 & 34.00 & 0.36 &      &  34.00  &   0.51  &   7.70E-15 &  34.00  &  0.51  &         \\
 34.90 & 0.42 & 6.43E-14 & 34.80 & 0.31 & 6.54E-15 &  34.90  &   0.42  &   8.40E-15 &  34.90  &  0.42  &    2.59E-15 \\
 35.90 & 0.27 & 1.09E-14 & 35.90 & 0.27 & 4.84E-15 &  35.90  &   0.27  &   1.35E-14 &  35.90  &  0.27  &         \\
 36.20 & 0.65 & 1.52E-13 & 36.80 & 0.61 & 9.27E-15 &  36.20  &   0.65  &   1.70E-15 &  36.20  &  0.65  &    1.24E-14 \\
 39.90 & 0.84 & 9.34E-14 & 39.60 & 0.24 & 8.97E-15 &  39.70  &   0.18  &        &  39.70  &  0.18  &    3.64E-15 \\
 40.60 & 0.43 & 7.22E-14 & 40.60 & 0.43 & 2.91E-14 &  40.60  &   0.43  &   2.96E-15 &  40.60  &  0.43  &    5.64E-15 \\
 42.00 & 0.32 & 2.49E-14 & 41.60 & 0.31 & 2.27E-14 &  42.00  &   0.32  &   6.09E-15 &  42.00  &  0.32  &    3.32E-15 \\
 42.80 & 0.19 & 2.36E-14 & 42.80 & 0.19 & 6.24E-15 &  43.20  &   0.32  &   2.20E-15 &  43.20  &  0.32  &    1.42E-15 \\
 43.60 & 0.20 & 1.74E-14 & 43.60 & 0.20 &      &  43.90  &   0.13  &   9.96E-15 &  43.90  &  0.13  &    8.62E-15 \\
     &    &      &     &    &      &  44.20  &   0.46  &   4.91E-15 &  44.20  &  0.46  &    1.74E-14 \\

\hline
\end{tabular}
\end{center}
\end{table}

\clearpage
\begin{table}
\begin{center}
\caption{Same as Table~\ref{tab:FeatList1}
            but for R Aql, R Cas and RS Per
            }
\label{tab:FeatList3}
\begin{tabular}{ c c c c c c c c c }
\hline\hline
            &  R Aql &             &          & R Cas &            &          &   RS Per&             \\
\hline
      $\lambda$&   $\gamma\lambda$  &    $P$    &  $\lambda$  &  $\gamma\lambda$  &   $P$    &   $\lambda$ &   $\gamma\lambda$ &    $P$    \\
       &      &       &  8.10 & 0.63  &   1.69E-12 &     &       &          \\
       &      &       &     &     &        &  9.50 &   0.74  &   6.29E-13   \\
    9.80 &  0.76  &  1.49E-12 & 10.10 & 0.80  &        & 10.10 &   0.80  &   8.58E-13   \\
   10.10$^{\star}$ &  2.52  &  5.44E-12 & 10.10$^{\star}$ & 2.52  &   5.51E-11 & 10.10$^{\star}$ &   2.52  &   8.87E-13   \\
   10.80 &  1.03  &  7.88E-13 & 10.80 & 0.86  &        & 10.50 &   0.99  &   4.24E-13   \\
   11.05 &  0.89  &       &     &     &        & 11.05 &   0.88  &   1.80E-13   \\
   11.30 &  1.08  &  8.97E-13 & 11.30 & 1.07  &   5.70E-12 & 11.30 &   1.07  &   5.66E-13  \\
   15.20 &  0.30  &       & 15.20 & 0.30  &   3.59E-13 & 15.20 &   0.30  &   5.15E-14   \\
   15.80 &  0.28  &       & 15.80 & 0.28  &   1.68E-13 & 15.80 &   0.28  &   5.18E-14   \\
   16.10 &  0.29  &       & 16.10 & 0.29  &        & 16.10 &   0.29  &   2.56E-14   \\
   16.80 &  0.25  &  1.31E-14 & 16.50 & 0.25  &        & 16.50 &   0.25  &   1.25E-14   \\
   17.80$^{\star}$ & 11.76  &  1.04E-12 & 17.80$^{\star}$ &11.76  &        & 17.80$^{\star}$ &  11.76  &   3.63E-12   \\
   17.80 &  0.27  &  3.84E-14 & 17.80 & 0.27  &   2.35E-14 & 17.80 &   0.27  &   1.75E-14   \\
   18.10 &  0.27  &  5.60E-14 & 18.10 & 0.27  &   6.92E-14 & 18.10 &   0.27  &   5.88E-15   \\
   18.98 &  1.42  &  3.96E-13 & 18.98 & 1.42  &   3.34E-13 & 18.98 &   1.42  &   6.85E-14   \\
   19.36 &  1.45  &       & 19.36 & 1.45  &        & 19.36 &   1.45  &          \\
   22.50 &  0.41  &  1.20E-13 & 22.30 & 0.30  &   7.52E-14 & 22.30 &   0.30  &   7.34E-15   \\
   23.40 &  0.46  &  4.64E-14 & 22.90 & 0.38  &   1.37E-14 & 22.90 &   0.38  &   6.02E-15   \\
   23.81 &  1.82  &  3.33E-13 & 23.81 & 1.82  &        & 23.81 &   1.82  &   5.90E-14   \\
   25.50 &  1.95  &  3.72E-13 & 25.50 & 1.95  &        & 25.50 &   1.95  &   3.85E-14   \\
   26.50 &  0.44  &  3.53E-14 & 27.30 & 0.33  &   2.26E-14 & 26.70 &   0.40  &   1.80E-14   \\
   27.80 &  0.25  &  4.72E-14 & 27.80 & 0.25  &   1.37E-14 & 27.80 &   0.25  &   1.49E-14   \\
   28.10 &  0.46  &  8.91E-15 & 28.10 & 0.46  &   3.94E-14 & 28.20 &   0.17  &   1.19E-15   \\
   28.80 &  0.30  &  1.06E-14 & 28.80 & 0.30  &   2.84E-14 & 28.70 &   0.13  &          \\
   29.50 &  0.49  &  7.30E-14 & 29.30 & 0.26  &   3.54E-14 & 29.60 &   0.18  &   2.80E-15   \\
   30.50 &  0.41  &  4.60E-14 & 30.90 & 0.70  &   5.65E-14 & 30.50 &   0.32  &   4.01E-10   \\
   31.30 &  0.47  &  7.76E-14 & 31.30 & 0.47  &   2.79E-14 & 31.30 &   0.38  &          \\
   32.10 &  0.48  &  7.32E-14 & 32.40 & 0.53  &   7.33E-14 & 32.30 &   0.29  &          \\
   32.80 &  0.30  &  4.56E-14 & 32.80 & 0.30  &   3.37E-14 & 32.80 &   0.20  &          \\
   33.20 &  0.30  &  4.52E-14 & 33.20 & 0.30  &   3.13E-14 & 33.00 &   0.20  &          \\
   33.60 &  0.40  &  8.10E-14 & 33.60 & 0.55  &   4.71E-14 & 33.50 &   0.20  &          \\
   34.00 &  0.36  &  3.66E-14 & 34.00 & 0.51  &   4.75E-14 & 34.00 &   0.26  &   1.83E-15   \\
   34.80 &  0.31  &  3.17E-14 & 34.90 & 0.42  &   2.49E-14 & 35.10 &   0.37  &          \\
   35.90 &  0.27  &  1.71E-14 & 35.90 & 0.27  &        & 35.90 &   0.27  &          \\
   36.80 &  0.61  &  7.70E-14 & 36.20 & 0.65  &   5.39E-14 & 36.50 &   0.33  &          \\
   40.00 &  0.42  &  3.65E-14 & 39.90 & 0.84  &   4.38E-14 & 39.90 &   0.42  &          \\
   40.70 &  0.43  &  2.88E-14 & 40.60 & 0.43  &   2.66E-14 & 40.60 &   0.43  &          \\
   42.00 &  0.32  &  1.03E-14 & 42.00 & 0.32  &   1.88E-14 & 41.80 &   0.31  &          \\
   43.00 &  0.19  &  4.28E-15 & 43.20 & 0.32  &   1.91E-14 & 43.20 &   0.32  &          \\
   43.60 &  0.20  &       & 43.70 & 0.39  &   3.19E-14 & 43.70 &   0.39  &          \\
   44.50 &  0.20  &  2.58E-15 & 44.70 & 0.27  &   3.15E-14 & 44.75 &   0.27  &   2.28E-15   \\

\hline
\end{tabular}
\end{center}
\end{table}

\clearpage
\begin{table}
\begin{center}
\caption{Same as Table~\ref{tab:FeatList1}
         but for RV Cam, RW Cep and RW Cyg
         }
\label{tab:FeatList4}
\begin{tabular}{ c c c c c c c c c }
\hline\hline
            &   RV Cam&            &          &  RW Cep&             &          &  RW Cyg &            \\
\hline
      $\lambda$&   $\gamma\lambda$  &    $P$   &   $\lambda$ &   $\gamma\lambda$ &    $P$    &   $\lambda$ &   $\gamma\lambda$  &   $P$    \\
    8.10 &   0.63  &        &  8.10 &    0.73  &    3.66E-13 &   8.10  &  0.63  &    3.25E-13  \\
       &       &        &  9.50 &    0.74  &    1.21E-12 &       &      &          \\
   10.10 &   0.81  &   2.06E-13 & 10.20 &    0.61  &    1.09E-12 &  10.10  &  0.81  &    2.04E-12  \\
   10.10$^{\star}$ &   2.52  &   1.70E-12 & 10.10$^{\star}$ &    2.52  &    3.73E-12 &  10.10$^{\star}$  &  2.52  &    2.15E-11  \\
   10.50 &   1.00  &        & 10.50 &    0.79  &    5.52E-13 &  10.50  &  1.00  &    8.99E-14  \\
   11.05 &   0.89  &   1.02E-13 & 11.10 &    1.17  &    1.72E-12 &  11.05  &  0.89  &    7.65E-13  \\
   11.30 &   1.08  &   3.50E-14 & 11.30 &    1.19  &         &  11.30  &  1.08  &    1.64E-12  \\
   15.20 &   0.30  &        & 15.20 &    0.30  &         &  15.20  &  0.30  &    5.14E-14  \\
   15.80 &   0.28  &   2.75E-14 & 15.80 &    0.28  &         &  15.80  &  0.28  &    7.84E-14  \\
   16.10 &   0.29  &        & 16.10 &    0.29  &         &  16.10  &  0.29  &    4.49E-14  \\
   16.50 &   0.25  &        & 16.50 &    0.25  &         &  16.50  &  0.25  &    1.26E-14  \\
   17.80$^{\star}$ &  11.76  &   1.28E-12 & 17.80$^{\star}$ &   11.76  &         &  17.80$^{\star}$  & 11.76  &    7.93E-12  \\
   17.80 &   0.27  &   1.93E-14 & 17.80 &    0.27  &    2.22E-14 &  17.80  &  0.27  &    3.28E-14  \\
   18.10 &   0.27  &   2.16E-14 & 18.10 &    0.27  &         &  18.10  &  0.27  &    7.75E-14  \\
   18.98 &   1.42  &   1.41E-15 & 18.98 &    1.42  &    9.89E-15 &       &      &          \\
   19.36 &   1.45  &   7.95E-15 & 19.36 &    1.45  &         &  19.30  &  0.29  &    4.09E-14  \\
   22.30 &   0.30  &        & 22.30 &    0.30  &         &  22.30  &  0.30  &    1.34E-14  \\
   22.90 &   0.38  &   3.71E-15 & 22.90 &    0.38  &         &  22.90  &  0.38  &          \\
   23.81 &   1.82  &        & 23.81 &    1.82  &         &  23.81  &  1.82  &          \\
   25.50 &   1.95  &        & 25.51 &    1.95  &         &  25.50  &  1.95  &          \\
   26.70 &   0.40  &        & 26.70 &    0.40  &    1.75E-15 &  26.70  &  0.40  &    2.87E-14  \\
   27.80 &   0.25  &        & 27.80 &    0.25  &    2.97E-14 &  27.80  &  0.25  &    2.42E-14  \\
   28.20 &   0.17  &   6.21E-15 & 28.45 &    0.17  &    2.19E-14 &  28.10  &  0.25  &    1.48E-14  \\
   28.70 &   0.13  &        & 28.70 &    0.13  &    9.91E-15 &  28.70  &  0.13  &    2.13E-14  \\
   29.30 &   0.18  &   2.91E-15 & 29.60 &    0.18  &    2.12E-14 &  29.60  &  0.18  &    1.95E-14  \\
   30.50 &   0.32  &   7.91E-15 & 30.50 &    0.32  &    2.94E-14 &  30.50  &  0.32  &    4.16E-14  \\
   31.30 &   0.38  &        & 31.30 &    0.38  &    2.66E-14 &  31.30  &  0.38  &    3.05E-14  \\
   32.30 &   0.29  &   1.28E-14 & 32.30 &    0.29  &    3.26E-14 &  32.30  &  0.29  &    2.72E-14  \\
   32.80 &   0.20  &   4.07E-15 & 32.80 &    0.20  &    1.79E-14 &  32.80  &  0.20  &    1.57E-14  \\
   33.20 &   0.20  &   1.05E-14 & 33.20 &    0.20  &    1.52E-14 &  33.20  &  0.20  &    1.71E-14  \\
   34.00 &   0.26  &   1.45E-14 & 33.40 &    0.10  &    5.89E-15 &  33.50  &  0.10  &    4.06E-14  \\
   34.20 &   0.15  &   5.52E-15 & 34.20 &    0.15  &    6.19E-15 &  34.20  &  0.15  &    3.02E-15  \\
   35.00 &   0.26  &   7.88E-10 & 35.00 &    0.26  &    1.67E-14 &  34.90  &  0.21  &    1.60E-14  \\
   35.80 &   0.27  &   5.24E-15 & 36.10 &    0.22  &    1.76E-14 &  35.80  &  0.27  &    1.97E-14  \\
   36.80 &   0.22  &   2.11E-15 & 36.80 &    0.22  &    2.75E-14 &  36.30  &  0.22  &    2.52E-14  \\
   39.90 &   0.42  &   3.30E-15 & 39.90 &    0.42  &    1.40E-14 &  39.90  &  0.24  &    3.81E-15  \\
   40.60 &   0.43  &   1.18E-14 & 40.60 &    0.43  &    2.08E-14 &  40.60  &  0.43  &    1.81E-14  \\
   41.50 &   0.31  &   5.34E-15 & 41.50 &    0.31  &    8.27E-15 &  41.50  &  0.31  &    6.84E-15  \\
   43.40 &   0.20  &   3.89E-15 & 43.00 &    0.10  &    3.48E-15 &  43.00  &  0.10  &    2.03E-15  \\
   44.00 &   0.33  &   1.34E-14 & 44.00 &    0.33  &    9.50E-15 &  44.00  &  0.33  &    4.17E-15  \\
   44.50 &   0.13  &   4.70E-15 & 44.50 &    0.13  &    3.63E-15 &  44.50  &  0.13  &    1.28E-15  \\

\hline
\end{tabular}
\end{center}
\end{table}

\clearpage
\begin{table}
\begin{center}
\caption{Same as Table~\ref{tab:FeatList1}
         but for RX Boo, S Per and SU Per
         }
\label{tab:FeatList5}
\begin{tabular}{ c c c c c c c c c }
\hline\hline
             & RX Boo &             &         & S Per  &             &         & SU Per &             \\
\hline
      $\lambda$ &  $\gamma\lambda$  &    $P$    &  $\lambda$ &   $\gamma\lambda$  &   $P$     &  $\lambda$ &   $\gamma\lambda$ &    $P$    \\
    9.50  &  0.74  &   2.68E-12 &  9.40 &   0.70  &    3.50E-12  &  9.40  &   0.70  &         \\
   10.10  &  0.81  &   2.26E-12 &  9.70 &   0.87  &    8.61E-13  &  9.90  &   0.74  &   3.22E-13  \\
   10.10$^{\star}$  &  2.52  &   1.74E-11 & 10.10$^{\star}$ &   2.52  &    8.21E-12  & 10.10$^{\star}$  &   2.52  &   3.01E-12  \\
   10.50  &  1.00  &   1.32E-13 & 10.60 &   0.80  &    2.24E-12  & 10.70  &   1.28  &   6.75E-13  \\
   11.05  &  0.89  &   1.25E-12 & 11.20 &   1.68  &    2.85E-12  &      &       &         \\
   11.30  &  1.08  &   3.50E-12 & 11.30 &   1.02  &    2.77E-12  & 11.30  &   0.68  &   1.52E-13  \\
   15.20  &  0.30  &   2.08E-13 & 15.20 &   0.30  &          & 15.10  &   0.27  &   6.67E-10  \\
   15.80  &  0.28  &   1.21E-13 & 15.80 &   0.28  &          & 16.00  &   0.26  &   7.38E-15  \\
   16.10  &  0.29  &   2.12E-14 & 16.10 &   0.29  &          & 16.30  &   0.42  &   3.68E-14  \\
   16.50  &  0.25  &        & 16.50 &   0.25  &          & 16.80  &   0.25  &   1.16E-14  \\
   17.80$^{\star}$  & 11.76  &   1.09E-11 & 17.80$^{\star}$ &  11.76  &    2.03E-12  & 17.80$^{\star}$  &  11.76  &   5.55E-13  \\
   17.80  &  0.27  &   6.62E-14 & 17.80 &   0.27  &    1.61E-13  & 17.40  &   0.26  &   2.55E-14  \\
   18.10  &  0.27  &   1.25E-13 & 18.10 &   0.27  &    1.04E-13  &      &       &         \\
   19.30  &  0.29  &   4.60E-13 & 18.80 &   0.28  &    2.00E-13  & 19.00  &   0.43  &         \\
   19.80  &  0.45  &   6.82E-13 & 19.30 &   0.29  &    2.27E-13  & 19.40  &   0.44  &   2.47E-14  \\
   22.30  &  0.30  &        & 22.30 &   0.30  &    4.03E-14  & 22.00  &   0.50  &         \\
   22.90  &  0.38  &        & 22.90 &   0.38  &          & 23.00  &   0.45  &   6.22E-15  \\
   23.81  &  1.82  &        & 23.81 &   1.82  &          & 23.81  &   0.61  &         \\
   25.50  &  1.95  &        & 25.50 &   1.95  &          &      &       &         \\
   26.70  &  0.40  &   5.99E-14 & 26.70 &   0.40  &    4.24E-10  &      &       &         \\
   27.80  &  0.25  &   8.16E-14 & 27.80 &   0.25  &    3.67E-14  & 27.80  &   0.17  &   6.08E-15  \\
   28.10  &  0.25  &   7.46E-14 & 28.10 &   0.25  &    1.34E-14  & 28.10  &   0.17  &   6.85E-15  \\
   28.70  &  0.13  &   2.90E-14 & 28.70 &   0.13  &    1.31E-14  & 28.80  &   0.17  &   3.26E-15  \\
   29.60  &  0.18  &   5.50E-14 & 29.60 &   0.27  &    4.98E-14  & 29.20  &   0.13  &         \\
   30.50  &  0.32  &   7.22E-14 & 30.50 &   0.32  &    5.44E-14  & 30.80  &   0.28  &   5.58E-15  \\
   31.30  &  0.38  &   7.97E-14 & 31.30 &   0.38  &    4.72E-14  & 31.20  &   0.28  &   1.02E-14  \\
   32.30  &  0.29  &   5.23E-14 & 32.30 &   0.29  &    4.45E-14  & 31.90  &   0.38  &   2.26E-14  \\
   32.80  &  0.20  &   2.56E-14 & 32.80 &   0.20  &    1.78E-14  & 32.50  &   0.34  &   1.62E-14  \\
   33.20  &  0.20  &   3.00E-14 & 33.00 &   0.25  &    3.83E-14  & 33.10  &   0.25  &   5.15E-15  \\
   33.80  &  0.10  &   1.18E-14 & 33.60 &   0.20  &    2.59E-14  & 33.70  &   0.25  &   6.30E-15  \\
   34.00  &  0.26  &   2.08E-14 & 34.00 &   0.26  &    1.54E-14  & 34.00  &   0.15  &   2.61E-15  \\
   34.90  &  0.21  &        & 34.90 &   0.21  &    2.41E-15  & 34.90  &   0.31  &   5.92E-15  \\
   35.80  &  0.27  &   2.31E-14 & 36.00 &   0.27  &    1.72E-14  & 35.90  &   0.27  &   6.93E-15  \\
   36.30  &  0.22  &   2.58E-14 & 36.70 &   0.22  &    1.85E-14  & 36.20  &   0.33  &   7.93E-15  \\
   39.90  &  0.24  &   1.17E-14 & 39.70 &   0.24  &    6.68E-15  & 39.90  &   0.84  &         \\
   40.60  &  0.43  &   3.13E-14 & 40.60 &   0.43  &    2.26E-14  & 40.60  &   0.43  &   4.47E-15  \\
   42.10  &  0.32  &   1.65E-15 & 41.50 &   0.19  &    1.86E-14  & 41.90  &   0.25  &   4.65E-15  \\
   43.00  &  0.10  &        & 43.10 &   0.19  &    2.07E-14  & 43.20  &   0.32  &   1.75E-14  \\
   44.00  &  0.33  &        & 44.00 &   0.33  &    2.85E-14  & 44.00  &   0.20  &   3.22E-15  \\
   44.50  &  0.13  &        & 44.50 &   0.13  &    1.82E-14  & 44.60  &   0.40  &   1.22E-14  \\

\hline
\end{tabular}
\end{center}
\end{table}

\clearpage
\begin{table}
\begin{center}
\caption{Same as Table~\ref{tab:FeatList1}
         but for SV Peg, SV Psc and TY Dra
         }
\label{tab:FeatList6}
\begin{tabular}{ c c c c c c c c c }
\hline\hline
             & SV Peg &            &          & SV Psc &              &        &  TY Dra &             \\
\hline
      $\lambda$ &  $\gamma\lambda$  &    $P$   &   $\lambda$ &   $\gamma\lambda$ &    $P$     &  $\lambda$&    $\gamma\lambda$ &    $P$    \\
    9.50 &   0.74  &        &  9.50 &   0.74  &   1.39E-13 &       &       &           \\
    9.70 &   1.21  &   1.82E-12 & 10.10 &   0.81  &        &  10.10  &   0.81  &   1.93E-13    \\
   10.10$^{\star}$ &   2.52  &   5.74E-12 & 10.10$^{\star}$ &   2.52  &   2.99E-12 &  10.10$^{\star}$  &   2.52  &   5.90E-12    \\
   10.50 &   1.00  &   6.25E-13 & 10.50 &   1.00  &        &  10.50  &   1.00  &           \\
   11.05 &   0.89  &   6.39E-14 & 11.05 &   0.89  &        &  11.05  &   0.89  &           \\
   11.30 &   1.08  &   2.09E-12 & 11.30 &   1.08  &   1.35E-13 &  11.30  &   1.08  &           \\
   15.30 &   0.28  &   1.39E-13 & 15.30 &   0.28  &   4.52E-14 &  15.30  &   0.28  &           \\
   15.80 &   0.26  &   8.93E-14 &     &       &        &       &       &           \\
   16.20 &   0.24  &   1.24E-13 & 16.20 &   0.10  &   2.28E-14 &       &       &           \\
   16.80 &   0.25  &   8.00E-14 & 16.80 &   0.25  &   1.42E-14 &  16.40  &   0.17  &           \\
   17.50 &   0.18  &   7.82E-14 & 17.50 &   0.18  &   1.39E-14 &  17.70  &   0.19  &   2.24E-14    \\
   17.80$^{\star}$ &  11.76  &   2.88E-12 & 17.80$^{\star}$ &  11.76  &   1.05E-12 &  17.80$^{\star}$  &  11.76  &   1.67E-12    \\
   17.90 &   0.19  &   8.34E-14 & 17.90 &   0.19  &   2.35E-14 &  18.00  &   0.19  &   1.48E-14    \\
   19.00 &   0.43  &   7.30E-14 & 19.00 &   0.43  &   3.42E-14 &  18.85  &   0.57  &   9.38E-14    \\
   19.36 &   0.44  &   2.07E-13 & 19.36 &   0.44  &   5.34E-14 &  19.36  &   0.44  &   1.27E-14    \\
   22.00 &   0.50  &        & 22.20 &   0.57  &   2.72E-14 &  22.20  &   0.57  &   1.51E-14    \\
   23.00 &   0.45  &        & 23.00 &   0.45  &   1.95E-14 &  23.00  &   0.45  &   5.40E-15    \\
       &       &        &     &       &        &  23.40  &   0.53  &   6.41E-15    \\
   23.81 &   0.61  &        & 24.00 &   0.54  &   2.72E-14 &  24.00  &   0.54  &   4.58E-15    \\
       &       &        & 25.00 &   0.56  &   2.38E-14 &  25.00  &   0.56  &   5.08E-15    \\
       &       &        & 26.80 &   0.52  &   1.98E-14 &  26.80  &   0.52  &           \\
   27.80 &   0.25  &   2.34E-14 & 27.80 &   0.25  &   1.17E-14 &  27.80  &   0.25  &   5.97E-15    \\
   28.40 &   0.55  &   5.58E-14 & 28.40 &   0.55  &        &  28.30  &   0.34  &   7.89E-15    \\
   28.60 &   0.60  &   2.93E-14 & 28.60 &   0.60  &   9.04E-10 &  28.60  &   0.30  &   7.64E-15    \\
   29.70 &   0.76  &   7.26E-14 & 29.50 &   0.31  &   9.71E-10 &  29.50  &   0.31  &   5.71E-15    \\
   30.50 &   0.55  &   4.99E-14 & 30.50 &   0.55  &        &  30.50  &   0.55  &   2.56E-14    \\
   31.20 &   0.28  &   2.11E-14 & 30.80 &   0.28  &        &  31.00  &   0.28  &   1.87E-14    \\
   32.50 &   0.34  &   1.14E-14 & 32.20 &   0.14  &        &  32.20  &   0.14  &   4.55E-15    \\
   32.80 &   0.74  &   5.04E-14 & 32.80 &   0.74  &        &  32.70  &   0.25  &   5.73E-15    \\
       &       &        & 33.30 &   0.20  &        &  33.00  &   0.15  &   4.37E-15    \\
   33.60 &   0.55  &   3.65E-14 & 33.60 &   0.55  &        &  33.70  &   0.56  &   1.29E-14    \\
   34.00 &   0.15  &   3.52E-15 & 34.00 &   0.15  &        &  34.00  &   0.15  &   2.89E-15    \\
   34.90 &   0.31  &   4.20E-15 & 35.10 &   0.16  &        &  35.10  &   0.47  &   1.02E-14    \\
   35.90 &   0.27  &        & 35.80 &   0.27  &        &  35.90  &   0.59  &   6.32E-10    \\
   36.20 &   0.33  &   6.89E-15 & 36.60 &   0.33  &        &  36.60  &   0.33  &   5.01E-10    \\
   39.90 &   0.30  &   4.81E-15 & 39.90 &   0.30  &        &  39.80  &   0.18  &   5.01E-10    \\
   40.40 &   0.12  &   1.29E-15 & 40.40 &   0.12  &        &  40.40  &   0.12  &           \\
   41.90 &   0.44  &   2.18E-14 & 41.70 &   0.25  &        &  41.70  &   0.25  &           \\
   43.20 &   0.32  &   1.53E-15 & 42.80 &   0.19  &        &  43.20  &   0.32  &           \\
   43.80 &   0.20  &        & 43.80 &   0.20  &        &  43.80  &   0.20  &           \\
   44.20 &   0.40  &   2.00E-14 & 44.20 &   0.40  &        &  44.30  &   0.27  &   9.94E-15    \\

\hline
\end{tabular}
\end{center}
\end{table}

\clearpage
\begin{table}
\begin{center}
\caption{Same as Table~\ref{tab:FeatList1}
         but for VX Sgr, W Hor and W Hya
         }
\label{tab:FeatList7}
\begin{tabular}{ c c c c c c c c c }
\hline\hline
            &  VX Sgr &            &          &   W Hor &            &          &   W Hya&             \\
\hline
      $\lambda$&   $\gamma\lambda$  &    $P$   &   $\lambda$ &   $\gamma\lambda$  &   $P$    &   $\lambda$ &   $\gamma\lambda$ &    $P$    \\
    8.10  &  0.63  &        &  8.10  &   0.63  &        &  8.50 &   0.64 &    2.36E-11\\
   10.10$^{\star}$  &  2.52  &   2.12E-10 & 10.10$^{\star}$  &   2.52  &   7.22E-12 & 10.10$^{\star}$ &   2.52 &    1.65E-10\\
   10.50  &  1.00  &        & 10.50  &   1.00  &        & 10.50 &   1.00 &        \\
   11.05  &  0.89  &        & 11.05  &   0.89  &   1.51E-13 & 11.05 &   0.89 &    1.57E-12\\
   11.50  &  1.10  &   1.73E-11 & 11.50  &   1.10  &   6.59E-13 & 11.50 &   1.10 &    3.12E-11\\
   15.10  &  0.27  &   1.69E-13 & 15.10  &   0.27  &   6.31E-14 & 15.10 &   0.34 &    2.13E-12\\
   15.80  &  0.21  &        & 15.80  &   0.21  &   1.79E-14 & 15.80 &   0.36 &    2.85E-12\\
   16.40  &  0.04  &        &      &       &        & 16.20 &   0.04 &    3.52E-13\\
   16.90  &  0.41  &   4.45E-13 & 16.40  &   0.04  &        & 16.90 &   0.41 &    1.41E-12\\
   17.80$^{\star}$  & 11.76  &        & 17.80$^{\star}$  &  11.76  &   1.20E-12 & 17.80$^{\star}$ &  11.76 &        \\
   18.00  &  0.27  &   5.61E-13 & 18.00  &   0.27  &   1.44E-14 & 18.00 &   0.27 &    3.06E-13\\
   19.00  &  0.43  &   3.47E-13 & 19.00  &   0.43  &   5.17E-14 & 19.00 &   0.43 &    4.66E-13\\
   19.40  &  0.26  &   3.03E-13 & 19.40  &   0.26  &   4.13E-14 & 19.40 &   0.26 &    2.50E-13\\
   22.40  &  0.17  &   7.83E-14 & 22.40  &   0.17  &   6.89E-15 & 22.40 &   0.17 &        \\
   22.80  &  0.31  &   1.35E-13 & 22.80  &   0.31  &   1.57E-14 & 22.80 &   0.31 &        \\
   24.20  &  0.18  &        & 23.80  &   0.18  &   8.02E-15 & 23.81 &   0.18 &        \\
   24.70  &  0.22  &        & 24.70  &   0.22  &   1.26E-14 & 24.70 &   0.22 &        \\
   25.20  &  0.19  &        & 25.20  &   0.19  &   1.03E-14 & 25.20 &   0.19 &        \\
   26.70  &  0.32  &        & 26.70  &   0.32  &   2.04E-14 & 26.70 &   0.32 &        \\
   27.80  &  0.25  &   7.05E-14 & 27.50  &   0.29  &   1.17E-14 & 27.50 &   0.29 &    2.47E-14\\
   28.10  &  0.46  &        & 28.10  &   0.46  &   2.55E-14 & 28.10 &   0.46 &    1.20E-13\\
   28.60  &  0.43  &        & 28.80  &   0.17  &   2.75E-15 & 28.80 &   0.17 &    2.79E-14\\
   29.40  &  0.49  &   1.14E-13 & 29.70  &   0.36  &   8.70E-15 & 29.70 &   0.36 &    1.29E-13\\
   30.50  &  0.37  &   1.16E-13 & 30.50  &   0.50  &   1.79E-14 & 30.50 &   0.50 &    1.97E-13\\
   31.20  &  0.19  &   6.60E-14 & 31.20  &   0.33  &   1.74E-14 & 31.20 &   0.33 &    1.30E-13\\
   32.20  &  0.24  &   1.02E-13 & 32.20  &   0.39  &   2.95E-14 & 32.20 &   0.39 &    2.68E-13\\
   32.80  &  0.30  &   8.85E-14 & 32.80  &   0.39  &   1.64E-14 & 32.90 &   0.39 &    1.90E-13\\
   33.20  &  0.30  &   1.00E-13 & 33.20  &   0.40  &   1.80E-14 & 33.20 &   0.40 &    7.72E-14\\
   33.60  &  0.55  &   1.23E-13 & 33.60  &   0.55  &   7.19E-15 & 33.60 &   0.55 &    1.69E-13\\
   34.00  &  0.51  &   8.95E-14 & 34.00  &   0.51  &   8.92E-15 & 34.00 &   0.51 &    1.16E-13\\
   34.90  &  0.42  &   1.07E-13 & 35.10  &   0.32  &   7.27E-15 & 35.10 &   0.32 &    8.71E-14\\
   35.90  &  0.27  &   2.32E-14 & 35.90  &   0.27  &   6.44E-15 & 35.90 &   0.27 &    9.22E-14\\
   36.20  &  0.65  &   1.92E-13 & 36.50  &   0.38  &   4.44E-15 & 36.50 &   0.38 &    1.24E-13\\
   39.70  &  0.18  &   2.55E-14 & 39.90  &   0.18  &   3.04E-15 & 39.80 &   0.24 &    2.96E-14\\
   40.60  &  0.43  &   1.06E-13 & 40.50  &   0.43  &   3.57E-10 & 40.70 &   0.24 &    4.36E-14\\
   42.00  &  0.32  &   3.22E-14 & 41.80  &   0.31  &   4.37E-15 & 41.80 &   0.31 &    3.75E-14\\
   43.20  &  0.32  &   3.42E-14 & 43.10  &   0.19  &        & 43.10 &   0.19 &    4.04E-14\\
   43.90  &  0.13  &   1.02E-14 & 43.60  &   0.20  &        & 43.90 &   0.13 &    1.80E-14\\
   44.20  &  0.46  &   1.46E-14 & 44.70  &   0.34  &        & 44.70 &   0.34 &    6.97E-14\\

\hline
\end{tabular}
\end{center}
\end{table}

\clearpage
\begin{table}
\begin{center}
\caption{Same as Table~\ref{tab:FeatList1}
         but for W Per, X Her and X Oph
         }
\label{tab:FeatList8}
\begin{tabular}{ c c c c c c c c c }
\hline\hline
             &   W Per&            &          &  X Her &             &          & X Oph   &             \\
\hline
      $\lambda$ &  $\gamma\lambda$  &    $P$   &   $\lambda$ &   $\gamma\lambda$ &    $P$    &   $\lambda$ &   $\gamma\lambda$  &   $P$     \\
    8.10  &  0.63  &   2.33E-13 &   8.10  &   0.63  &        &       &       &         \\
        &      &        &       &       &        &   9.50  &   0.17  &    9.20E-14 \\
   10.10  &  0.81  &   3.03E-13 &   9.80  &   0.51  &   8.13E-13 &  10.10  &   0.81  &    3.31E-13 \\
   10.10$^{\star}$  &  2.52  &   6.78E-12 &  10.10$^{\star}$  &   2.52  &   7.46E-12 &  10.10$^{\star}$  &   2.52  &    3.04E-12 \\
   10.50  &  1.00  &        &  10.50  &   1.00  &        &  10.50  &   1.00  &    6.46E-13 \\
   11.05  &  0.89  &        &  11.05  &   0.89  &        &  11.05  &   0.89  &    8.40E-13 \\
   11.50  &  1.10  &   6.41E-13 &  11.50  &   1.10  &   3.51E-13 &  11.50  &   1.10  &    1.10E-12 \\
   15.10  &  0.34  &   1.16E-14 &  15.10  &   0.34  &        &  15.10  &   0.34  &    7.55E-14 \\
   15.80  &  0.36  &   3.08E-15 &  15.90  &   0.17  &        &  15.90  &   0.17  &    2.84E-14 \\
   16.20  &  0.04  &   1.07E-15 &  16.10  &   0.07  &        &  16.10  &   0.07  &    3.23E-10 \\
   16.90  &  0.41  &        &  16.40  &   0.15  &        &  16.40  &   0.15  &         \\
        &      &        &  17.00  &   0.23  &   1.38E-13 &  17.50  &   0.11  &         \\
   17.80$^{\star}$  & 11.76  &   4.17E-12 &  17.80$^{\star}$  &  11.76  &   1.15E-11 &  17.80$^{\star}$  &  11.76  &         \\
   18.00  &  0.27  &   3.18E-14 &  18.20  &   0.27  &   1.48E-13 &  18.20  &   0.27  &    3.69E-14 \\
   18.50  &  0.67  &   7.54E-14 &  18.80  &   0.20  &   2.60E-13 &  18.80  &   0.20  &    2.52E-14 \\
   19.40  &  0.26  &   1.66E-14 &  19.80  &   0.74  &   7.13E-13 &  19.40  &   0.26  &    2.85E-14 \\
   22.40  &  0.17  &   2.22E-15 &  22.40  &   0.17  &   2.95E-14 &  22.40  &   0.17  &    3.84E-10 \\
   22.80  &  0.31  &   3.68E-15 &  22.80  &   0.31  &   1.87E-15 &  22.80  &   0.31  &         \\
   23.81  &  0.18  &   3.34E-15 &  23.81  &   0.18  &   4.22E-15 &  23.81  &   0.18  &         \\
   24.70  &  0.22  &   1.35E-14 &  24.80  &   0.22  &   1.44E-14 &  24.80  &   0.22  &         \\
   25.20  &  0.19  &   1.11E-14 &  25.30  &   0.19  &   3.29E-14 &  25.30  &   0.19  &         \\
   26.70  &  0.32  &   3.00E-14 &  26.70  &   0.32  &   3.38E-14 &  26.70  &   0.32  &         \\
   27.50  &  0.29  &   8.63E-15 &  27.50  &   0.29  &   3.66E-14 &  27.50  &   0.29  &    1.48E-14 \\
   28.10  &  0.46  &   3.58E-15 &  28.30  &   0.42  &   2.33E-14 &  28.10  &   0.46  &    5.23E-10 \\
   28.80  &  0.17  &   9.27E-15 &  28.80  &   0.17  &        &  28.80  &   0.17  &    1.73E-14 \\
   29.70  &  0.36  &   1.43E-14 &  29.70  &   0.36  &   3.22E-14 &  29.70  &   0.36  &    6.28E-14 \\
   30.70  &  0.23  &   5.06E-15 &  30.70  &   0.23  &   3.74E-15 &  30.70  &   0.23  &    1.59E-14 \\
   31.20  &  0.33  &   1.15E-14 &  31.20  &   0.33  &   2.70E-14 &  31.20  &   0.33  &    4.02E-14 \\
   32.20  &  0.39  &   3.16E-14 &  32.20  &   0.39  &   3.45E-14 &  32.20  &   0.39  &    5.63E-14 \\
   32.90  &  0.39  &   3.06E-14 &  32.70  &   0.29  &   1.76E-14 &  32.70  &   0.29  &    2.09E-14 \\
   33.20  &  0.40  &   3.70E-15 &  33.10  &   0.30  &   1.99E-14 &  33.10  &   0.30  &    2.51E-14 \\
   33.60  &  0.40  &   2.00E-14 &  33.60  &   0.40  &   1.28E-14 &  33.60  &   0.40  &   2.54E-14  \\
   34.10  &  0.36  &   1.18E-14 &  34.10  &   0.36  &   7.87E-15 &  34.10  &   0.36  &   1.87E-14  \\
   35.10  &  0.32  &   1.63E-14 &  34.90  &   0.31  &   1.52E-14 &  34.90  &   0.31  &   2.27E-14  \\
   35.90  &  0.27  &   2.30E-15 &  35.90  &   0.27  &   5.39E-15 &  35.90  &   0.27  &   1.36E-14  \\
   36.70  &  0.28  &   2.51E-14 &  36.70  &   0.28  &   4.62E-15 &  36.70  &   0.28  &   1.81E-14  \\
   39.80  &  0.18  &   3.67E-15 &  39.80  &   0.18  &   5.70E-15 &  39.80  &   0.18  &   3.43E-15  \\
   40.70  &  0.24  &   7.06E-10 &  40.70  &   0.24  &   2.81E-15 &  40.70  &   0.24  &   1.44E-14  \\
   41.80  &  0.31  &   7.89E-15 &  41.80  &   0.31  &        &  41.70  &   0.25  &   2.98E-15  \\
   43.20  &  0.19  &   2.00E-15 &  43.20  &   0.19  &   2.97E-15 &  43.00  &   0.19  &   9.74E-15  \\
   43.60  &  0.20  &        &  43.60  &   0.20  &        &  43.60  &   0.20  &   7.02E-15  \\
   44.70  &  0.34  &        &  44.70  &   0.34  &   4.84E-15 &  44.70  &   0.34  &   2.08E-14  \\

\hline
\end{tabular}
\end{center}
\end{table}

\clearpage
\begin{table}
\begin{center}
\caption{ Same as Table~\ref{tab:FeatList1}
         but for YZ Per, Z Cas and Z Cyg
         }
\label{tab:FeatList9}
\begin{tabular}{ c c c c c c c c c }
\hline\hline
             &  YZ Per&             &         &  Z Cas &             &          & Z Cyg  &             \\
\hline
      $\lambda$ &  $\gamma\lambda$  &    $P$    &  $\lambda$ &   $\gamma\lambda$ &    $P$    &   $\lambda$ &   $\gamma\lambda$ &    $P$    \\
       &       &         &       &       &        &   8.50  &   1.02  &   1.82E-12 \\
    9.30 &   0.03  &         &   9.50  &   0.17  &        &   9.50  &   0.85  &   1.12E-12 \\
   10.10 &   0.81  &    3.27E-13 &  10.10  &   0.81  &   3.65E-15 &  10.10  &   0.81  &        \\
   10.10$^{\star}$ &   2.52  &    3.42E-12 &  10.10$^{\star}$  &   2.52  &   7.58E-13 &  10.10$^{\star}$  &   2.52  &   6.12E-12 \\
   10.50 &   1.00  &         &  10.50  &   1.00  &        &  10.50  &   1.00  &        \\
   11.05 &   0.89  &    1.11E-13 &  11.05  &   0.89  &   9.14E-14 &  11.05  &   0.89  &        \\
   11.50 &   1.10  &    1.73E-13 &  11.50  &   1.10  &   3.01E-13 &  11.50  &   1.10  &   2.57E-13 \\
   15.30 &   0.18  &    5.37E-15 &  15.50  &   0.28  &   1.41E-14 &  15.10  &   0.18  &        \\
   15.90 &   0.17  &         &  15.90  &   0.17  &   9.26E-15 &  15.50  &   0.28  &        \\
   16.20 &   0.15  &         &  16.20  &   0.15  &   2.34E-15 &  16.20  &   0.15  &   1.37E-15 \\
   16.40 &   0.15  &         &  16.70  &   0.15  &   6.69E-15 &  16.50  &   0.20  &        \\
   17.60 &   0.13  &         &  17.80  &   0.21  &   1.80E-14 &  17.50  &   0.13  &   1.14E-14 \\
   17.80$^{\star}$ &  11.76  &         &  17.80$^{\star}$  &  11.76  &   8.19E-13 &  17.80$^{\star}$  &  11.76  &   1.28E-12 \\
   18.20 &   0.27  &    2.37E-14 &  18.30  &   0.27  &   2.44E-14 &  18.00  &   0.16  &   1.90E-14 \\
   18.70 &   0.20  &    1.71E-14 &  19.00  &   0.17  &   1.67E-14 &  19.00  &   0.17  &   4.50E-14 \\
   19.40 &   0.17  &    3.92E-15 &  19.40  &   0.17  &   6.25E-15 &  19.70  &   0.68  &   1.28E-13 \\
   22.40 &   0.17  &    4.73E-15 &  22.40  &   0.17  &        &  22.10  &   0.40  &   5.21E-14 \\
   22.80 &   0.31  &    1.19E-14 &  22.80  &   0.31  &        &  23.20  &   0.31  &   4.60E-14 \\
   23.50 &   0.63  &    2.18E-14 &  23.30  &   0.28  &   6.30E-15 &  23.70  &   0.57  &   7.49E-14 \\
   24.10 &   0.54  &    1.74E-14 &  24.10  &   0.33  &        &       &       &        \\
   24.50 &   0.18  &    4.55E-15 &  24.50  &   0.18  &        &  24.40  &   0.44  &   5.87E-14 \\
   25.00 &   0.60  &    2.55E-14 &  25.30  &   0.30  &        &  25.20  &   0.42  &   4.77E-14 \\
   26.00 &   0.20  &    9.75E-15 &  27.00  &   0.24  &        &  26.30  &   0.63  &   2.86E-14 \\
   27.50 &   0.29  &    9.47E-15 &  27.50  &   0.33  &   2.63E-15 &  27.40  &   0.33  &   1.49E-14 \\
   28.20 &   0.25  &    8.17E-15 &  28.20  &   0.17  &   5.70E-15 &  28.20  &   0.30  &   2.07E-14 \\
   28.80 &   0.17  &         &  28.90  &   0.13  &   1.22E-15 &  29.10  &   0.22  &   1.16E-14 \\
   29.70 &   0.36  &    1.69E-14 &  29.70  &   0.36  &   1.13E-14 &  29.70  &   0.18  &   5.41E-15 \\
   30.70 &   0.23  &         &  30.50  &   0.23  &   2.42E-15 &  30.50  &   0.18  &   9.99E-15 \\
   31.20 &   0.33  &    1.49E-14 &  31.20  &   0.33  &   1.09E-14 &  31.00  &   0.23  &   1.01E-14 \\
   32.20 &   0.39  &    1.47E-14 &  32.00  &   0.14  &   3.10E-15 &  32.30  &   0.15  &   4.94E-15 \\
   32.50 &   0.29  &    1.13E-14 &  32.50  &   0.29  &   1.37E-14 &  32.80  &   0.30  &   4.82E-15 \\
   33.10 &   0.30  &    9.67E-15 &  33.10  &   0.30  &   1.03E-14 &  33.20  &   0.15  &   4.17E-15 \\
   33.60 &   0.40  &    1.30E-14 &  33.60  &   0.40  &   1.19E-14 &  33.50  &   0.25  &   6.05E-15 \\
   34.20 &   0.26  &    4.79E-15 &  34.00  &   0.15  &   6.33E-15 &  34.20  &   0.15  &   7.52E-15 \\
   34.90 &   0.31  &    1.57E-14 &  34.90  &   0.16  &   4.85E-15 &  34.90  &   0.42  &   6.34E-15 \\
   35.90 &   0.27  &    4.45E-15 &  35.90  &   0.27  &   4.42E-15 &  35.90  &   0.27  &   4.16E-15 \\
   36.70 &   0.28  &    7.45E-15 &  36.30  &   0.27  &   2.30E-10 &  36.90  &   0.17  &   3.99E-15 \\
   40.00 &   0.30  &    3.11E-15 &  39.80  &   0.12  &        &  39.90  &   0.24  &   3.73E-15 \\
   40.70 &   0.24  &         &  40.70  &   0.24  &   7.11E-15 &  40.60  &   0.43  &   4.86E-15 \\
   41.70 &   0.25  &    3.96E-15 &  41.70  &   0.13  &   6.23E-15 &  41.70  &   0.19  &   9.46E-15 \\
   43.10 &   0.13  &    2.93E-15 &  42.70  &   0.26  &        &  42.70  &   0.19  &   3.79E-15 \\
   43.90 &   0.07  &    3.73E-10 &  43.60  &   0.20  &        &  43.20  &   0.13  &   2.27E-15 \\
   44.70 &   0.34  &         &  44.70  &   0.34  &   9.59E-15 &  44.70  &   0.13  &   2.89E-15 \\

\hline
\end{tabular}
\end{center}
\end{table}

\clearpage
\begin{table}
\begin{center}
\caption{ The Mean Wavelength and its Range (variance)
         as well as the Mean FWHM and its Range
         of Each Crystalline Silicate Feature
         Derived in This Work
         Compared with that of
         Molster et al.\ (2002b). The quantities in all the columns are in $\mum$.
         }
\label{tab:lambdaminmax}
\begin{tabular}{ c c c c c c | c c c c c c}
\hline\hline
           &          & This work  &       &          &         &         &          & Molster et al.&\ (2002b)        &          &           \\
\hline
   $\lambdamean$ &  $\lambdamin$ & $\lambdamax$ &  $\gammamean$  &   $\gammamin$  &  $\gammamax$   & $\lambdamean$&  $\lambdamin$ & $\lambdamax$  &  $\gammamean$ &    $\gammamin$ &  $\gammamax$    \\
    8.14  &   7.80   &  8.50   &  0.70  &   0.63   &  1.02    &    8.30 &    8.20  &   8.40   &  0.42 &    0.41  &   0.43    \\
    8.99  &   8.70   &  9.14   &  0.81  &   0.69   &  1.04    &    9.14 &    9.12  &   9.17   &  0.30 &    0.24  &   0.68    \\
    9.46  &   9.20   &  9.50   &  0.53  &   0.03   &  0.85    &    9.45 &    9.45  &   9.46   &  0.19 &    0.15  &   0.25    \\
   10.01  &   9.70   & 10.20   &  0.82  &   0.51   &  1.21    &    9.80 &    9.77  &   9.84   &  0.17 &    0.14  &   0.29    \\
   10.10$^{\star}$  &  10.10   & 10.10   &  2.52  &   2.52   &  2.52    &   10.10$^{\star}$ &    9.59  &  10.61   &  2.56 &    1.30  &   3.77    \\
   10.55  &  10.50   & 10.80   &  0.98  &   0.53   &  1.30    &   10.70 &   10.57  &  10.90   &  0.28 &    0.11  &   0.66    \\
   11.06  &  11.05   & 11.20   &  0.88  &   0.01   &  1.68    &   11.05 &   11.04  &  11.06   &  0.05 &    0.03  &   0.11    \\
   11.36  &  11.30   & 11.50   &  1.07  &   0.57   &  1.70    &   11.40 &   11.33  &  11.50   &  0.48 &    0.38  &   0.86    \\
   15.19  &  15.00   & 15.50   &  0.28  &   0.03   &  0.46    &   15.20 &   15.00  &  15.42   &  0.26 &    0.13  &   0.73    \\
   15.81  &  15.50   & 16.00   &  0.28  &   0.12   &  1.19    &   15.90 &   15.69  &  16.06   &  0.43 &    0.24  &   0.65    \\
   16.14  &  16.00   & 16.40   &  0.21  &   0.04   &  0.42    &   16.20 &   16.10  &  16.37   &  0.16 &    0.08  &   0.62    \\
   16.60  &  16.40   & 16.90   &  0.25  &   0.04   &  0.68    &   16.50 &   16.49  &  16.50   &  0.11 &    0.11  &   0.11    \\
   17.80$^{\star}$  &  17.80   & 17.80   & 11.76  &  11.76   & 11.76    &   17.50$^{\star}$ &   16.79  &  18.46   &  2.10 &    0.81  &   3.66    \\
   17.60  &  16.90   & 17.80   &  0.24  &   0.03   &  0.61    &   17.50 &   17.43  &  17.61   &  0.18 &    0.13  &   0.36    \\
   18.08  &  17.70   & 18.40   &  0.26  &   0.16   &  0.44    &   18.00 &   17.90  &  18.16   &  0.48 &    0.28  &   1.24    \\
   18.92  &  18.50   & 19.30   &  0.69  &   0.17   &  1.42    &   18.90 &   18.43  &  19.17   &  0.62 &    0.36  &   1.20    \\
   19.41  &  19.30   & 19.80   &  0.74  &   0.17   &  1.45    &   19.50 &   19.36  &  19.75   &  0.40 &    0.14  &   0.86    \\
   22.30  &  22.00   & 22.50   &  0.34  &   0.10   &  1.06    &   22.40 &   22.26  &  22.51   &  0.28 &    0.16  &   0.55    \\
   22.92  &  22.71   & 23.40   &  0.43  &   0.10   &  1.74    &   23.00 &   22.82  &  23.14   &  0.48 &    0.28  &   0.72    \\
   23.65  &  23.30   & 23.90   &  0.73  &   0.18   &  1.82    &   23.70 &   23.45  &  23.81   &  0.79 &    0.54  &   1.29    \\
   23.83  &  22.90   & 24.20   &  0.95  &   0.18   &  1.82    &   23.89 &   23.88  &  23.90   &  0.18 &    0.13  &   0.25    \\
   24.64  &  24.40   & 24.80   &  0.27  &   0.18   &  0.73    &   24.50 &   24.16  &  24.65   &  0.42 &    0.16  &   1.04    \\
   25.32  &  24.80   & 25.51   &  1.13  &   0.19   &  1.95    &   25.00 &   24.83  &  25.14   &  0.32 &    0.25  &   0.53    \\
   26.68  &  26.00   & 27.30   &  0.41  &   0.20   &  1.11    &   26.80 &   26.71  &  26.93   &  0.37 &    0.21  &   0.47    \\
   27.71  &  27.40   & 27.80   &  0.28  &   0.17   &  0.54    &   27.60 &   27.46  &  27.79   &  0.49 &    0.28  &   1.18    \\
   28.17  &  28.00   & 28.50   &  0.37  &   0.08   &  0.64    &   28.20 &   27.97  &  28.45   &  0.42 &    0.23  &   0.90    \\
   28.76  &  28.60   & 29.10   &  0.25  &   0.13   &  0.60    &   28.80 &   28.68  &  28.88   &  0.24 &    0.19  &   0.42    \\
   29.52  &  29.20   & 29.85   &  0.33  &   0.13   &  0.79    &   29.60 &   29.37  &  29.90   &  0.89 &    0.58  &   1.99    \\
   30.62  &  30.40   & 30.90   &  0.43  &   0.18   &  0.70    &   30.60 &   30.48  &  30.77   &  0.32 &    0.18  &   0.81    \\
   31.22  &  30.80   & 31.30   &  0.37  &   0.19   &  0.89    &   31.20 &   31.12  &  31.27   &  0.24 &    0.21  &   0.36    \\
   32.24  &  31.90   & 32.50   &  0.36  &   0.14   &  0.63    &   32.20 &   32.06  &  32.51   &  0.46 &    0.24  &   0.75    \\
   32.75  &  32.50   & 32.90   &  0.32  &   0.20   &  0.74    &   32.80 &   32.56  &  33.03   &  0.60 &    0.36  &   1.00    \\
   33.10  &  32.71   & 33.30   &  0.32  &   0.15   &  0.74    &   32.97 &   32.96  &  32.99   &  0.20 &    0.11  &   0.28    \\
   33.61  &  33.40   & 34.00   &  0.42  &   0.10   &  0.60    &   33.60 &   33.45  &  33.71   &  0.70 &    0.52  &   1.15    \\
   34.06  &  34.00   & 34.40   &  0.34  &   0.15   &  0.51    &   34.10 &   33.93  &  34.36   &  0.36 &    0.17  &   0.74    \\
   34.94  &  34.80   & 35.10   &  0.33  &   0.16   &  0.47    &   34.90 &   34.67  &  35.35   &  1.36 &    0.63  &   1.88    \\
   35.90  &  35.80   & 36.10   &  0.30  &   0.22   &  0.65    &   35.90 &   35.76  &  36.20   &  0.53 &    0.37  &   0.88    \\
   36.52  &  36.20   & 36.90   &  0.41  &   0.17   &  0.65    &   36.50 &   36.44  &  36.72   &  0.39 &    0.25  &   0.97    \\
   39.84  &  39.60   & 40.00   &  0.41  &   0.12   &  0.89    &   39.80 &   39.44  &  40.36   &  0.74 &    0.21  &   2.57    \\
   40.59  &  40.40   & 40.70   &  0.36  &   0.12   &  0.49    &   40.50 &   40.34  &  40.80   &  0.93 &    0.53  &   1.53    \\
   41.83  &  41.50   & 42.20   &  0.30  &   0.13   &  0.51    &   41.80 &   41.52  &  42.11   &  0.72 &    0.42  &   1.73    \\
   43.07  &  42.70   & 43.40   &  0.24  &   0.10   &  0.45    &   43.00 &   42.55  &  43.07   &  0.89 &    0.51  &   1.59    \\
   43.76  &  43.20   & 44.00   &  0.22  &   0.07   &  0.39    &   43.80 &   43.30  &  44.05   &  0.78 &    0.41  &   3.01    \\
   44.53  &  44.20   & 45.00   &  0.32  &   0.13   &  0.54    &   44.70 &   44.39  &  45.13   &  0.58 &    0.42  &   1.16    \\

\hline
\end{tabular}
\end{center}
\end{table}
\clearpage
\clearpage

\clearpage

\clearpage
\clearpage

\clearpage

%


\clearpage
\clearpage
\clearpage

\clearpage

\clearpage

\clearpage

\clearpage
\clearpage
\clearpage

\clearpage
\clearpage


\clearpage

\clearpage

\clearpage

\clearpage
\clearpage
\clearpage

\bsp	
\label{lastpage}
\end{document}